\documentclass[11pt,pagebackref]{article}
\usepackage{graphicx} 

 \pdfoutput=1
\usepackage{jheppub} 

\usepackage{color}
\usepackage[utf8]{inputenc}  
\usepackage[english]{babel}
\usepackage{graphicx}
\usepackage{tikz}
\usepackage{tikz-cd}
\usepackage{circuitikz}
\usepackage{makecell}

\usepackage{amsmath,amsthm,amssymb}
\usepackage{appendix}
\usepackage{slashed}
\usepackage{cite}
\usepackage{multirow}

\usepackage{color}
\usepackage{empheq}

\makeatletter
\newcommand*{\textoverline}[1]{$\overline{\hbox{#1}}\m@th$}

\definecolor{myblue}{rgb}{.8, .8, 1}
\definecolor{myred}{rgb}{1,.8,.8}
\definecolor{mygreen}{rgb}{.8,1,.8}
\newlength\mytemplen
\newsavebox\mytempbox

\makeatletter
\newcommand\mybluebox{%
    \@ifnextchar[
       {\@mybluebox}%
       {\@mybluebox[0pt]}}

\def\@mybluebox[#1]{%
    \@ifnextchar[
       {\@@mybluebox[#1]}%
       {\@@mybluebox[#1][0pt]}}

\def\@@mybluebox[#1][#2]#3{
    \sbox\mytempbox{#3}%
    \mytemplen\ht\mytempbox
    \advance\mytemplen #1\relax
    \ht\mytempbox\mytemplen
    \mytemplen\dp\mytempbox
    \advance\mytemplen #2\relax
    \dp\mytempbox\mytemplen
    \colorbox{myblue}{\hspace{1em}\usebox{\mytempbox}\hspace{1em}}}

\newcommand\mygreenbox{%
    \@ifnextchar[
       {\@mygreenbox}%
       {\@mygreenbox[0pt]}}
       
\def\@mygreenbox[#1]{%
    \@ifnextchar[
       {\@@mygreenbox[#1]}%
       {\@@mygreenbox[#1][0pt]}}

\def\@@mygreenbox[#1][#2]#3{
    \sbox\mytempbox{#3}%
    \mytemplen\ht\mytempbox
    \advance\mytemplen #1\relax
    \ht\mytempbox\mytemplen
    \mytemplen\dp\mytempbox
    \advance\mytemplen #2\relax
    \dp\mytempbox\mytemplen
    \colorbox{mygreen}{\hspace{1em}\usebox{\mytempbox}\hspace{1em}}}

\newcommand\myredbox{%
    \@ifnextchar[
       {\@myredbox}%
       {\@myredbox[0pt]}}

\def\@myredbox[#1]{%
    \@ifnextchar[
       {\@@myredbox[#1]}%
       {\@@myredbox[#1][0pt]}}

\def\@@myredbox[#1][#2]#3{
    \sbox\mytempbox{#3}%
    \mytemplen\ht\mytempbox
    \advance\mytemplen #1\relax
    \ht\mytempbox\mytemplen
    \mytemplen\dp\mytempbox
    \advance\mytemplen #2\relax
    \dp\mytempbox\mytemplen
    \colorbox{myred}{\hspace{1em}\usebox{\mytempbox}\hspace{1em}}}

\makeatother

\newcommand*\dd{\mathop{}\!\mathrm{d}}

\newcommand*\de{\mathop{}\!\partial}

\newcommand*\al{\alpha}
\newcommand*\be{\beta}
\newcommand*\ga{\gamma}
\newcommand*\di{\delta}
\newcommand*\ep{\epsilon}

\newcommand*\vphi{\varphi}

\newcommand*\la{\lambda}

\newcommand*\om{\omega}
\newcommand*\thi{\theta}

\newcommand*\RR{\mathbb{R}}

\newcommand*\ZZ{\mathbb{Z}}

\newcommand*\calA{\mathcal{A}}
\newcommand*\calB{\mathcal{B}}

\newcommand*\calF{\mathcal{F}}

\newcommand*\calH{\mathcal{H}}

\newcommand*\calL{\mathcal{L}}

\newcommand*\calO{\mathcal{O}}

\newcommand*\calS{\mathcal{S}}
\newcommand*\calT{\mathcal{T}}

\newcommand*\calV{\mathcal{V}}

\newcommand*\Arf{{\mathrm{Arf}}}
\newcommand*\frg{\mathfrak{g}}
\newcommand*\Tr{\mathop{\rm tr}}
\newcommand*\im{\mathop{\mathrm{Im}}}
\newcommand*\re{\mathop{\mathrm{Re}}}

\title{\boldmath Anomalies and Persistent Order in the Chiral Gross-Neveu model}

\author[a,c]{Riccardo Ciccone,}
\author[b,c]{Lorenzo Di Pietro,}
\author[a,c]{Marco Serone}

\affiliation[a]{SISSA,\\ via Bonomea 265, 34136 Trieste, Italy,}
\affiliation[b]{Dipartimento di Fisica, Università di Trieste,\\ Strada Costiera 11, I-34151 Trieste, Italy}
\affiliation[c]{INFN, Sezione di Trieste,\\ via Valerio 2, 34127 Trieste, Italy}

\emailAdd{rciccone@sissa.it}
\emailAdd{ldipietro@units.it}
\emailAdd{serone@sissa.it}

\abstract{We study the $2d$ chiral Gross-Neveu model at finite temperature $T$ and chemical potential $\mu$. 
The analysis is performed by relating the theory to a $SU(N)\times U(1)$ Wess-Zumino-Witten model with appropriate levels
and global identifications necessary to keep track of the fermion spin structures. 
At $\mu=0$ we show that a certain $\ZZ_2$-valued 't Hooft anomaly forbids the system to be trivially gapped when 
fermions are periodic along the thermal circle for any $N$ and any $T>0$. 
We also study the two-point function of a certain composite fermion operator which allows us to determine 
the remnants for $T>0$ of the inhomogeneous chiral phase configuration found at $T=0$ for any $N$ and any $\mu$.
The inhomogeneous configuration decays exponentially at large distances for anti-periodic fermions while it persists for $T>0$ and any $\mu$ for periodic fermions, as expected
from anomaly considerations. A large $N$ analysis confirms the above findings.}

\begin{document} 
\maketitle
\flushbottom

\section{Introduction}

Understanding the phase diagram of $4d$ strongly coupled quantum field theories (QFTs)  at finite temperature and density is notoriously hard,
because of the limited availability of analytical tools. Even numerical methods, notably lattice Monte Carlo methods, are hard to implement at finite density
due to the sign problem. For this reason, it is interesting to learn lessons from specific QFTs in lower dimensions, where more analytic control is available. 
A particularly interesting model of this kind is the chiral Gross-Neveu (cGN) model \cite{Gross:1974jv}, a $2d$ theory of $N$ massless Dirac fermions $\psi_a$ $(a=1,\ldots, N)$ with four-fermion interactions preserving a $U(N)_V\times U(1)_A$ global symmetry. The theory is UV-free, it undergoes dynamical mass generation and is gapless in the IR.\footnote{The cGN model should not be confused with  its better-known version (GN model) which has maximal $O(2N)$ vector symmetry, is gapped in the IR and shows spontaneous breaking of a $\mathbb{ Z}_2$ chiral symmetry.}

In the large $N$ limit, the phase diagram of the cGN model at finite temperature $T$ and in the presence of a $U(1)_V$ fermion number chemical potential $\mu$ 
has been shown to exhibit ``chiral spiral'' configurations with a spatially periodic phase with $\mu$-dependent period at low $T$ for any $\mu>0$  \cite{Basar:2009fg}, (see \cite{Wolff:1985av,Barducci:1994cb} for earlier results obtained assuming a translation invariant vacuum). 
At $N=\infty$ the $U(1)_A$ symmetry is spontaneously broken  (i.e. we have a strict long-range ordered phase), evading the
no-go theorem forbidding the breakdown of continuous symmetries in $d=2$ dimensions  \cite{Mermin:1966fe,Hohenberg:1967zz,Coleman:1973ci}.
For $\mu=0$ we just have the breaking of $U(1)_A$, whereas for $\mu \neq 0$  a linear combination of the $U(1)_A$ symmetry and of spatial translations is spontaneously broken.

More recently, it has been shown that the above inhomogeneous configurations persist at {\it finite} $N$, for any $N$,
at $T=0$ \cite{Ciccone:2022zkg}. The analysis in \cite{Ciccone:2022zkg} was based on non-abelian bosonization, relating the cGN theory to a $U(N)_1$ Wess-Zumino-Witten (WZW) model deformed by current-current interactions. In this description, the appearance of the inhomogeneous phase is surprisingly simple since it is encoded in the compact scalar $\phi$ describing the $U(1)\subset U(N)$ sector of the theory, which is free.  At finite $N$ strict long-range order is forbidden and the chiral wave corresponds to a quasi long-range ordered gapless phase  \cite{Berezinsky:1970fr,Berezinsky:1972rfj,Kosterlitz:1973xp}.

The aim of this work is to extend the results of \cite{Ciccone:2022zkg} to finite temperature. 
At finite $N$ and $T\neq 0$, the theory reduces to a quantum mechanical system.
According to common lore, no ordered phase for both continuum or discrete symmetries can arise in quantum mechanics. 
However, it is known that this common lore can fail giving rise to the phenomenon of persistent order, i.e. situations in which the ordered phase
persists up to arbitrarily high temperatures.
Persistent order can occur when a theory is affected by 't Hooft anomalies which are not broken upon compactification 
to the thermal circle. By now several examples of theories featuring a form of persistent order have been discussed, see e.g. \cite{Gaiotto:2017yup,Komargodski:2017dmc,Tanizaki:2017mtm}.
Most models with persistent order feature one-form symmetries \cite{Gaiotto:2014kfa}, but this is not a necessary requirement, provided fields are appropriately twisted \cite{Tanizaki:2017qhf}.\footnote{The persistent order discussed here should be distinguished from the more drastic phenomenon of symmetry non restoration at arbitrarily high temperatures
\cite{Weinberg:1974hy} where the non restoration of the symmetry is not related to 't Hooft anomalies or to the presence of (real or complex) chemical potentials/twists. 
To our knowledge no UV complete, unitary, and local QFT which features this phenomenon has been found so far. See e.g. \cite{Chai:2020onq,Chai:2021djc,Chai:2021tpt} for recent attempts. Persistent order can also occur in quantum phase transitions at zero temperature, as a function of a coupling constant, as in the example of deconfined criticality \cite{PhysRevB.70.144407,doi:10.1126/science.1091806}.}
A notable example of how twists can drastically change the thermal sum over states in a Hilbert space is provided by $(-1)^F$, which in supersymmetric theories turn the partition functions to the Witten index \cite{Witten:1982df}.

An important role in our work is played by a certain $\ZZ_2$-valued 't Hooft anomaly which is present in the cGN model for any $N$.
The anomaly affects a dihedral subgroup symmetry $D_8$ of the fermion theory, which involves charge conjugation and $\ZZ_2$ subgroups of the $U(1)_V\times U(1)_A$  axial symmetries. It manifests as a projective realization of $D_8$ on the Hilbert space.
When fermions are in the Neveu-Schwarz (NS) spin structure (anti-periodic), $D_8$ is realized linearly and for any $T>0$ the quasi long-range ordered phase is destroyed. However, 
for fermions in the Ramond (R) spin structure (periodic), $D_8$ is projectively realized and as a consequence the theory cannot be trivially gapped. 
The different realization of the action of a symmetry between NS and R spin structures is
not unexpected: free NS and R fermions on $S^1$ lead to a unique and degenerate ground state, respectively. Hence, anomalies are expected to persist 
more easily in R spin structures, see e.g. \cite{Delmastro:2021xox} for a recent analysis.\footnote{We have not investigated if and how, by turning on appropriate backgrounds, the anomaly 
becomes visible in the NS sector. We also do not discuss in this paper the nature of the anomaly by means of the general classification in terms of cobordism groups \cite{Kapustin:2014dxa}.}

{The chemical potential $\mu$ explicitly breaks charge conjugation, but luckily enough its presence only affects the free theory sector of the theory, so we can check that it does not modify the
spin structure dependence above.} We explicitly show the presence of persistent order by computing the two-point function $\langle {\cal O}_F {\cal O}_F\rangle$, where $\calO_F=\det(\psi^\dag_+\psi_-)$
is a determinant operator, $\psi_\pm^a$ being the two Weyl components of the Dirac fermions $\psi^a$. This operator is special because it is $U(1)_A$ charged, but
it is neutral under the subgroup $\ZZ_N^A\subset U(1)_A$. $\ZZ_N^A$ neutrality is important, because it allows us to compute the large distance behavior of the correlator for $T\ll M$, where $M$
is the mass gap of the strongly coupled sector of the theory (the $SU(N)_1$ sector in the WZW model description). We find the following behavior for the two-point function when $|x| T \to \infty$:
    \begin{equation}
  \begin{split}  \label{eq:NSRFinal}
    \langle \calO_F(x)\calO^\dag_F(0)\rangle_{NS} & \sim e^{-\frac{|x|}{\xi_T}}e^{2iN\mu'x}\,, \\
       \langle \calO_F(x)\calO^\dag_F(0)\rangle_{R} & \sim e^{2i N\mu' x}\,,
  \end{split}  \end{equation}
where $\xi_T^{-1}\propto N T$ and $\mu'\propto \mu$  (see section \ref{sec:FiniteT} for the precise relation and further details).
The chiral spiral modulation is visible only at intermediate scales in the NS case, provided that $N\mu'\gtrsim \xi_T^{-1}$, while it does not decay and  is visible at 
arbitrarily large distances and finite temperature in the R case. In the last case, we see that ${\cal O}_F$ develops a vacuum expectation value.
As a consequence, the quasi long-range ordered phase for $U(1)_A$ turns into strict order at finite temperature $T>0$ for R fermions.
The approximations made to get \eqref{eq:NSRFinal}  do not hold for $T\geq M$, yet we expect an ordered phase for {\it any} $T>0$ due to the anomaly.

{We emphasize that the very same anomaly affects the GN model for any $N$. As a consequence, persistent order of the chiral $\mathbb{Z}_2$ symmetry should occur 
when fermions are periodic on $S^1$. This is not totally unexpected, as the persistence of the order for periodic fermions, 
under the assumption of homogeneity, was already observed in the GN model at large $N$ for $\mu=0$ in \cite{Wolff:1985av}.}

We start in section \ref{sec:JJbarRev} by reviewing the results of \cite{Ciccone:2022zkg} for a non-compact Euclidean space-time $M_2=\mathbb{R}^2$.
Care has to be paid to global considerations when applying non-abelian bosonization in the presence of compact spaces with homologically non-trivial cycles. 
We discuss these issues in section \ref{sec:freefermibosonization}, where we find the precise form of the bosonic theory ${\cal B}$ dual to the cGN model.
The precise form depends on whether $N$ is even or odd, but in both cases it can be written as certain orbifolds of the $SU(N)_1$ and $U(1)_N$ WZW models, which also involve
the Arf invariant, see \eqref{eq:FpNodd} and \eqref{eq:FSUN}.

Anomalies are discussed in section \ref{sec:anomalies}. After reviewing a known $\ZZ_N$ 't Hooft anomaly present in the $SU(N)_1$ model in isolation \cite{Tanizaki:2018xto,Yao:2018kel}, 
we discuss in section \ref{subsec:compacttheta} a generalization of a $\ZZ_2$ anomaly found in \cite{Alavirad:2019iea} in the compact boson which involves momentum, winding and charge conjugation $\ZZ_2$ symmetries. We then show in section \ref{subsec:FermPersOrder} that the above anomaly is responsible for a symmetry extension of $\ZZ_2^3$ to the dihedral group $D_8$.\footnote{See e.g.  \cite{Lanzetta:2022lze,Gaiotto:2017yup,Thorngren:2021yso} for works where $D_8$ appears as an anomaly extension.}
However, the realization of $D_8$ in the fermionic theory is linear or projective depending on the spin structure along a given cycle. 
In section \ref{sec:FiniteT} we study where and when the chiral spiral and persistent order reveal themselves in two-point functions of local operators in the cGN model.
Since locally the bosonic theory is a product of a strongly coupled $SU(N)$ sector and a free compact scalar $\phi$, two-point functions of fermion operators
which upon bosonization are expressed purely in terms of $\phi$ can be exactly computed in the large distance limit when $T\ll M$.
This is in particular the case of the determinant operator ${\cal O}_F$ introduced above.
The ending result of this computation is \eqref{eq:NSRFinal}. 
In section \ref{sec: largeN} we show how the chiral spiral and persistent order arise in the large $N$ limit for $R$ fermions, providing a good consistency check
of the previous results. We conclude in section \ref{sec:conclusions}.

Several appendices complement the work. In appendix \ref{app:gaugings} we review discrete gaugings between bosonic theories and how to bosonize a fermionic theory on a general $2d$ closed manifold.
Basic facts about the free compact scalar and its relation to rational $U(1)$ WZW models when its squared radius is
rational are reviewed in appendix \ref{app:u1boson}. We prove \eqref{eq:spinunitaryrelation} in appendix \ref{app:spinunitary} by using the relation between chiral algebras in $2d$ and Chern-Simons theories in $3d$ \cite{Witten:1988hf}. 
In appendix \ref{app:PotentialAbBos} we show the appearance of $N$ classical vacua for the deformed $SU(N)_1$ WZW model, parametrized by $\langle {\rm tr}\, U\rangle$, and determine
the exact value for $N=2$ using integrability \cite{Lukyanov:1996jj}. We finally collect in appendix \ref{app:twistedapp} technical details for the computation of the two-point function $\langle {\cal O}_F {\cal O}_F^\dag\rangle$ performed in
section \ref{sec:FiniteT}.

\section{The chiral Gross-Neveu model as $J\bar{J}$ deformations of WZW models}
\label{sec:JJbarRev}

In this section, we briefly review the physical context of our set-up and summarize the findings of \cite{Ciccone:2022zkg} valid on $M_2=\mathbb{R}^2$.
The chiral Gross-Neveu (cGN) model \cite{Gross:1974jv} is a field theory of $N$ massless Dirac fermions in $2d$ interacting according to
\begin{equation}\label{eq:LcGN}
    \calL_{\text{cGN}}=i\psi_{+a}^\dag\de_-\psi^a_+-i\psi_{-a}^\dag\de_+\psi_-^a{+}\frac{\la_s}{N}|\psi_{-a}^\dag\psi_+^a|^2-\frac{\la_v}{N^2}(\psi^\dag_{+a}\psi_+^a)(\psi^\dag_{-b}\psi^b_-),
\end{equation}
where $x^\pm=(x^1\mp i x^2)/2$, $a=1,\dots,N$ are flavor indices, and $\psi_\pm^a$ are the two Weyl components of the $N$ Dirac fermions.
Using Fierz identities \eqref{eq:LcGN} can be rewritten as 
\begin{equation}\label{eq:LcGN2}
    \calL_{cGN}=i\psi_{+a}^\dag\de_-\psi_+^a-i\psi_{-a}^\dag\de_+\psi_-^a+ \frac{\la}{N}J_+^AJ_-^A +\frac{\la'}{N^2}J_+J_-\,,
\end{equation}
where\footnote{Note that \eqref{eq:currents} corrects a typo in eq.(5) of \cite{Ciccone:2022zkg}, where the $\pm$ in front of the currents was missing.} 
\begin{equation} \label{eq:currents}
    J_\pm^A={\pm} \psi_{a\pm}^\dag\left(T^A\right)^a{}_b\psi^b_\pm\,,\qquad J_\pm={\pm}\psi_{a\pm}^\dag\psi_\pm^a\,,
\end{equation}
are the $SU(N)$ and $U(1)$ chiral currents, respectively, with $T^A$, $A=1,\dots,N^2-1$, being the generators of $SU(N)$ in the fundamental representation, chosen with the normalization $\Tr(T^AT^B)=\frac12\delta^{AB}$. The couplings $\la_s,\,\la_v,\,\la,\,\la'$ are related by 
\begin{equation}\label{eq:fierzedcouplings}
    \la=2\la_s\,,\qquad \la'=\la_v+\la_s\,.
\end{equation}
The last term in \eqref{eq:LcGN} can be neglected at leading order in a large $N$ analysis, but it should be kept at finite $N$.
The $O(2N)_L\times O(2N)_R$  global symmetry of the $N$ free Dirac fermions
is broken by the interaction to\footnote{The theory has in addition a time reversal symmetry $\ZZ_2^T$ which does not play any role in the analysis that follows and hence will be neglected.}
\begin{equation}\label{eq:glsymfermions}
    G(N\text{-flavor cGN} )=\frac{U(N)_V\times U(1)_A}{\ZZ_2} \rtimes \ZZ_2^C \,,  
\end{equation} 
where $\ZZ_2^C$ is charge conjugation, see section \ref{subsec:globalsymms} for an explanation of how charge conjugation arises from $O(2N)_L\times O(2N)_R$.
The symmetry action on the fields is
\begin{equation}
\begin{cases}
    U(N)_V:& \qquad \psi^a_\pm\mapsto \mathcal{V}^a{}_b\psi^b_\pm\,,\qquad\, \mathcal{V}\in U(N)\,,\\
    U(1)_A: &\qquad  \psi^a_\pm\mapsto {e}^{\pm i \alpha}\psi^a_\pm\,,\qquad  \alpha\sim\alpha+2\pi\,,\\
    \ZZ_2^C: &\qquad \psi^a_\pm\mapsto \psi_{\pm a}^{\dagger}\,.
    \end{cases}
    \label{eq:FermTrans}
\end{equation}
The $\ZZ_2$ quotient in \eqref{eq:glsymfermions} arises because the diagonal $\ZZ_2$ in $U(N)_V\times U(1)_A$, with $\mathcal{V}=-\mathbf{1}$ and $\al=\pi$, is trivial on fields. 
The corresponding ``off-diagonal'' $\ZZ_2$ is identified with fermion parity $\ZZ_2^F$, $\psi_\pm^a\mapsto -\psi_\pm^a$.

Using non-abelian bosonization \cite{Witten:1983ar}, the cGN model \eqref{eq:LcGN2} can be expressed as a $J\bar{J}$ deformation of a $U(N)_1$ WZW model. Locally this model is described
by an $SU(N)$ matrix $U$ and a free compact scalar $\phi$ parametrizing the $U(1)$ factor, 
\begin{equation}\label{eq:bosonizedthirring}
    {\cal L}[U,\phi]={\cal L}_0[U,\phi] + \frac{\lambda}N  {J^A_+J^A_-} + \frac{\lambda'}{N^2}{J_+J_-}~,
\end{equation}
where 
\begin{align}
{\cal L}_0[U,\phi]  & =  \frac{1}{8\pi} \partial_+ \phi \,\partial_- \phi + \frac1{8\pi} {\rm tr} (\partial_+ U^\dag \partial_- U+\partial_- U^\dag \partial_+ U)+ {\cal L}_{\text{WZ}}^{SU(N)_1}~,
\end{align}
is the Lagrangian of the undeformed theory, with ${\cal L}_{\text{WZ}}^{SU(N)_1}$ being the level $k=1$ $SU(N)$ Wess-Zumino term, and
\begin{equation}
    J^A_+  =\frac{i}{2\pi}\Tr\left( U^\dag (\partial_+U)T^A\right)\,,\qquad J^A_-=\frac{i}{2\pi}\Tr\left((\partial_-U) U^\dag T^A\right)\,,\qquad 
  J_\pm  =- \frac{\sqrt{N}}{4\pi}\partial_\pm\phi~,
    \label{eq:bosecurrents}
\end{equation}
are the bosonized $SU(N)$ and $U(1)$ currents.  
Under the continuous global symmetries in \eqref{eq:glsymfermions} the fields transform as
\begin{equation}
U(N)_V:\begin{cases}U\mapsto \mathcal{V}^\dag U \mathcal{V}\,,\\
\phi\mapsto\phi \,,\\
\tilde\phi\mapsto\tilde\phi+\frac{4}{\sqrt{N}}\arg\det(\mathcal{V})\,,\end{cases}\qquad 
U(1)_A: \begin{cases} U\mapsto U\,,\\ \phi\mapsto\phi-{2}{\sqrt{N}}\alpha\,,\\ \tilde\phi\mapsto\tilde\phi\,, \end{cases} 
\end{equation}
while under charge conjugation we have
\begin{equation}\label{eq:z2cz2t}
\ZZ_2^C:\begin{cases}U\mapsto U^*\,,\\
\,\phi\mapsto- \phi\,,\\
\,\tilde\phi\mapsto-\tilde\phi\,,\end{cases}
\end{equation}
where $\tilde\phi$ is the scalar dual to $\phi$. The deformation 
rescales the radius of the compact scalar $\phi$ by a factor 
\begin{equation}\label{eq:scalarradius}
    R_0\longrightarrow R=R_0\sqrt{1+\frac{\la'}{2\pi N}}\,,
\end{equation}
where $R_0$ is the radius before the deformation.

The key simplification that makes non-abelian bosonization useful occurs when we add a chemical potential for the $U(1)_V$ charge.
Indeed, the corresponding Lagrangian term is mapped to
\begin{equation}\label{eq:Lmu0}
    {\cal L}_\mu= \mu(\psi^\dag_{+a}\psi_+^a+\psi^\dag_{-a}\psi_-^a)\longrightarrow -\mu\frac{\sqrt{N}}{2\pi}\Big(1+\frac{\la'}{2\pi N}\Big)^{-\frac 12}\partial_1\phi
\end{equation}
and does not depend on the interacting $SU(N)$ degrees of freedom. The additional term (\ref{eq:Lmu0}) immediately implies that on the vacuum  $\langle \partial_1 \phi\rangle\neq 0$
and a chiral spiral configuration follows for any finite $N$ \cite{Ciccone:2022zkg}. This is detectable from the large distance behavior of the two-point function
of $\psi_{-a}^\dag\psi_+^a(x)$, assuming that in the deformed $SU(N)_1$ theory $\Tr U$ develops a non-vanishing expectation value,
\begin{align}\label{eq:twopoint00}
   \langle \psi_{-a}^\dag\psi_+^a(x)\psi_{+b}^\dag\psi_-^b(0)\rangle \underset{|x|\to\infty}{\longrightarrow}  
      |\langle\Tr U\rangle|^2 e^{2i\mu'x^1}|x|^{-\frac{2}{N(1+\lambda'/2\pi N)}}\,,
\end{align}
where we have defined
\begin{equation}\label{eq:muprime}
    \mu'=\mu \left(1+\frac{\la'}{2\pi N}\right)^{-1/2}\,.
\end{equation}
While on $M_2=\mathbb{R}^2$ the compact scalar $\phi$ and the $SU(N)$ matrix $U$ are effectively decoupled from each other, this is not the case in compact space due to global identifications and because $M_2$ can have different spin structures. We discuss in the next sections how the above results generalize when $M_2$ is a compact space. 
We also provide an explanation for the assumption $\langle \Tr U\rangle\neq  0$ in terms of 't Hooft anomaly matching conditions.

\section{Non-abelian bosonization revisited}
\label{sec:freefermibosonization}

In this section, we rediscuss non-abelian bosonization from a modern point of view, paying attention to global aspects. These are important when discussing the cGN model, like any other theory, on non-trivial spaces. Most of the considerations in this section are independent of the $J\bar J$ deformation, so we focus in what follows on the correspondence between free fermions and undeformed WZW models.

It is well-known that $2N$ free massless Majorana fermions bosonize to the (diagonal) ${Spin}(2N)_1$ WZW model \cite{Witten:1983ar}. 
The precise correspondence, valid on arbitrary spin manifolds, requires some specifications. For example, a fermionic theory depends on the spin structure on $M_2$, while the ${Spin}(2N)_1$ WZW model, being a bosonic theory, cannot. The spin structure dependence of the fermionic theory is attached to the bosonic one by stacking the latter with the
topological theory given by the Arf invariant. We review the basics of this construction in appendix \ref{app:gaugings}. In the notation of appendix \ref{app:gaugings}, if we denote by
$\calB$ the diagonal ${Spin}(2N)_1$ WZW model, we take as the free fermion theory $\calF'$ and the two theories are related as \cite{Fukusumi:2021zme}\footnote{In general, there are two fermionizations of a bosonic theory $\calB$ with respect to a given $\ZZ_2$ symmetry, denoted by $\calF$ and $\calF'=\calF\times\Arf$ in appendix \ref{app:gaugings}. To check which is the correct choice for a fixed $\calB$ one can look at boundary states, as done in \cite{Fukusumi:2021zme}. In our setup we deal with local operators on manifolds without boundary and there is no real distinction between $\calF$ and $\calF'$. However, certain $\ZZ_2$ anomalies distinguish the two theories. This will be discussed in detail in section \ref{sec:anomalies}.}  
\begin{equation}\label{eq:fermionization0}
    \calF'=({\calB\times\Arf})/{\ZZ_2^L}\times\Arf\,,
\end{equation}
which is a combination of \eqref{eq:fermionization} and \eqref{eq:fermionizationarf}.
The $\ZZ_2^L$ symmetry in \eqref{eq:fermionization0} is 
a $\ZZ_2$ subgroup of the center of the $Spin(2N)_L$ symmetry of the ${Spin}(2N)_1$ WZW model. Its action on the $\Arf$ theory is given in \eqref{eq:fermionizationarf}.
In turn, the $\ZZ_2$ subgroup of  $Z[Spin(2N)_L]$ is the one that assigns charge $-1$ to operators transforming in the left-handed spinor representations 
 (recall that $Z[Spin(2N)]$ equals $\ZZ_2\times\ZZ_2$ for even $N$ and $\ZZ_4$ for odd $N$).
The inverse relation between $\calB$ and $\calF'$ is given by \eqref{eq:BFprimeRel}.

It is also well-known that there are different ways to perform bosonization of free fermions, according to the amount of global symmetry of the bosonized theory one wishes to keep manifest. While the bosonized theory $\calB$ is unique, it can be described using different variables. For instance, in abelian bosonization one bosonizes each Dirac fermion independently and keeps manifest only the Cartan subgroup $U(1)^{N}\subset Spin(2N)$. Alternatively, one can perform non-abelian bosonization keeping manifest a $U(N)\subset Spin(2N)$ or the whole $Spin(2N)$
symmetry. Given that we will eventually restrict to the unbroken symmetries \eqref{eq:glsymfermions} left after the $J\bar J$ deformation, we focus on 
the $U(N)$ description. Nevertheless, the latter must be equivalent to the more general $Spin(2N)$ one. We argue 
that
\begin{equation}\label{eq:spinunitaryrelation}
    \calB={Spin}(2N)_1 =\begin{cases}\dfrac{SU(N)_1 \times U(1)_N}{\ZZ_{N/2}^L}& \quad N\text{ even}\,,\\
    \dfrac{SU(N)_1 \times U(1)_{4N} }{\ZZ_N^L}
    & \quad N\text{ odd}\,,
    \end{cases}
\end{equation}
where $U(1)_{k'}$ denotes the compact boson with squared radius $R^2=k'$.\footnote{In our conventions, the self-$T$-dual radius for the compact boson is $R=\sqrt{2}$.} 
For both $N$ even and odd, the $\ZZ_n$ quotients are the diagonal ones between the one contained in $Z[SU(n)_L]$ and the $\ZZ_n\subset U(1)_L$.\footnote{One could alternatively use the action on the anti-holomorphic sector $(R)$. The ending result is the same.} The $\ZZ_2$ subgroup of $Z[Spin(2N)_L]$ that gets gauged according to \eqref{eq:fermionization0} is mapped via \eqref{eq:spinunitaryrelation} to $\ZZ_2 \subset U(1)_L$ for $N$ odd. For $N$ even, it is given by the diagonal $\ZZ_N$ generator
of $Z[SU(N)_L]$ and $\ZZ_N\subset U(1)_L$, which in the $\ZZ_{N/2}^L$-gauged theory has order $2$.
The discrepancy in (\ref{eq:spinunitaryrelation}) between the even and odd case is due to the fact that the $R^2=k'$ compact boson is not a diagonal RCFT when $k'$ is odd. See appendix \ref{app:u1boson} for details, appendix \ref{app:spinunitary} for a proof of \eqref{eq:spinunitaryrelation} when $M_2 = T^2$, and appendix \ref{app:spinbkg} for the generalization to the case of nontrivial $\ZZ_2^L\subset Z[Spin(2N)_L]$ backgrounds.

The relation \eqref{eq:fermionization0} can also be written in a form where only $U(1)_N$ factors are involved in the bosonic theory, for any $N$.
Indeed, when $N$ is odd, we can gauge $\ZZ_{2}^L$ before $\ZZ_{N}^L$. Since the action of $\ZZ_N^L$ is identical to that of $\ZZ_N^P$ in the $U(1)_{4N}$ theory (see \eqref{eq:continuousglobal} and the sentence below), we can use \eqref{eq:U14nvsU1n} to get
\begin{equation}\label{eq:FpNodd}
{\calF'} = \frac{\calB'\times\Arf}{\widehat \ZZ_2^{W}}\,, \qquad \calB' = \dfrac{SU(N)_1\times U(1)_{N}}{\widehat \ZZ_N}\,, \qquad \text{$N$ odd}\,,
\end{equation}
where $\widehat \ZZ_N$ is the diagonal between $Z[SU(N)_L]$ and $\widehat \ZZ_N^W\subset U(1)_W$.
For $N$ even, the action of $\ZZ_{N/2}^L$ is identical to that of $\ZZ_{N/2}^P$ in the $U(1)_N$ theory (see again \eqref{eq:continuousglobal}), so we can write
\begin{equation}
{\calF'} = \left(\dfrac{SU(N)_1\times U(1)_{N}}{\ZZ_{N/2}'}\times \Arf\right)  \Big/\ZZ_2^L\times \Arf\,, \quad \text{$N$ even}\,,
\label{eq:FSUN}
\end{equation}
where $\ZZ_{N/2}'$ is the diagonal between  $Z[SU(N)_L]$ and $\ZZ_{N/2}^P\subset U(1)_P$.

\subsection{Global symmetries} \label{subsec:globalsymms}

The group-like symmetries of the $Spin(2N)_1$ WZW theory are given by
\begin{equation}\label{eq:spin2nwzwsymm}
    G(Spin(2N)_1)=    \dfrac{Spin(2N)_L\times Spin(2N)_R}{Z(Spin(2N)_V)}\rtimes \ZZ_2^K \,,
\end{equation}
where the subscript $V$ denotes the diagonal $Spin(2N)_V\subset Spin(2N)_L\times Spin(2N)_R$.
Explicitly, their action on the matrix field $g$ is
\begin{equation}
    \begin{cases}
        Spin(2N)_L:& g\mapsto L g\,,\\
        Spin(2N)_R:& g\mapsto g R^t\,,\\
        Spin(2N)_V:& g\mapsto V g V^t\,.
    \end{cases}
\end{equation}
The factor $\ZZ_2^K$ is a $\ZZ_2$ outer automorphism of the algebra that exchanges the spinor and the conjugate spinor representations.
It does not act on the matrix field of the theory because it sits in a real representation of the symmetry algebra. 
For odd $N$ $\ZZ_2^K$ can be identified with complex conjugation.

The symmetry group of $2N$ free Majorana fermions $\xi^a$ ($a=1,\ldots, 2N)$ is 
\begin{equation}
\begin{aligned}
    G(2N\,\text{Majorana})&=O(2N)_L\times O(2N)_R \\&=\begin{cases}
    (SO(2N)_L\times SO(2N)_R)\rtimes (\ZZ_2^{C_L}\times \ZZ_2^{C_R}) &N\text{ odd}\,,\\
    (SO(2N)_L\times SO(2N)_R)\rtimes (\ZZ_2^{K_L}\times \ZZ_2^{K_R}) &N\text{ even}\,,
    \end{cases}
    \end{aligned}
    \label{eq:Gfermion}
\end{equation}
where we have also broken down $O(2N)_{L,R}$ into their connected and disconnected parts.  The matrix $-\mathbf{1}$ is inside $SO(2N)$, therefore fermion parity $\ZZ_2^F$ is the diagonal subgroup of $\ZZ_2^{F_L}\times\ZZ_2^{F_R}= Z(SO(2N)_L\times SO(2N)_R)$. Note that there are differences between $N$ even and odd. For $N$ odd, we can define the matrix 
$C=\left(\begin{smallmatrix}\mathbf{1}_N &0\\ 0 &- \mathbf{1}_N \end{smallmatrix}\right)$, that has determinant $-1$. $C$ does not commute with $SO(2N)$ rotations, hence the semidirect product, leaving $SO(2N)$ as a normal subgroup of $O(2N)$. Its physical meaning is charge conjugation.\footnote{On Dirac fermions $\psi_\pm^a=\xi_\pm^{a}+i\xi_\pm^{a+N}$ the matrix $C$ acts as \eqref{eq:FermTrans}.}
For $N$ even, instead, these charge conjugation matrices are inside $SO(2N)$. We can however define a reflection matrix $K$, for instance as the unit matrix with the first entry replaced by $-1$. For $N$ odd $C$ is equivalent to $K$ via an $SO(2N)$ rotation, so we can always use $K$ to extend $SO(2N)$ to $O(2N)$. The explicit action of \eqref{eq:Gfermion} on the fermions is
\begin{equation}
    \begin{cases}
        O(2N)_L: & \xi_+^a\mapsto \xi_+^a\,,\qquad\qquad\! \xi_-^a\mapsto L^a{}_b\xi_-^b\,,\\
        O(2N)_R: & \xi_+^a\mapsto R^a{}_b\xi_+^a\,,\qquad \xi_-^a\mapsto \xi_-^a\,.
    \end{cases}
\end{equation}

In order for the bosonization procedure to be consistent, upon gauging $\ZZ_2^F$ in the fermionic theory we should obtain the symmetry group \eqref{eq:spin2nwzwsymm}. 
The symmetry $\ZZ_2^\vee$ dual of $\ZZ_2^F$, is the factor that extends $SO(2N)_L$ to $Spin(2N)_L$. This is easily seen by noting that the spinorial characters
of $Spin(2N)_1$ are in the twisted sectors of the $\ZZ_2^F$ gauged theory. The symmetry group after the gauging is then 
\begin{equation}
    G(Spin(2N)_1)=\frac{Spin(2N)_L\times Spin(2N)_R}{Z(Spin(2N)_V)}\rtimes \ZZ_2^K\,,
\end{equation}
where $\ZZ_2^K$ is the diagonal between $\ZZ_2^{K_L}$ and $\ZZ_2^{K_R}$,\footnote{The orthogonal combination becomes a non-invertible symmetry \cite{Komargodski:2020mxz}, which we neglect from now on.} and coincides with \eqref{eq:spin2nwzwsymm}. 

It is useful to also report the manifest global symmetries in the $SU(N)$ description and its explicit realization in the fields.
The global symmetry group of the $SU(N)_1$ WZW model is
\begin{equation}\label{eq:sunwzwsymm}
    G(SU(N)_1)=\begin{cases}
        \dfrac{SU(2)_L\times SU(2)_R}{\ZZ_2^V} & N=2\,,\\
        \dfrac{SU(N)_L\times SU(N)_R}{\ZZ_N^V}\rtimes \ZZ_2^C & N>2\,,
    \end{cases}
\end{equation}
where $\ZZ_N^V=Z(SU(N)_V)$ is the center of the diagonal $SU(N)_V\subset SU(N)_L\times SU(N)_R$ and $\ZZ_2^C$ is charge conjugation. 
Notice that for $N=2$ there is no charge conjugation symmetry acting on the matrix field $U$ because the latter is in the bifundamental representation of $SU(2)_L\times SU(2)_R$, which is pseudo-real. 

The $U(1)$ compact boson has global symmetry group
\begin{equation}
    G_R=(U(1)_P\times U(1)_W)\rtimes \ZZ_2^{C}\,, 
\end{equation}
with $U(1)_P$, $U(1)_W$ and $\ZZ_2^C$ respectively the momentum and winding symmetries, and charge conjugation. 
The action of the symmetry group on the $SU(N)$ field $U$ and on the compact scalar $\phi$ and its dual $\tilde\phi$ is in total 
\begin{equation}\label{eq:continuousglobal}
        \begin{cases}
        SU(N)_L: & U\mapsto L U\,,\\
        SU(N)_R: & U\mapsto U R^\dag\,,\\
        SU(N)_V: & U\mapsto V U V^\dag\,,\\
    \end{cases}\qquad
    \begin{cases}
     U(1)_P:& \phi\mapsto \phi+ \al_P R\,,\qquad \tilde\phi\mapsto\tilde\phi\,,\\
  U(1)_W:& \phi\mapsto \phi\,,\qquad\qquad\quad \tilde\phi\mapsto\tilde\phi+2\al_W/R\,,
    \end{cases}
\end{equation}
where $\al_P\sim\al_P+2\pi$ and $\al_W\sim\al_W+2\pi$. Charge conjugation  acts as in \eqref{eq:z2cz2t}.  

For $R^2=k'\equiv 2p'/p$, $\gcd(p,p')=1$,  the linear combinations of $U(1)_P$ and $U(1)_W$ given by $\alpha_P = p \alpha_L$, $\alpha_W = p'\alpha_L$  or $\alpha_P = p \alpha_R$, $\alpha_W = -p'\alpha_R$ define $U(1)_{L,R}$ chiral symmetry transformations on $\phi$ and $\tilde\phi$. The chiral $\ZZ_k^L$ actions discussed correspond to take $\al_L=2\pi l/k$, $l=0,\ldots k-1$.\footnote{Recall that 
for both $N$ even and odd, we effectively have $p=1$, see appendix \ref{app:u1boson}.}

\section{Anomalies and persistent order}
\label{sec:anomalies}

We have discussed in section \ref{sec:freefermibosonization} the bosonization of the chiral Gross-Neveu model in absence of the $J\bar{J}$ deformation, i.e.
of free fermions. The deformation has a mild effect in the $U(1)$ sector, where it just changes the radius of the compact scalar, and a 
non-trivial effect to the $SU(N)$ sector, where it gives rise to a strongly coupled gapped theory.
In this section, we use 't Hooft anomaly matching conditions to determine key IR properties of the theory. {In this section we set $\mu=0$.}

We start by discussing 't Hooft anomalies of the bosonic $SU(N)_k$ WZW model in isolation. 
It has been shown that $SU(N)_k$ WZW  models  have a mixed $\ZZ_N$ anomaly between $PSU(N)_V$ and $\ZZ_N^L$ symmetries for $k=0$ mod $N$ \cite{Tanizaki:2018xto,Yao:2018kel}.\footnote{Another notable 't Hooft anomaly of $SU(N)_k$ WZW models is a $\ZZ_2$ mixed global-gravitational one present for $N$ even and $k$ odd \cite{Gepner:1986wi,Furuya:2015coa,Numasawa:2017crf,Lin:2021udi}. This anomaly is at the root of the different treatment of even and odd $N$ in section \ref{sec:freefermibosonization}, but it will not play further roles in what follows.}
The anomaly has been computed by showing that, in presence of background gauge fields $P$ and $A_L$ for the $PSU(N)$ and $\ZZ_N^L$ groups, the partition function is not 
background gauge invariant. 

An alternative way of detecting the anomaly is to study the bosonization of QED with $N$ Dirac fermions, i.e. gauging $U(1)_V$. This theory  
is expected to flow in the IR to the $SU(N)_1$ WZW model. Using anomaly matching conditions, we can then determine the anomalies from the UV theory.
The axial $U(1)_A$ symmetry is broken down to $\ZZ_{2N}$ by the ordinary chiral anomaly. The sum over the spin structures is performed by gauging $\ZZ_2^F\subset \ZZ_{2N}$, so in the bosonic theory we are left with a $\ZZ_N$ chiral symmetry. In the presence of a non-trivial $PSU(N)$ bundle we can have fractional $U(1)$ fluxes,
\begin{equation}
\int_{M_2} \frac{F}{2\pi} = \frac{1}{N}\,,
\label{eq:PSUNbundles}
\end{equation}
where $F$ is the field strength for the $U(1)_V$ gauge field. In this backgroud the fermion measure is not invariant under a chiral $\ZZ_N$ transformation and $\ZZ_N$ is completely broken.
The IR manifestation of this anomaly is precisely the mixed anomaly between $PSU(N)_V$ and $\ZZ_N^L$ discussed above. 

Yet another way to get this anomaly is by using the correspondence between 3d $G_k$ Chern-Simons on $M_2\times S^1$ and the WZW coset $G_k/G_k$ on $M_2$ (for $G$ simply connected) obtained by dimensional reduction \cite{Blau:1993tv}. The coset is implemented by gauging the vector $G$ symmetry of $G_k$ in the two-dimensional theory. 
For $G_k=SU(N)_k$, the 3d Chern-Simons theory has a global $\ZZ_N$ one-form symmetry $\widetilde{\ZZ}_N^{(1)}$, implemented by topological defect lines. 
This symmetry has a 't Hooft anomaly unless $k=0$ mod $N$ \cite{Gaiotto:2014kfa}.
Upon compactification to $M_2$, $\widetilde{\ZZ}_N^{(1)}$ gives both the $\ZZ_N^L$ global zero-form symmetry and the $\ZZ_N^{(1)}$ one-form symmetry of the $SU(N)$ two-dimensional gauge theory. These symmetries are implemented in the two-dimensional theory by topological defect lines and by topological local operators, respectively. The topological defect lines for $\ZZ_N^L$ are Wilson lines for the two-dimensional gauge theory, and are thus charged under $\ZZ_N^{(1)}$ unless they have vanishing $N$-ality. Conversely, the topological local operators that realize the one-form symmetry of the gauge theory are charged under $\ZZ_N^L$. We conclude that there is a mixed 't Hooft anomaly between $\ZZ_N^L$ and $\ZZ_N^{(1)}$ in the $SU(N)_1/SU(N)_1$ coset WZW model. In the limit of infinite gauge coupling, the gauge fields turn into non-dynamical backgrounds for the non-gauged $SU(N)_1$ WZW model. In a non-trivial $SU(N)$ gauge field configuration obtained by means of a non-trivial $\ZZ_N^{(1)}$ background to form a non-trivial $PSU(N)_V$ background, the 't Hooft anomaly between $\ZZ_N^L$ and $\ZZ_N^{(1)}$ reproduces the anomaly of the $SU(N)_1$ WZW model.

We now consider the $J\bar{J}$-deformed $SU(N)_1$ WZW model. As discussed in the previous section,
the deformation preserves a $PSU(N)_V\times\ZZ_N^L$ subgroup of the symmetry group $(SU(N)_L\times SU(N)_R)/\ZZ_N^V$ of the undeformed theory. 
The above mixed $PSU(N)_V\times \ZZ_N^L$ anomaly forbids the $J\bar{J}$-deformed theory to be trivially gapped in the IR. 
The most natural possibility is to assume that the deformed $SU(N)_1$ theory has $N$ vacua and displays spontaneous breaking of $\ZZ_N^L$. 
This is supported from the study of the classical potential arising from the $J\bar J$ deformation in abelian bosonization, where one finds $N$ degenerate minima, see appendix \ref{app:PotentialAbBos}.\footnote{As further evidence, using the correspondence between modular invariants of 2d rational CFTs and topological interfaces in 3d TQFTs \cite{Kapustin:2010if}, it has been 
conjectured in \cite{Gaiotto:2020iye} that the ground state of the $J\bar{J}$-deformed $SU(N)_1$ theory is exactly $N$-fold degenerate on $M_2=T^2$.}

We can use \eqref{eq:FpNodd}-\eqref{eq:FSUN} and the knowledge that the deformed $SU(N)_1$ flows in the IR to $N$ gapped vacua connected by the $\ZZ_N^L$ symmetry, to argue about the low energy effective theory of the cGN model. More specifically, we want to understand the fate of the vacuum at finite temperature. For this, it will be important to take a closer look at the possible discrete 't Hooft anomalies associated to the free $U(1)$ compact scalar.

\subsection{A $\ZZ_2$-valued anomaly of the bosonic theory} \label{subsec:compacttheta}

In this subsection, we discuss a certain $\ZZ_2$-valued 't Hooft anomaly of the bosonic theory $\calB$ in \eqref{eq:spinunitaryrelation}.
It involves the charge conjugation $\ZZ_2^C$ and other two $\ZZ_2$ symmetries which are not broken by the $J\bar{J}$ deformations.
The anomaly manifests itself as follows: when we gauge one of the two $\ZZ_2$ factors, the remaining global $\ZZ_2$ and $\ZZ_2^C$ are realized projectively in the twisted sector of the 
Hilbert space of the theory. 

This anomaly is a generalization of an anomaly found in the compact boson in \cite{Alavirad:2019iea}, so it is useful to first review the anomaly for the compact boson in isolation. In this case, the $\ZZ_2$ symmetries involved are, in addition to $\ZZ_2^C$, the $\ZZ_2^P$ and $\ZZ_2^W$ subgroups of the momentum and winding symmetries \eqref{eq:continuousglobal}.
When we gauge, say, $\ZZ_2^W$, the vertex operators \eqref{eq:vertexOp} with odd $m$ are projected out, but new operators with half-integer $e$ appear from the twisted sector. The 
physical vertex operators in the gauged theory are then ${\cal V}_{e+\ell/2,2m}$, $e,m\in \mathbb{Z}$, $\ell=0,1$. 
Let $P$ and $C$ be the topological lines implementing the $\ZZ_2^P$ and $\ZZ_2^C$ actions in the Hilbert space.
On vertex operators ${\cal V}_{e+\frac \ell 2,2m}$ we have
\begin{equation}\label{eq:lsmanomaly}
P^2 = (-1)^\ell \,, \qquad CP= (-1)^\ell PC \,, 
\end{equation}
where we used the action of charge conjugation given by
\begin{equation}\label{eq:Z2CactionCS}
\ZZ_2^C:  \qquad {\cal V}_{p,q} \rightarrow  {\cal V}_{-p,-q}  \,,
\end{equation}
for any fractional or integer $p$, $q$.
We see that the symmetry $\ZZ_2^C\times \ZZ_2^P$ acts linearly in the untwisted sector, but projectively in the twisted sector of the orbifolded theory.

We can refine the analysis by considering the torus partition function in presence of nontrivial backgrounds $W=[W_a,W_b]$ and $S=[S_a,S_b]$ for the winding $\ZZ_2^W$ and its dual $\widehat\ZZ_2^W$ symmetry{, where $T_a,T_b=0,1$ label the holonomy of the $\ZZ_2$ gauge field on the corresponding cycle}. 
The partition function of the orbifolded theory is given by \eqref{eq:Bgaugingpf}:
\begin{equation}\label{eq:Z2R'}
    Z_{2R}'[S_a,S_b]=\frac12\sum_wZ_R[w_a,w_b]e^{i\pi\int S\cup w}\,.
\end{equation}
Eq.\eqref{eq:lsmanomaly} implies that
\begin{equation}
     P^2=(-1)^{w_a}\,, \qquad         PC =(-1)^{w_a}CP \,, \qquad w_a=0,1\,,
   \end{equation}
from which we get
\begin{equation}\label{eq:pcwProj}
\begin{aligned}
    Z_{2R}^{(PC)'}[S_a,S_b]&=\frac12\sum_w Z_R^{(PC)}[w_a,w_b]e^{i\pi\int S\cup w} =\frac12\sum_w(-1)^{w_a}Z_R^{(CP)}[w_a,w_b]e^{i\pi\int S\cup w}\\
    &=\frac12\sum_w Z_R^{(CP)}[w_a,w_b]e^{i\pi\int w\cup[S_a,S_b+1]} = Z_{2R}^{(CP)'}[S_a,S_b+1]\,,\\
    Z_{2R}^{(P^2)'}[S_a,S_b]&=\frac12\sum_w Z_R^{(P^2)}[w_a,w_b]e^{i\pi\int S\cup w}=\frac12\sum_w(-1)^{w_a}Z_R[w_a,w_b]e^{i\pi\int S\cup w}\\
    &=\frac12\sum_wZ_R[w_a,w_b]e^{i\pi\int w\cup[S_a,S_b+1]} =Z_{2R}'[S_a,S_b+1]\,,
    \end{aligned}
\end{equation}
where $Z^{(A)}$ indicates the partition function with the insertion of the charge operator associated to the topological line $A$ in the trace.
While $\ZZ_2^P$ and $\ZZ_2^C$ acts projectively, the combination of $\ZZ_2^P$, $\ZZ_2^C$, and $\widehat\ZZ_2^{W}$ acts linearly, since from \eqref{eq:pcwProj} we have
\begin{equation}\label{eq:CBproje}
    P^2=\widehat{W}\,,\qquad PC=\widehat{W}CP\,.
\end{equation}
We see that $\ZZ_2^P\times\ZZ_2^C$ is centrally extended by the dual symmetry $\widehat\ZZ_2^W$ to the group $D_8$, the dihedral group of order 8, defined as 
\begin{equation}\label{eq:d8b}
 D_8=\langle P, C, \widehat{W}| P^4=C^2=1,\, CPC=P^3,\, P^2=\widehat{W}\rangle\,.
\end{equation}
The $D_8$ symmetry group of the gauged theory is anomaly-free. Upon gauging its $\widehat\ZZ_2^W$ subgroup, we reobtain the original theory with the anomalous $\ZZ_2^P\times\ZZ_2^C\times\ZZ_2^W$ symmetry. The extension by $\widehat\ZZ_2^W$ and the anomaly involving $\ZZ_2^W$ get exchanged by their gauging \cite{Bhardwaj:2017xup,Wang:2017loc}.
If we gauge $\ZZ_2^P$ instead of $\ZZ_2^W$, all the considerations above apply with the replacement $P\leftrightarrow W$.

We are now ready to discuss a similar 't Hooft anomaly in the bosonic model $\calB$. We first discuss $N$ odd and consider the theory $\calB'$ in the formulation \eqref{eq:FpNodd}.
The operator of $\calB'$ are $\ZZ_N^L$ gauge-invariant products of the operators of ${\cal S}$ in the $SU(N)_1$ sector and of the vertex operators ${\cal V}$ in the $U(1)_N$ sector.\footnote{In order to avoid clutter in the formulas, we remove the hat from $\ZZ_N^L$ and $\ZZ_2^W$ which appear in \eqref{eq:FpNodd}.} 
As far as our considerations are concerned, it is enough to classify the $SU(N)_1$ operators according to their $N$-ality, so we denote by ${\cal S}_{n,\bar n}$ the $SU(N)_1$ operators
with $L$ and $R$ $N$-ality $n$ and $\bar n$ under $SU(N)_L$ and $SU(N)_R$, respectively ($n,\bar n = 0\ldots, N-1$).
The diagonal $\ZZ_N$ symmetry acts on the $SU(N)$ and $U(1)$ operators as follows:
\begin{equation}
\begin{split}
\ZZ_N^L:     {\cal S}_{n,\bar n}& \rightarrow e^{-\frac{2i\pi n}{N}}   {\cal S}_{n,\bar n} \,, \\
\ZZ_N^W:     {\cal V}_{e,m} &\rightarrow e^{\frac{2i\pi m}{N}}   {\cal V}_{e,m}  \,.
\end{split}
\end{equation}
Charge conjugation acts on the $U(1)$ sector as in \eqref{eq:Z2CactionCS} and on $SU(N)_1$ operators as 
\begin{equation}\label{eq:Z2CactionSU}
\ZZ_2^C:     {\cal S}_{n,\bar n} \rightarrow  {\cal S}_{N-n,N- \bar n} \,.
\end{equation}
After gauging $\ZZ_N^L$, the physical operators of the theory are
\begin{equation}
    {\cal O}_{n,e,m}^{(k)} = {\cal S}_{n+k,n} {\cal V}_{e+\frac{k}{N},m}\,, \qquad n+k = m \;\; \text{mod} \;\; N\,,
\end{equation}
where $k=0$ represents the untwisted sector and $k=1,\ldots N-1$ the twisted sectors of the orbifolded theory.
The second global symmetry involved is the $\ZZ_2^W$ appearing in \eqref{eq:FpNodd}. Its action on the fields is\footnote{Note that for $N$ odd $e^{i \pi N^2} = e^{i \pi N}$, but the choice of shift in \eqref{eq:Z2WNodd} ensures that the $\ZZ_2^W$ has a chiral action on the fields.}
\begin{equation}\label{eq:Z2WNodd}
    \ZZ_2^W:     {\cal V}_{e+\frac{k}{N},m} \rightarrow e^{i\pi N^2 m} {\cal V}_{e+\frac{k}{N},m}  \,, \qquad  {\cal S}_{n+k,n} \rightarrow  {\cal S}_{n+k,n} \,.
\end{equation}
When we gauge  $\ZZ_2^W$, the physical operators read
\begin{equation}\label{eq:GIopNodd}
    {\cal O}_{n,e,m}^{(k,\ell)} = {\cal S}_{n+k,n} {\cal V}_{e+\frac{k}{N}+\frac{N^2\ell}{2},2m}\,, \qquad n+k = 2m \;\; \text{mod} \;\; N\,,
\end{equation}
where $\ell=0$ and $\ell=1$ represents the $\ZZ_2^W$ untwisted and twisted sector, respectively.
Finally, the third $\ZZ_2$ global symmetry involved is a $\ZZ_2^P\subset U(1)_P$ which acts as\footnote{This is obtained by taking $\alpha_P = N \pi$ in \eqref{eq:continuousglobal}, since after gauging $\ZZ_N^W$ $\alpha_P \sim \alpha_P + 2\pi N$.}
\begin{equation}\label{eq:Z2PNodd}
    \ZZ_2^P:     {\cal V}_{e+\frac{k}{N}+\frac{N^2\ell }{2},2m} \rightarrow e^{i\pi N (e+\frac{k}{N}+\frac{N^2\ell }{2})} {\cal V}_{e+\frac{k}{N}+\frac{N^2\ell}{2},2m}  \,, \qquad  {\cal S}_{n+k,n} \rightarrow  {\cal S}_{n+k,n} \,.
\end{equation}
Using eqs.\eqref{eq:Z2CactionCS}, \eqref{eq:Z2CactionSU} and \eqref{eq:Z2PNodd}, it is immediate to verify that on the operators \eqref{eq:GIopNodd}, 
\begin{equation}\label{eq:lsmanomalyNodd}
P^2 = (-1)^\ell \,, \qquad CP= (-1)^\ell  PC  \,.
\end{equation}
Similarly to the compact scalar case, $\ZZ_2^C\times \ZZ_2^P$ acts linearly in the $\ell=0$ untwisted sector and projectively in the $\ell=1$ twisted sector.
The analysis from \eqref{eq:Z2R'} until \eqref{eq:d8b} applies with obvious changes. Due to the anomaly the $\ZZ_2^P\times\ZZ_2^C$ is centrally extended by the dual symmetry $\widehat\ZZ_2^W$ to the group $D_8$ in \eqref{eq:d8b}.

We now discuss the theory $\calB$ with $N$ even in \eqref{eq:spinunitaryrelation}. As we will see, the role played by $\ZZ_2^P$ and $\ZZ_2^W$ above will be respectively played by 
a $\ZZ_2^W$ and a $\ZZ_2^L$ symmetry. The action of the diagonal $\ZZ_{N/2}^L$ symmetry on the $SU(N)$ and $U(1)$ operators is as follows:\footnote{Note that the action on the $U(1)_N$ sector is the same as a $\ZZ_N^P$ momentum shift.}
\begin{equation}
\ZZ_{N/2}^L:   \quad    {\cal V}_{e,m} \rightarrow e^{\frac{4i\pi}{N}(e+\frac{N}{2} m)}   {\cal V}_{e,m}  \,, \qquad  {\cal S}_{n,\bar n} \rightarrow e^{-\frac{4i\pi n}{N}}   {\cal S}_{n,\bar n} \,.
\end{equation}
Charge conjugation acts as before. After gauging $\ZZ_{N/2}^L$, the physical operators  are
\begin{equation}\label{eq:GIopNeven}
    {\cal O}_{n,e,m}^{(k)} = {\cal S}_{n+2k,n} {\cal V}_{e+k,m+\frac{2k}{N}}\,, \qquad n= e\;\; \text{mod} \;\; \frac N2\,,
\end{equation}
where $k=0$ represents the untwisted sector and $k=1,\ldots N/2-1$ the twisted sectors of the orbifolded theory.
The second global symmetry involved is the $\ZZ_2^L$ appearing in \eqref{eq:fermionization0}:
\begin{equation}\label{eq:Z2LNeven}
    \ZZ_2^L:  \quad   {\cal V}_{e+k,m+\frac{2k}{N}} \rightarrow e^{\frac{2i\pi}{N}\big(e+k+\frac{N}{2}( m+\frac{2k}{N})\big)}  {\cal V}_{e+k,m+\frac{2k}{N}}  \,, \quad  {\cal S}_{n+2k,n} \rightarrow e^{-\frac{2i\pi (n+2k)}{N}}  {\cal S}_{n+2k,n} \,.
\end{equation}
Note that $\ZZ_2^L$ is not a $\ZZ_2$ action individually on ${\cal S}$ and ${\cal V}$, but only on the operators ${\cal O}_{n,e,m}^{(k)}$, where it acts as 
${\cal O}_{n,e,m}^{(k)}\rightarrow (-1)^m {\cal O}_{n,e,m}^{(k)}$.
When we further gauge  $\ZZ_2^L$, the physical operators read
\begin{equation}\label{eq:GIopNeven}
    {\cal O}_{e,m}^{(k,\ell)} = {\cal S}_{n+2k+\ell,n} {\cal V}_{e+k+\frac{\ell}{2},m+\frac{2k}{N}+\frac{\ell}{N}}\,, \qquad n = e+\frac N2 m \;\; \text{mod} \;\; N\,,
\end{equation}
where $\ell=0$ and $\ell=1$ represents the $\ZZ_2^L$ untwisted and twisted sector, respectively.
Finally, the third $\ZZ_2$ global symmetry involved is a $\ZZ_2^W\subset U(1)_W$ which acts as
\begin{equation}\label{eq:Z2WNeven}
    \ZZ_2^W:  \   {\cal V}_{e+k+\frac{\ell}{2},m+\frac{2k}{N}+\frac{\ell}{N}} \rightarrow e^{\frac{i\pi N}{2}  (m +\frac{2k}{N} +\frac{\ell }{N})} {\cal V}_{e+k+\frac{\ell}{2},m+\frac{2k}{N}+\frac{\ell}{N}}  \,, \  {\cal S}_{n+2k+\ell,n} \rightarrow  {\cal S}_{n+2k+\ell,n} \,.
\end{equation}
Using eqs.\eqref{eq:Z2CactionCS}, \eqref{eq:Z2CactionSU} and \eqref{eq:Z2WNeven}, we get
\begin{equation}\label{eq:lsmanomalyNeven}
W^2 = (-1)^\ell \,, \qquad CW= (-1)^\ell  WC  \,.
\end{equation}
The $\ZZ_2^C\times \ZZ_2^W$ symmetry acts linearly in the $\ell=0$ untwisted sector and projectively in the $\ell=1$ twisted sector.
The analysis from \eqref{eq:Z2R'} until \eqref{eq:d8b} applies again, with obvious changes. Due to the anomaly the $\ZZ_2^W\times\ZZ_2^C$ symmetry is centrally extended by the dual symmetry $\widehat\ZZ_2^L$ to the group $D_8$ in \eqref{eq:d8b}.

Note that the mixed $\ZZ_2^P\times\ZZ_2^W\times\ZZ_2^C$ anomaly \eqref{eq:lsmanomaly} reduces in 1d to the $\ZZ_2$ anomaly in quantum mechanics discussed in app. D of \cite{Gaiotto:2017yup}.
Indeed, the action for the compact scalar $\phi\sim\phi +2\pi R$ with a background $U(1)_W$ gauge field $A_\mu$ on $\RR\times S^1_L$ is
\begin{equation}
    S_{2d}[\phi,A_\mu]=\frac1{8\pi}\int\! dt\int_0^L\! dx (\de\phi)^2-\frac{i}{2\pi R}\int dt \! \int_0^L dx \ep^{\mu\nu}A_\mu\de_\nu\phi\,.
\end{equation}
We choose $A_\mu$ to be a flat gauge field with a nontrivial holonomy $\thi$ around the compact $S^1_L$,
$A_t=0$, $A_x=\thi/L$, with $\thi\sim\thi+2\pi$. If we neglect the 2d massive excitations and only keep the zero mode, $\phi(x,t) \approx \phi_0(t)$, we get
\begin{equation}\label{eq:action1d}
    S_{1d}[\phi_0,\thi]=\frac L{8\pi}\int dt\, \dot\phi_0^2 -\frac{i\thi}{2\pi R}\int dt\, \dot\phi_0\,.
\end{equation}
The action  \eqref{eq:action1d} inherits the following symmetries from the 2d theory:
\begin{equation}
    \begin{cases}
        \ZZ_2^P:&\phi_0\mapsto\phi_0+\pi R\,,\quad \text{for any $\thi$,}\\
        \ZZ_2^C:&\phi_0\mapsto-\phi_0\,,\quad\quad\quad\! \text{for $\thi=0,\pi$.}
    \end{cases}
\end{equation}
We then recover the situation described in \cite{Gaiotto:2017yup}: $\ZZ_2^P\times\ZZ_2^C$ is realized linearly at $\thi=0$ and projectively at $\theta=\pi$.
In the latter case we have a $\ZZ_2$-valued 't Hooft anomaly which forbids to have a unique gapped vacuum.

\subsection{Fermionization and persistent order}
\label{subsec:FermPersOrder}

In this section, we study the fate of the bosonic anomaly discussed above upon fermionization.
It is useful to first look at the simplest case $N=1$, i.e. the duality between a compact boson at $R=1$ and a free Dirac fermion. 
The $\ZZ_2$ bosonic symmetry to be gauged is $\ZZ_2^W$ as in the second relation in \eqref{eq:U14nvsU1n}. The fermion partition function is obtained from \eqref{eq:fermionizationpf},
\begin{equation}
    Z_{\calF'}[\rho_a,\rho_b]=\frac12\sum_{w}Z_1[w_a,w_b](-1)^{\Arf[w\cdot\rho]}\,.
\end{equation}
We determine the effect of \eqref{eq:CBproje} in the fermionic theory by adding the corresponding line operators for $P$ and $C$ in the partition function. We then get
\begin{equation}\label{eq:ZfPCZ}
    Z_{\calF'}^{(PC)}[\rho_a,\rho_b]=\frac12\sum_w(-1)^{w_a}Z_1^{(CP)}[w_a,w_b]e^{i\pi\Arf[w\cdot\rho]} =(-1)^{\rho_a}Z_{\calF'}^{(CP)}[\rho_a,\rho_b+1]\,,
    \end{equation}
where $\rho_b+1$ means that we are exchanging $\rho_b=0={\rm NS}$ and $\rho_b=1={\rm R}$ spin structures, and we have used the identity
\begin{equation}
    \int w\cup T+\Arf[w\cdot \rho]=\Arf[w\cdot (T\cdot \rho)]+ \Arf[T\cdot\rho]+\Arf[\rho]\,,
\end{equation}
applied to the case in which $T=[T_a,T_b]=[0,1]$, i.e. for $T$ having nontrivial holonomy only along the $b$-cycle. Similarly, we have
\begin{equation} \label{eq:Zflasteq}
    Z_{\calF'}^{(P^2)}[\rho_a,\rho_b]=\frac12\sum_w(-1)^{w_a}Z_1[w_a,w_b]e^{i\pi\Arf[w\cdot\rho]}=(-1)^{\rho_a}Z_{\calF'}[\rho_a,\rho_b+1]\,.
    \end{equation}
The $\ZZ_2^C\times\ZZ_2^P$ action is realized projectively in the fermionic theory by $(-1)^F$ due to the shift $\rho_b\rightarrow \rho_b+1$:
\begin{equation}\label{eq:fermionizedD8}
     P^2=(-1)^F\,, \qquad         PC =(-1)^F CP \,.
   \end{equation}
Like in the bosonic case discussed in the previous subsection, the resulting extended group is $D_8^F$, where $D_8^F$ is the 
group \eqref{eq:d8b} with $\widehat W \rightarrow (-1)^F$.
{Its action on the Weyl components of the free Dirac fermion is given by
\begin{equation}\label{eq:D8Faction}
    D_8^F:\quad \begin{cases}
        P: &\psi_{\pm}\mapsto i \psi_{\pm}\,,\qquad \;\psi_{\pm}^\dagger \mapsto -i \psi_{\pm}^\dagger\,,\\
        C: &\psi_{\pm}\mapsto \psi_{\pm}^\dag\,, \qquad \;\;\, \psi_{\pm}^\dagger\mapsto  \psi_{\pm}\,, \\
        (-1)^F: & \psi_{\pm}\mapsto -\psi_{\pm}\,, \qquad  \psi_{\pm}^\dagger\mapsto - \psi_{\pm}^\dag\,.
    \end{cases}
\end{equation}
As it can be seen, the symmetry \eqref{eq:D8Faction} is not broken by the $\lambda_s$ and $\lambda_v$ deformations in \eqref{eq:LcGN}}.
In fact, it is unbroken also under the $O(2N)$ deformation which defines the GN model.
In the fermionized theory, $P$ is a $\ZZ_4\subset U(1)_V$ transformation whereas $C$ is still charge conjugation.
However, while $D_8^F$ acts linearly in the NS sector, it acts projectively  in the R sector because of an additional sign which arises from the $(-1)^{\rho_a}$ factor in \eqref{eq:ZfPCZ} and \eqref{eq:Zflasteq}.
The projective action in the R sector is a manifestation of a $\ZZ_2$ anomaly involving the group $D_8^F$. 

The same fermionic theory $\calF'$ can be alternatively obtained by using the first relation in \eqref{eq:U14nvsU1n}, namely starting from the 
compact boson with $R^2=4$ and gauging $\ZZ_2^P$.
In this formulation, it is easy to see that the factor $(-1)^{\rho_a}$ disappears from the analogue of \eqref{eq:ZfPCZ} and \eqref{eq:Zflasteq}, being reabsorbed 
due to the presence of the additional Arf term. The opposite occurs to the fermion theory $\calF$.
We conclude that, depending on which projection we perform in the R sector (i.e. if we take $\calF$ or $\calF'$, see \eqref{eq:fermionizationarf}), either a $D_8^F$ involving $\ZZ_2^P$ {(with $P$ corresponding to a $\ZZ_4^V$ transformation)} or one involving $\ZZ_2^W$ {(with $W$ corresponding to a $\ZZ_4^A$ transformation)} is realized projectively in the R sector. No matter what we choose, there is no way to have both $D_8^F$'s linearly realized and hence a $\ZZ_2$ anomaly persists. The $\ZZ_2^P$ or $\ZZ_2^W$ acts on fermions as discrete $U(1)_V$ or $U(1)_A$ rotations, so this phenomenon is nothing else than the discrete version of the well-known fact that we can move the mixed $U(1)_V-U(1)_A$ anomaly 
by counterterms, but there is no way to get rid of it altogether. 

We now consider the fermionization of the bosonic theory \eqref{eq:spinunitaryrelation}. 
For $N$ odd, we start from the bosonic theory $\calB'$ to get the fermionic theory $\calF'$ using 
\eqref{eq:FpNodd}. The analysis is identical to the one performed for the $N=1$ compact boson. It is enough to replace $Z_1$ from \eqref{eq:ZfPCZ} to \eqref{eq:Zflasteq} with the partition function of the $\calB'$ theory, to infer that in $\calF'$ we have a $D_8^F$ symmetry which commutes with $SU(N)$ flavor and is anomalous because it is realized projectively in the Ramond sector. 
For $N$ even, we use \eqref{eq:FSUN} but we look at the fermionic theory $\calF$. We get
\begin{equation}\label{eq:ZfPCZNeven}
    Z_{\calF}^{(WC)}[\rho_a,\rho_b]=\frac12\sum_l(-1)^{l_a}Z_{\calB'}^{(CW)}[l_a,l_b]e^{i\pi\Arf[l\cdot\rho]} =(-1)^{\rho_a}Z_{\calF}^{(CW)}[\rho_a,\rho_b+1]\,,
    \end{equation}
where $l_a$ and $l_b$ are the holonomies of the $\ZZ_2^L$ gauge field around the two cycles $a$ and $b$ of $T^2$. 
The comment in the last paragraph holds also in this case, so we have chosen $\calF'$ and $\calF$ for $N$ odd and even, respectively, because these are the theories
where the above $D_8^F$'s symmetries are projectively realized in the R sector.

As we will discuss in detail in section \ref{sec:FiniteT}, the above $\ZZ_2$ anomaly has an important physical consequence. 
The fate of the quasi long-range order discussed in \cite{Ciccone:2022zkg} for the chiral Gross-Neveu model at finite temperature depends crucially on the 
fermion periodicity along the thermal circle. In the NS sector (antiperiodic fermions) we expect that as soon as $T\neq 0$ quasi long-range order disappears due to thermal 
fluctuations (as generally expected in 2d models due to quantum mechanical tunneling effects). On the other hand, in the R sector (periodic fermions) the presence of a $\ZZ_2$ anomaly
forbids a trivially gapped spectrum and a form of quasi-long range order is expected. Interestingly enough, this ordered phase should persist 
at {\it arbitrarily high} temperatures. {We expect that the same phenomenon occurs in the GN model.}

\section{Inhomogeneities at finite temperature}
\label{sec:FiniteT}

In a two-dimensional theory, quasi long-range ordered phases are not detected directly from the one-point function of an order parameter, but rather from the slow decay of the two-point function of a ``would-be order parameter'' \cite{Berezinsky:1970fr,Berezinsky:1972rfj}. To explore the existence of inhomogeneous phases at finite temperature, we 
consider a two-point function in the fermionic theory, and in particular its long distance behavior when the temperature is turned on. The goal of this calculation is two-fold. Firstly, we show the 
disappearance of quasi long-range order at finite temperature in the theory with antiperiodic conditions on the thermal cycle, and the persistence of order with periodic conditions. Secondly, we show that the addition of a finite chemical potential $\mu$ for the $U(1)_V$ symmetry induces a spatial modulation of the would-be order parameter, i.e. the finite-temperature version of the chiral spiral behavior found at zero temperature in \cite{Ciccone:2022zkg}. 

What are the possible operators that can play the role of the would-be order parameter? The minimal requirement is that the operator is charged under $U(1)_A$. The most obvious candidate is the lightest scalar operator in this category, namely $\psi^\dagger_{+,a} \psi_-^a$. This is the operator we used to detect the chiral spiral order in $\mathbb{R}^2$ in \cite{Ciccone:2022zkg}. However, upon bosonization, this operator is mapped to a primary operator of the form ${\rm tr}\,\calS_{1,1}\calV_{1,0} = {\rm tr}\,Ue^{i\phi/R}$ for $N$ even, or ${\rm tr}\,\calS_{1,1}\calV_{2,0}={\rm tr}\,Ue^{2i\phi/R}$ for $N$ odd,\footnote{In this section, for $N$ odd we use the description in terms of the ($J\bar{J}$ deformed) $U(1)_{4N}$ compact scalar, i.e. $\calB$ and not $\calB'$.} and its two-point function receives contributions from various correlation functions of the operator $\mathrm{tr}\,U$ in the deformed $SU(N)_1$ theory, which is gapped and strongly coupled at large distances. In $\mathbb{R}^2$ the only required information about the correlation function of $\mathrm{tr}\,U$ could be deduced from the spontaneous breaking of the $\mathbb{Z}^L_N$ center symmetry. At finite temperature, instead, we need to include a sum over the insertions of the charge operator along the Euclidean time-like circle, and this requires substantial more information about the $SU(N)_1$ sector which is hard to obtain. 
To circumvent this problem we consider instead the following baryon (or ``determinant'') operator in the fermionic theory 
\begin{equation}\label{eq:detop}
    \calO_F=\frac1{N!}\ep^{a_1\dots a_N}\ep_{b_1\dots b_N}\psi^\dag_{+a_1}\psi^{b_1}_-\cdots \psi^\dag_{+a_N}\psi_-^{b_N}\equiv\det(\psi^\dag_+\psi_-)\,,
\end{equation}
which under bosonization is mapped to 
\begin{equation}
\begin{split}
\calO_B & =\calS_{0,0}\calV_{N,0}\equiv {\bf 1}\,e^{Ni\phi/R}\,, \qquad \quad N \; \text{even}\,, \\ 
\calO_B & =\calS_{0,0}\calV_{2N,0}\equiv {\bf 1}\,e^{2Ni\phi/R}\,, \qquad \;  N\; \text{odd}\,,
\end{split}
\end{equation} 
where normal ordering of the operators is understood.
This choice has the advantage that the part of the correlator due to the $SU(N)_1$ sector greatly simplifies, reducing essentially to the contribution from (twisted) partition functions in the absence of nontrivial operator insertions, and most of the calculation can be performed in the free $U(1)_N$ or $U(1)_{4N}$ sector, where exact results can be obtained.

Let us clarify the type of long-distance behavior that we expect at finite temperature, and what we mean by the order being lost or persisting. In the derivation of the chiral spiral order in the zero temperature case \cite{Ciccone:2022zkg} an important role  was played by the spontaneous breaking of the $\mathbb{Z}_N^L$ symmetry in the strongly coupled sector of the bosonized theory. The long-distance behavior of the two-point function on $\mathbb{R}^2$ combined the approach to a constant value in the $SU(N)_1$ sector, associated to $\mathbb{Z}_N^L$ breaking, and the power-law decay of the free scalar sector, associated to the quasi long-range order. When we turn on the temperature, effectively at large distances we are in a one-dimensional system, and the order is expected to be destroyed due to tunneling. As a result, barring the presence of anomalies in the effective quantum mechanics, generically we expect both 
behaviors to turn into an exponential decay on a scale fixed by the temperature. We find that this expectation is met for antiperiodic conditions: in this case, the spatially modulated chiral spiral has an exponentially decaying amplitude. However, this exponential decay is absent instead for periodic conditions. We interpret this as the presence of an obstruction to tunneling that makes the order persist. This is ultimately a manifestation of the anomaly presented in the previous section, that survives at finite temperature in the presence of periodic conditions. {Even though the addition of a chemical potential explicitly breaks charge conjugation symmetry, we find that the exponential decay is absent also for $\mu\neq 0$.} 

We study the two-point function in the simultaneous limit of low temperatures $T\ll M$, where $M$ is the mass gap in the strongly coupled $SU(N)_1$ sector, and large spatial distances $|x| M \gg 1$. To perform the calculation, especially in the compact scalar sector, it is convenient to study the model on a squared torus $T^2$ with modulus $\tau = i \beta/L$ and then take the limit $L\to\infty$ where the spatial dimension becomes non-compact.

\subsection{$N$ odd}

The $\mathbb{Z}^L_N$-twisted $SU(N)_1$ partition function, in the thermodynamic limit $L\rightarrow \infty$ and for $\beta M\gg 1$, behaves as follows:
\begin{equation}\label{eq:pfunclimit}
    Z_{SU(N)_1+J\bar{J}}[t_a,t_b]\sim C\times \di_{t_b,0} +{\cal O}(e^{-\beta M})~\,, \qquad t_{a,b}=0,\ldots,N-1\,.
\end{equation}
Here $t_a$ denotes the $\ZZ_N^L$ holonomy along the circle of length $L$ and $t_b$ the one along the circle of length $\beta$. 
In the above limit the $t_a$-twisted sectors become degenerate, being spaced by a factor $\sim 1/L$, and hence $Z$ becomes independent of $t_a$.
On the other hand, when $t_b\neq 0$, the contribution of all the states in each sector (both untwisted and twisted) are exponentially suppressed with $L$. 
$C$ is a constant which we will not need to fix.

We can use the identities \eqref{eq:fermionization0}, \eqref{eq:spinunitaryrelation}, and \eqref{eq:spinbkgnodd} to express correlation functions in the fermion theory in terms of $SU(N)_1$ and $U(1)_{4N}$ correlation functions, see appendix \ref{app:spinbkg} for details. Thanks to \eqref{eq:pfunclimit}, the partition function and the two-point function of the determinant operator simplify to\footnote{In order to avoid cluttering of formulas, from now on we will often omit to write the order of the sub-leading terms. These will however appear in the final formulas \eqref{eq:Final2ptFNodd} and \eqref{eq:Final2ptFNeven}.}
\begin{align}\label{eq:OF2pt}
\begin{split}
&  Z_{\mathcal{F}'}[\rho]\sim\frac{C}{2N}\sum_{s\in H^1(T^2, \ZZ_2)} (-1)^{\Arf[s\cdot\rho]-\Arf[\rho]} \sum_{t_a=0}^{N-1}Z_{U(1)_{4N}+J\bar{J}}[T(t_a,s)]~,\\
   &  Z_{\calF'}[\rho]\langle \calO_F(z,\bar{z})\calO^\dag_F(0,0)\rangle_\rho 
   \sim ~\frac{C}{2N}\sum_{s\in H^1(T^2,\ZZ_2)}(-1)^{\Arf[s\cdot\rho]-\Arf[\rho]}\times \\ &\qquad\qquad\qquad\qquad \times \sum_{t_a=0}^{N-1}Z_{U(1)_{4N}+J\bar{J}}[T(t_a,s)] \, \langle e^{2iN\phi/R}(z,\bar{z})e^{-2iN\phi/R}(0,0)\rangle_{T(t_a,s)} \,.
\end{split}    
\end{align}
The subscript in the two-point function and the argument in square brackets in the partition function both denote that these quantities are computed with line insertions along the two cycles, corresponding to the two entries of the vector $T(t_a,s)\equiv 2(N+1)(t_a,0)+2N s $. This is the background appearing in \eqref{eq:spinbkgnodd} 
with $t_b$ set to $0$.  Thanks to \eqref{eq:pfunclimit} the gauging of the $\mathbb{Z}_N$ diagonal  between the WZW and the compact boson is implemented through a single sum over the remaining label $t_a$. Finally, the sum over the background $s$ for $\mathbb{Z}_2^P$ implements the fermionization, and the choice of spin structure is denoted by $\rho$. Note that the $J\bar{J}$ deformation in the $U(1)_{4N}$ sector modifies the value of the radius $R$ according to \eqref{eq:scalarradius}.

The quantities appearing in \eqref{eq:OF2pt}  can be computed exactly, since the summands are correlators in a free CFT, see appendix \ref{app:vertex} for details.
We will compute the two-point function of generic vertex operators of charge $e$, namely $e^{i e \phi/R}$, and then set $e=0$ to obtain $Z_{\mathcal{F}'}$ and $e = 2N$ to obtain $Z_{\mathcal{F}'}\langle\mathcal{O}_F \mathcal{O}_F^\dagger\rangle$. We consider correlation functions with operators inserted at equal Euclidean times and a spatial distance $x=z=\bar{z}$. The result is
\begin{align}
\begin{split}\label{eq:oddpreferm}
&\sum_{t_a}Z_{U(1)_{4N}+J\bar{J}}[T(t_a,s)]\langle e^{\frac{ie\phi}{R}}(x)e^{-\frac{ie\phi}{R}}(0)\rangle_{T(t_a,s)} =\\&\left|\frac{\thi'_1(0|iLT)}{\thi_1(-ixT|iLT)}\right|^{\frac{2e^2}{R^2}}\frac{Ne^{\frac{4i\tilde\mu N e x}{R^2}}}{|\eta(iLT)|^2} 
F_1^s\left(\frac{2\tilde\mu N^2 L}{\pi R^2}-\frac{2ieNxT}{R^2}\middle|\frac{2iLTN^2}{R^2}\right)F_2^s\left(0\middle|\frac{iLTR^2}{2}\right)\,.
\end{split}
\end{align}
In \eqref{eq:oddpreferm}
\begin{align}
\begin{split}
        & F_1^{(0,0)}(y|\tau) = F_2^{(0,0)}(y|\tau) =\theta_3(y|\tau) \\
        & F_1^{(0,1)}(y|\tau) = \theta_3(y|\tau)~,~~F_2^{(0,1)}(y|\tau) = \theta_2(y|\tau) \\
        & F_1^{(1,0)}(y|\tau) = \theta_4(y|\tau)~,~~F_2^{(1,0)}(y|\tau) = \theta_3(y|\tau) \\
        & F_1^{(1,1)} (y|\tau)= \theta_4(y|\tau)~,~~F_2^{(1,1)}(y|\tau) = \theta_2(y|\tau)~,
\end{split}
\end{align}
where $\theta_i$ are the theta functions on $T^2$ defined in \eqref{eq:ZfreeDirac},
and $\tilde\mu$ is related to the chemical potential $\mu$ appearing in the action \eqref{eq:Lmu0} by a rescaling,
\begin{equation}\label{eq:mutilde}
    \tilde\mu=\mu\,\sqrt{1+\frac{\la'}{2\pi N}}\,,
\end{equation}
such that $2\tilde\mu N$ is the chemical potential for the $U(1)_W$ current at radius $R$. The details of why the chemical potential induces a spatial modulation of the two-point function are very similar to the zero temperature case studied in \cite{Ciccone:2022zkg} and can be deduced from appendix \ref{app:vertex}.

The next step is to plug \eqref{eq:oddpreferm} back in \eqref{eq:OF2pt} and perform the fermionization sum. We set $u\equiv \frac{2\tilde\mu N^2 L}{\pi R^2}-\frac{2ieNxT}{R^2}$. The result of the sum is
\begin{align}\label{eq:finalresultgene}
    \begin{split}
        &Z_F[\rho]\langle \calO_F(x)\calO^\dag_F(0)\rangle_\rho \sim \frac{C}2\left|\frac{\thi'_1(0|iLT)}{\thi_1(-ixT|iLT)}\right|^{\frac{2e^2}{R^2}}\frac{e^{\frac{4i\tilde\mu N e x}{R^2}}}{|\eta(iLT)|^2}\times\\
        &\hspace{-0.4cm} \left[\theta_3\left(0\middle|\tfrac{iLTR^2}{2}\right)(\theta_3 + w_1^\rho \theta_4)\left(u\middle|\tfrac{2iLTN^2}{R^2}\right)+ w_2^\rho \theta_2\left(0\middle|\tfrac{iLTR^2}{2}\right)(\theta_3 + w_3^\rho \theta_4)\left(u\middle|\tfrac{2iLTN^2}{R^2}\right)\right]~,
\end{split}
\end{align}
where $w_{1,2,3}^\rho$ are the following signs which depend on the spin structure\footnote{We remind that $\rho=[\rho_a,\rho_b]$ denotes the periodicity along the (large) $a$ cycle and the (small) $b$ cycle, respectively.}
\begin{align}
\begin{split}
  w_1^{[\rm{NS,\, NS}]} &= +1\,,\,w_2^{[\rm{NS,\, NS}]} = +1 \,,\,w_3^{[\rm{NS,\, NS}]} = -1~,\\
   w_1^{[\rm{R,\, NS}]} &= +1\,,\ \ w_2^{[\rm{R,\, NS}]} = -1 \,,\ \ w_3^{[\rm{R,\, NS}]} = -1~,\\
w_1^{[\rm{NS,\, R}]} &= -1\,,\ \  w_2^{[\rm{NS,\, R}]} = +1 \,,\ \ w_3^{[\rm{NS,\, R}]} = +1~,\\
 w_1^{[\rm{R,\, R}]} &= -1\,,\quad\,  w_2^{[\rm{R,\, R}]} = -1 \,,\quad w_3^{[\rm{R,\, R}]} = -1~.
\end{split}
\end{align}
To take $L\to \infty$ we use the following asymptotic expansions for $t\to i\infty$,
\begin{equation}\label{eq:thetaasym}
\begin{aligned}
    \thi_1(u|it)&\underset{t\to i\infty}{\sim} 2e^{-\frac{\pi t}4}\sin(\pi u)\,,\qquad\quad\ \,  \thi_2(u|it)\underset{t\to i\infty}{\sim} 2e^{-\frac{\pi t}4}\cos(\pi u)\,,\\\thi_3(u|it)&\underset{t\to i\infty}{\sim} 1+2e^{-\pi t}\cos(2\pi u)\,,\quad\  \thi_4(u|it)\underset{t\to i\infty}{\sim} 1-2e^{-\pi t}\cos(2\pi u)\,,\\
    \thi'_1(0|it)&\underset{t\to i\infty}{\sim} 2\pi e^{-\frac{\pi t}4}\,,\qquad\qquad\quad\quad\ \ \ \ \, 
    \eta(it)\underset{t\to i\infty}{\sim} e^{-\frac{\pi t}{12}}\,.
\end{aligned}    
\end{equation}
Plugging these asymptotics in \eqref{eq:finalresultgene}, we obtain
\begin{align}\label{eq:2ptFunGen}
\begin{split}
          Z_{\mathcal{F}'}[\rho]&\langle \calO_F(x)\calO^\dag_F(0)\rangle_\rho \sim C\left|\frac{\pi}{\sinh(\pi x T)}\right|^{\frac{2e^2}{R^2}}e^{\frac{4i\tilde\mu Ne x}{R^2}}e^{\frac{\pi L T}6} \times\\
         &\hspace{-0.1cm}\begin{cases}
             1 + \mathcal{O}(e^{-\beta M}),& \rho=[{\rm any,\, NS}]\,,\\
             2e^{-\frac{2\pi L T N^2}{R^2}}\cosh\Big(\frac{4\pi e N xT}{R^2} +\frac{4i\tilde{\mu}N^2 L}{R^2}\Big)+ \mathcal{O}(e^{-\beta M}),& \rho=[{\rm any,\, R}]\,.
         \end{cases}
\end{split}
\end{align}
Note that upon taking $L\to\infty$ we have lost dependence on the periodicity condition along the cycle of length $L$, but we retain dependence on the periodicity on the cycle of length $\beta = T^{-1}$.
For $|xT|\ll 1$, the $\sinh$ factor in \eqref{eq:2ptFunGen} reproduces the correct power-like decay of the correlator on $\mathbb{R}^2$.
Evaluating \eqref{eq:2ptFunGen} for $e=0$ gives
\begin{equation}\label{eq:2ptFunGene0}
    \begin{aligned}
         Z_{\mathcal{F}'}[\rho]&\sim Ce^{\frac{\pi L T}6}\begin{cases}
             1+O(e^{-\be M}),& \rho=[{\rm any,\, NS}]\,,\\
             2e^{-\frac{2\pi L T N^2}{R^2}}\cos\big(\frac{4\tilde\mu N^2 L}{R^2}\big)+ \mathcal{O}(e^{-\beta M}) ,& \rho=[{\rm any,\, R}]\,.
         \end{cases}
    \end{aligned}
\end{equation}
We can now take the ratio between \eqref{eq:2ptFunGen} and \eqref{eq:2ptFunGene0} to get the final result for the two-point function.
In the large-distance limit $|x| T \to \infty$ we have
\begin{equation} \label{eq:Final2ptFNodd}
 \langle \calO_F(x)\calO^\dag_F(0)\rangle_\rho  \underset{|x| T \to \infty}{\sim} (2\pi)^{\frac{8 N^2}{R^2}}e^{\frac{8i\tilde\mu x N^2 }{R^2}} \begin{cases}
        e^{-\frac{8 N^2\pi |x| T}{R^2}}+O(e^{-\be M})\,,& \rho=[{\rm any,\, NS}]\,,\\
        \frac12 +O(e^{-\be M}e^{-\frac{8 N^2\pi |x| T}{R^2}}) \,,& \rho=[{\rm any,\,R}]~.      
    \end{cases}
\end{equation}
The large distance, constant behavior of the two-point function is an indication of persistent order and the presence of a disconnected term in the correlator.
We have a strict long-range order governed by $\langle {\cal O}_F \rangle$, which is non-vanishing for spin structures $\rho=[{\rm any,\,R}]$.

Summarizing, we see the following behavior: 
\begin{itemize}
    \item for thermal fermions, i.e. fermions with antiperiodic conditions $\rho=[{\rm any\,,\ NS}]$, the correlation function of the operator $\calO_F=\det(\psi^\dag_+\psi_-)$ vanishes at large $|x|$ exponentially with the temperature. We have
    \begin{equation}\label{eq:NS2ptFinal}
    \langle \calO_F(x)\calO^\dag_F(0)\rangle_{NS}\sim e^{-\frac{|x|}{\xi_T}}e^{2iN\mu'x}\,,\qquad \xi_T^{-1}={2N}{\pi T}\left(1+\frac{\la'}{2\pi N}\right)\,,
    \end{equation}
    where we used \eqref{eq:scalarradius} with $R_0 = \sqrt{4N}$ and the definition of $\mu'$ in \eqref{eq:muprime}.  
    We see that the modulation is visible only at intermediate scales, provided that $N\mu'\gtrsim \pi \xi_T^{-1}$.
    \item for periodic fermions, i.e. with conditions $\rho=[{\rm any\,,\ R}]$, the same correlation function does not decay, and the spatial modulation due to the chemical potential $\mu$ is visible at large distances even at finite temperature,
    \begin{equation}\label{eq:R2ptFinal}
    \langle \calO_F(x)\calO^\dag_F(0)\rangle_{R}\sim e^{2i N\mu' x}\,.
    \end{equation}
\end{itemize}

\subsection{$N$ even}
The $SU(N)_1$ partition function with insertion of $\mathbb{Z}_{N/2}^L$ and possibly also $\mathbb{Z}_2^L$ topological lines on both cycles behaves as follows in the limit
\begin{align}\label{eq:evensimpl}
\begin{split}
Z_{SU(N)_1+J\bar{J}}[2t_a,2t_b]&\sim C\times \di_{t_b,0}~,\\
Z_{SU(N)_1+J\bar{J}}[2t_a+s_a,2t_b+s_b]&\sim C\times \di_{t_b,0}Z^\pm[s_a,s_b]\,,
\end{split}
\end{align}
where
\begin{equation}
Z^\pm[s_a,s_b]=(\pm1)^{s_as_b}~,
\end{equation}
is a sign that remains ambiguous due to the anomaly in $\mathbb{Z}_2^L$ in the $SU(N)_1$ theory. The independence of $t_a$ and the projection to $t_b = 0$ have the same explanation as in the case of $N$ odd (cf. appendix \ref{app:spinbkg}), and again $C$ denotes an undetermined constant.

Plugging \eqref{eq:evensimpl} in the fermionization formula that can be derived from \eqref{eq:fermionization0}, \eqref{eq:spinunitaryrelation}, and \eqref{eq:spinbkgneven}, we obtain
\begin{align}
\begin{split}
    Z_{\mathcal{F}'}[\rho]&\sim\frac{C}{N}\sum_{s\in H^1(T^2,\ZZ_2)} (-1)^{\Arf[s\cdot\rho]-\Arf[\rho]} Z^\pm[s] \sum_{t_a=0}^{N/2-1}Z_{U(1)_{N}+J\bar{J}}[T(t_a,s)]\,,\\
    Z_{\mathcal{F}'}[\rho]&\langle \calO_F(z,\bar{z})\calO^\dag_F(0,0)\rangle_\rho\sim\frac{C}{N}\sum_{s\in H^1(\ZZ_2)}(-1)^{\Arf[s\cdot\rho]-\Arf[\rho]}Z^\pm[s] \times
    \\
    & \sum_{t_a=0}^{N/2-1}Z_{U(1)_{N}+J\bar{J}}[T(t_a,s)] \langle e^{iN\phi/R}(z,\bar{z})e^{-iN\phi/R}(0,0)\rangle_{T(t_a,s)} \,,
\end{split}
\end{align}
where the notation follows the same conventions as in the case of $N$ odd, explained below \eqref{eq:OF2pt}, and $T(t_a,s) = 2(t_a,0)+s$ is the background appearing in \eqref{eq:spinbkgneven} with the renaming $[k_\psi,\ell_\psi]=s$, $k= 2t_a$, $\ell=t_b$, with the latter set to $0$. Also for $N$ even the radius $R$ in the $U(1)_N$ sector is modified according to \eqref{eq:scalarradius}.

Like we did in the previous section, we proceed by performing exactly the part of the calculation that involves the free scalar sector. Again, we compute for generic charge $e$ of the vertex operator, and this will give the partition function for $e=0$, and the product of the partition function with the two-point function for $e=N$. The result is
\begin{align}
\begin{split}
    &\sum_{t_a}Z_{U(1)_{N}+J\bar{J}}[T(t_a,s)]\langle e^{ie\phi/R}(x)e^{-ie\phi/R}(0)\rangle_{T(t_a,s)}=\\&\left|\frac{\thi'_1(0|iLT)}{\thi_1(-ixT|iLT)}\right|^{\frac{2e^2}{R^2}}\frac{Ne^{\frac{2i\tilde\mu N e x}{R^2}}}{2|\eta(iLT)|^2} F_1^s\left(\frac{\tilde\mu N^2 L}{2\pi R^2}-\frac{ieNxT}{R^2}\middle|\frac{iLTN^2}{2R^2}\right)F_2^s\left(0\middle|\frac{iLTR^2}{2}\right)\,,
\end{split}
\end{align}
where 
\begin{align}
\begin{split}
        & F_1^{(0,0)}(y|\tau) = F_2^{(0,0)}(y|\tau) =\theta_3(y|\tau)~, \\
        & F_1^{(0,1)}(y|\tau) = F_2^{(0,1)}(y|\tau) = \theta_2(y|\tau)~, \\
        & F_1^{(1,0)}(y|\tau) = F_2^{(1,0)}(y|\tau) = \theta_4(y|\tau)~, \\
        & F_1^{(1,1)}(y|\tau) = F_2^{(1,1)}(y|\tau) = \theta_1(y|\tau)~,
\end{split}
\end{align}
and $\tilde \mu$ as in \eqref{eq:mutilde}. Note that the $s=(1,1)$ contribution vanishes as $\theta_1(0|\tau)=0$. This means that $Z^{\pm}[s]$ contributes always $1$ and can be neglected. 
Performing the fermionization sum over $s$ we get
\begin{align}
        Z_{\mathcal{F}'}[\rho]&\langle \calO_F(x)\calO^\dag_F(0)\rangle_\rho\sim \frac{C}2\left|\frac{\thi'_1(0|iLT)}{\thi_1(-ixT|iLT)}\right|^{\frac{2e^2}{R^2}}\frac{e^{\frac{2i\tilde\mu N e x}{R^2}}}{|\eta(iLT)|^2} \left[\thi_3\left(0\middle|\tfrac{iLTR^2}{2}\right)\thi_3\left(u\middle|\tfrac{iL T N^2}{2R^2}\right) \right.\nonumber \\
       & \hspace{1cm} \left. + w_1^\rho \thi_4\left(0\middle|\tfrac{iLTR^2}{2}\right)\thi_4\left(u\middle|\tfrac{iL T N^2}{2R^2}\right)
        +w_2^\rho\thi_2\left(0\middle|\tfrac{iLTR^2}{2}\right)\thi_2\left(u\middle|\tfrac{iL T N^2}{2R^2}\right)\right]\,,
        \end{align}
    where 
    \begin{equation}
        u\equiv \frac{\tilde\mu N^2 L}{2\pi R^2}-\frac{ieNxT}{R^2}\,,
    \end{equation}
    and $w_{1,2}^\rho$ are the following signs that depend on the spin structure:
    \begin{align}
    \begin{split}
    w_1^{[\rm{NS,\, NS}]}&=+1\,,\,w_2^{[\rm{NS,\, NS}]}=+1~,\\
    w_1^{[\rm{R,\, NS}]}&=+1\,,\ \ w_2^{[\rm{R,\, NS}]}=-1~,\\
    w_1^{[\rm{NS,\, R}]}&=-1\,,\ \ w_2^{[\rm{NS,\, R}]}=+1~,\\
    w_1^{[\rm{R,\, R}]}&=-1\,,\quad\,w_2^{[\rm{R,\, R}]}=-1~.\\
    \end{split}
    \end{align}
We use \eqref{eq:thetaasym} to take the limit $L\to\infty$ and we obtain
\begin{align}
\begin{split}
         Z_{\mathcal{F}'}[\rho]&\langle \calO_F(x)\calO^\dag_F(0)\rangle_\rho \sim C\left|\frac{\pi}{\sinh(\pi x T)}\right|^{\frac{2e^2}{R^2}}e^{\frac{2i\tilde\mu Ne x}{R^2}}e^{\frac{\pi L T}6}\times \\
         &  \begin{cases}
             1+\mathcal{O}(e^{-\beta M}),& \rho=[{\rm any\,,\ NS}]\,,\\
             2e^{-\frac{\pi L T N^2}{2R^2}}\cosh\Big(\frac{2\pi e N xT}{R^2} +\frac{i\tilde{\mu}N^2 L}{R^2}\Big)+\mathcal{O}(e^{-\beta M}),& \rho=[{\rm any\,,\ R}]\,.
         \end{cases}
\end{split}
\end{align}
Plugging $e= 0$, the result for the partition function is
\begin{align}
\begin{split}
         Z_{\mathcal{F}'}[\rho]&\sim C e^{\frac{\pi L T}{6}}\begin{cases}
             1+O(e^{-\be M}),& \rho=[{\rm any\,,\ NS}]\,,\\
             2e^{-\frac{\pi L TN^2}{2R^2}}\cos\Big(\frac{\tilde\mu N^2 L}{R^2}\Big) + O(e^{-\be M}),& \rho=[{\rm any\,,\ R}]\,.
         \end{cases}
\end{split}
\end{align}
Taking the ratio between $e=N$ and $e=0$ we obtain for the two-point function in the $|x|T \to \infty$ limit:
\begin{align}\label{eq:Final2ptFNeven}
    \langle \calO_F(x)\calO^\dag_F(0)\rangle_\rho \underset{|x| T \to \infty}{\sim} (2\pi)^{\frac{2N^2}{R^2}}e^{\frac{2i\tilde\mu N^2 x}{R^2}}
    \begin{cases}
        e^{-\frac{2\pi |x| N^2 T}{R^2}}+\mathcal{O}(e^{-\beta M})\,,& \rho=[{\rm any\,,\ NS}]\,,\\
        \frac12 +\mathcal{O}(e^{-\beta M} e^{-\frac{2\pi |x| N^2 T}{R^2}})\,, & \rho=[{\rm any\,,\ R}]~.     \end{cases}
\end{align}
Using \eqref{eq:scalarradius} with $R_0 = \sqrt{N}$ and the definition of $\mu'$ in \eqref{eq:muprime}, we see that \eqref{eq:Final2ptFNeven} exactly reproduces the behavior found in the case of $N$ odd, summarized in \eqref{eq:NS2ptFinal}-\eqref{eq:R2ptFinal}.

\section{Persistent order at large $N$}
\label{sec: largeN}

The limit of large $N$ gives us another way to control the dynamics of the chiral Gross-Neveu theory at finite temperature and density, alternative to the bosonization approach that we studied until now. We show that using this tool we find results consistent with the analysis in the previous sections. In particular, we find different patterns of symmetry restoration induced by the temperature, depending on the choice of periodicity conditions on the fermions. The observable we compute is the grand canonical free energy density $F$, defined in terms of the partition function by
\begin{equation}
 \log Z =- N \int d^2x\, F~.
\end{equation}
We obtain $F$ at finite temperature and density, as a function of a condensate for the fermion bilinear. The existence of a chiral-spiral order is then detected by comparing the values of the partition function with a homogeneous and a inhomogeneous condensate. Note that the fact that we can have a non-zero one-point function breaking the chiral symmetry, both at zero and nonzero temperature, is due to the large $N$ limit. In the following we take the limit of large $N$ with $\lambda_s$ fixed and setting $\lambda_v = 0$.

\subsection{Antiperiodic conditions}

The calculation of $F$ at large $N$ with antiperiodic conditions for the fermions was performed in \cite{Ciccone:2022zkg}, reproducing the result of \cite{Schon:2000he,Thies:2003kk,Schnetz:2004vr,Basar:2009fg} with a different method. Let us recall the main ingredients that will be useful to repeat the calculation in the case of periodic conditions. We introduce a complex Hubbard-Stratonovich (HS) field $\Delta$, which equals the complex fermion bilinear on-shell. The large $N$ free energy takes the form
\begin{equation}
    F=
    \frac{\overline{|\Delta|^2}}{\lambda_s}-\Tr\log\left(\slashed{\partial}+  \Delta P_+ + \Delta^*P_-\right)\,,
\end{equation}
where $\Delta$ here denotes the condensate of the HS field, and $P_\pm$ are 2d chiral projectors. We make the choice to encode the $\mu$ dependence by twisting appropriately the periodicity conditions for the fermions. We consider the ansatz
\begin{equation}
\Delta = M e^{2i q x}~,
\end{equation}
and minimize $F$ as a function of $M$ and $q$.

The trace can be computed diagrammatically using the expansion
\begin{equation}
 \Tr\log\left(\slashed{\partial}+  \Delta P_+ + \Delta^*P_-\right) = \sum_{k=1}^\infty \frac{(-1)^{k+1}}{k}\times\mathrm{Tr}\left[\slashed{\partial}^{-1}M(e^{2iqx}P_++e^{-2iqx}P_-)\right]^k~.
\end{equation}
Since the $P_+$ insertion carries momentum $q$ and the $P_-$ insertion momentum $-q$, only diagrams with the same numbers of $P_+$ and $P_-$ insertions survive in the sum, coming from terms with $k = 2n$ even. Moreover, the $P_+$ and $P_-$ insertions must alternate in the loop, to avoid a trivial cancelation due to chirality. Denoting the insertions of $P_+$ and $P_-$ with red and blue circles, respectively, we obtain
\begin{equation}
    -\Tr\left[ \slashed{\partial} ^{-1} M(e^{2iqx}P_++e^{-2iqx}P_-)\right]^{2n} =2
    \tikz[baseline=.05ex]{
    \node[circle,minimum size=30pt,draw](c) at (0,0){};
    \draw (c.west)--(c.west) node[currarrow,sloped,midway] {};
    \draw (c.north west) node[rotate=0] {\color{red} $\oplus$};
    \draw (c.north west) node[anchor=south east](t) {$1_+$};
    \draw (c.north)--(c.north) node[currarrow,midway] {};
    \draw (c.north east) node[rotate=0] {\color{blue} $\ominus$};
    \draw (c.north east) node[anchor=south west] {$1_-$};
     \draw (c.east)--(c.east) node[currarrow,sloped,midway,rotate=180] {};
     \draw (c.south west) node[rotate=0] {\color{blue} $\ominus$};
    \draw (c.south west) node[anchor=north east] {$n_-$};
     \draw (c.south east) node[anchor=north, rotate=30](b) {$...$};
     }~.
\end{equation}
Setting $p_{1\mp}=p_1\mp2q$ we find
\begin{equation}
    \sum_{n=1}^\infty\frac{1}{n}
    \tikz[baseline=.05ex]{
    \node[circle,minimum size=30pt,draw](c) at (0,0){};
    \draw (c.west)--(c.west) node[currarrow,sloped,midway] {};
    \draw (c.north west) node[rotate=0] {\color{red} $\oplus$};
    \draw (c.north west) node[anchor=south east](t) {$1_+$};
    \draw (c.north)--(c.north) node[currarrow,midway] {};
    \draw (c.north east) node[rotate=0] {\color{blue} $\ominus$};
    \draw (c.north east) node[anchor=south west] {$1_-$};
     \draw (c.east)--(c.east) node[currarrow,sloped,midway,rotate=180] {};
     \draw (c.south west) node[rotate=0] {\color{blue} $\ominus$};
    \draw (c.south west) node[anchor=north east] {$n_-$};
     \draw (c.south east) node[anchor=north, rotate=30](b) {$...$};
     } =\int\frac{d^2p}{(2\pi)^2}\log\left(1+\frac{M^2(p_1^2+p_2^2-q^2+2iqp_2)}{(p_{1-}^2+p_2^2)(p_{1+}^2+p_2^2)}\right).
\end{equation}
At $T>0$ the symbol $\int dp_2$ actually refers to the discrete sum over Matsubara frequencies $2\pi T \sum_{n\in\mathbb{Z}}$, with $p_2 = \pi(2n+1)T$ in the antiperiodic case. The twisted periodicity induced by $\mu\neq 0$ can be reabsorbed by defining $p_{2+}(q) = \pi(2n+1)T + i (\mu -q)$. The integration over the spatial component $p_1$ of the momentum gives
\begin{equation}
        F=
        \frac{M^2}{\lambda_s} -T\sum_{n\in\mathbb{Z}}\left(\sqrt{M^2+p_{2+}^2(q)}-\sqrt{p_{2+}^2(q)} \right) \,.
\end{equation}
In order to compute the value at the minimum, we need to introduce a regularization in the sum by putting a cutoff $p_{2,\text{max}}=\pi(2n_{\text{max}}-1)T$, and reabsorb the divergence in the renormalization of the coupling $\lambda_s$. As result $\lambda_s$ gets replaced in the equation by the parameter $M_0$, the mass of the fermion at $T=0$ and $\mu=0$. We obtain
\begin{equation}\label{eq:Fsumq}
\begin{aligned}
   F&= \frac{M^2}{2\pi}\left(\log\left(\frac{4\pi T}{M_0}\right)+\psi\left(\frac12 + \frac{i(\mu-q)}{2\pi T}\right)\right)\\
   &\qquad\qquad\qquad-T\sum_{n\in\mathbb{Z}}\left(\sqrt{p_{2}^2+M^2}-\sqrt{p_{2+}^2}-\frac{M^2}{2\sqrt{p_{2+}^2}}\right)\,. 
   \end{aligned}\end{equation}

\begin{figure}[t]
    \centering
    \begin{tikzpicture}
    \node[anchor=south west,inner sep=0] (image) at (0,0) {\includegraphics[width=0.45\textwidth]{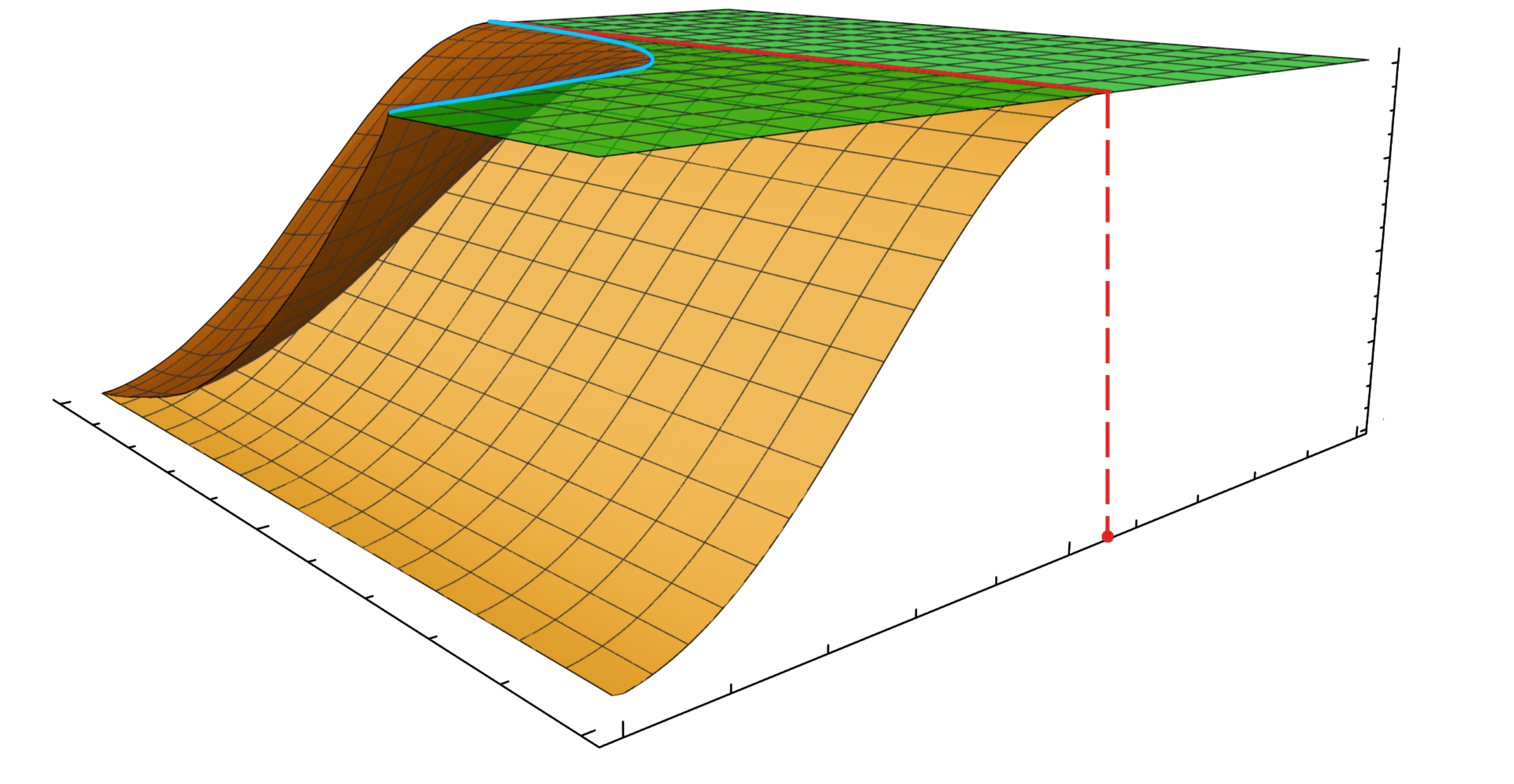}};
    \begin{scope}[x={(image.south east)},y={(image.north west)}]
        \node [anchor=north east,scale=0.8] at (0.04,0.48) {$0.0$};
        \node [anchor=north east,scale=0.8] at (0.38,0.07) {$1.0$};
        \node [anchor=north west,scale=0.8] at (0.41,0.07) {$0.0$};
        \node [anchor=north west,scale=0.8,text=red] at (0.72,0.31) {$T_c$};
        \node [anchor=north,scale=0.8] at (0.89,0.42) {$1.0$};
        \node [anchor=west,scale=0.8] at (0.9,0.45) {$-0.08$};
        \node [anchor=west,scale=0.8] at (0.905,0.57) {$-0.06$};
        \node [anchor=west,scale=0.8] at (0.91,0.69) {$-0.04$};
        \node [anchor=west,scale=0.8] at (0.915,0.81) {$-0.02$};
        \node [anchor=west,scale=0.8] at (0.925,0.93) {$0.00$};
        \node [anchor=north east] at (0.18,0.3) {$\mu$};
        \node [anchor=north west] at (0.66,0.25) {$T$};
        \node [anchor=east] at (0.15,.9) {$F$};
    \end{scope}
    \end{tikzpicture}
    \caption{The large $N$ free energy $F$ as a function of  $\mu$ and $T$, in units of $M_0$. At low $T$, the chiral spiral configuration (yellow) is favored with respect to the homogeneous configuration (brown) and the chirally symmetric phase (green). At $T=T_c$, there is a second order phase transition (red line) dividing the chiral spiral phase from the chirally symmetric phase. Assuming homogeneity, one finds a different phase transition line (blue line).}
    \label{fig:freeenergycomparison}
\end{figure}
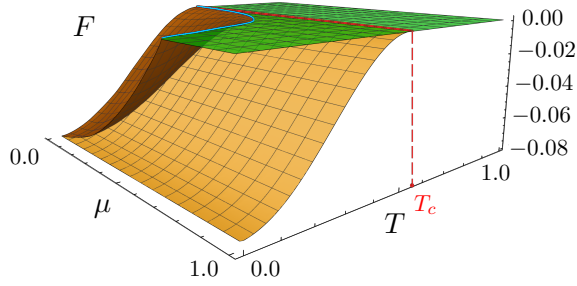

This expression is minimized for $q=\mu$, where
\begin{equation}\label{eq:Fsum}
   F\vert_{q=\mu}= \frac{M^2}{2\pi}\left(\log\left(\frac{\pi T}{M_0}\right)-\gamma\right)-T\sum_{n\in\mathbb{Z}}\left(\sqrt{p_{2}^2+M^2}-|p_{2}|-\frac{M^2}{2|p_{2}|}\right)\,, \end{equation}
where $\ga$ is the Euler-Mascheroni constant.
Notably, the result is $\mu$ independent. In figure \ref{fig:freeenergycomparison} the value of $F$ at this minimum is compared with the result one obtains by setting $q=0$ and minimizing only with respect to $M$, i.e. for a homogeneous condensate.
The figure shows that the inhomogeneous condensate is always favored, whenever a condensate exists. Above the critical temperature $T_c=\frac{e^\gamma}{\pi} M_0$ (the precise value depends on the regularization and renormalization scheme) the chiral symmetry is fully restored and there is no condensate: the order is lost at high temperatures.

\subsection{Periodic conditions}

The initial steps of the calculation with periodic conditions are identical to the ones showed above in the antiperiodic case. We therefore arrive at the same formula
\eqref{eq:Fsum} with the important difference that now $p_{2+}(q) = 2\pi n T + i(\mu-q)$ and the sum is over $n\in\mathbb{Z}$.

\begin{figure}[t!]
    \centering
    \scalebox{0.45}{{\includegraphics[trim= {2cm 9cm 0 10cm}]{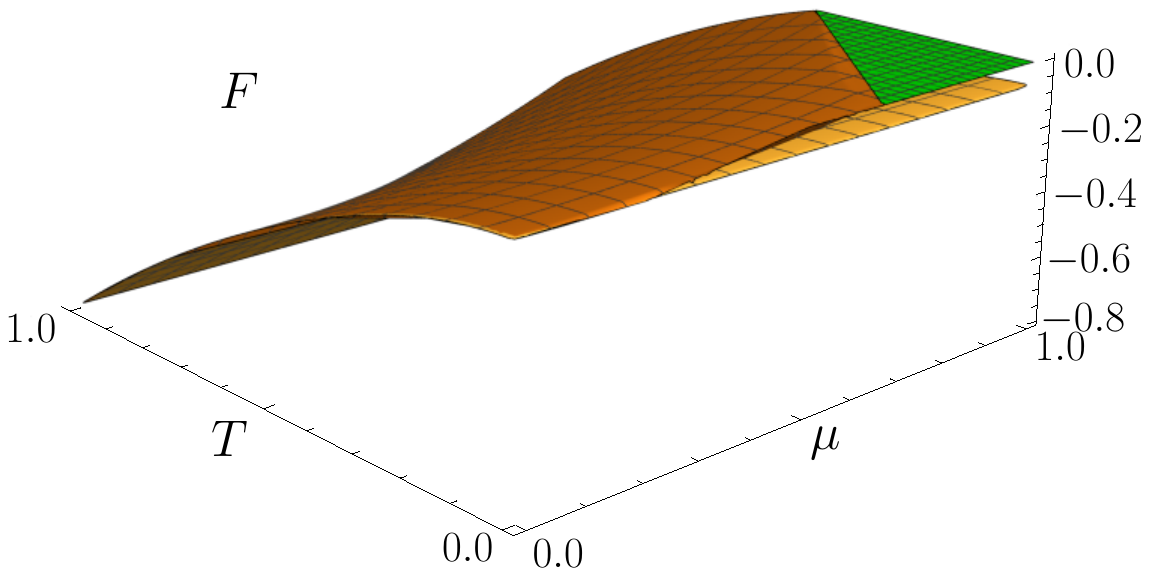}}}
    \caption{The large $N$ free energy $F$ (for periodic fermions) as a function of $\mu$ and
$T$, in units of $M_0$. The massless configuration (green) is favored only at high $\mu$ with respect to the homogeneous configuration (orange), but the spiral configuration (yellow) is always favored with respect to both of them. As a consequence, symmetry is never restored at high temperature.}
    \label{fig:comparisonperiodic}
\end{figure}

It is convenient to separate the contribution from the zero-mode in the sum,
\begin{align}
\begin{split}
    F & =\frac{M^2}{\la_s}-T\left(\sqrt{M^2-(\mu-q)^2}-\sqrt{-(\mu-q)^2}\right) \\
    &\hspace{2cm}-2T\re\sum_{n>0}\left(\sqrt{M^2+p_{2+}^2(q)}-\sqrt{p_{2+}^2(q)}\right)~.
\end{split}    
\end{align}
Like in the antiperiodic case, we regularize the sum by introducing a cutoff $p_{2,\text{max}}=2\pi n_{\text{max}}T$ and we absorb the divergence in coupling $\lambda_s$, which gets replaced by $M_0$. This gives
\begin{align}
    F&=\frac{M^2}{2\pi}\left(\log\left(\frac{4\pi T}{M_0}\right)+\re\psi\left(1+\frac{i(\mu-q)}{2\pi T}\right)\right)-T\left(\sqrt{M^2-(\mu-q)^2}-\sqrt{-(\mu-q)^2}\right) \nonumber \\
    &\qquad-2T\re\sum_{n>0}\left(\sqrt{M^2+p_{2+}^2(q)}-\sqrt{p_{2+}^2(q)}-\frac{M^2}{2\sqrt{p_{2+}^2(q)}}\right)\,.
\end{align}
We now compare the minimum of $F$ over $q$ and $M$ with the one computed assuming translational invariance ($q=0$). In the former case,
$q$-minimization is obtained for $q=\mu$, as in the antiperiodic setup. The extremization with respect to $M$ gives the equation 
\begin{align}
\begin{split}
        &\log\left(\frac{4\pi T}{M_0}\right)+\re\psi\left(1+\frac{i(\mu-q)}{2\pi T}\right)-\pi T\re\frac1{\sqrt{M^2-(\mu-q)^2}}\\
        &-2\pi T\re\sum_{n>0}\left(\frac1{\sqrt{M^2+p_2^2(q)}}-\frac1{\sqrt{p_2^2(q)}}\right)=0\,,
\end{split}  
\end{align}
that we can solve numerically for $M$, in the two cases $q=\mu$ and $q=0$. Plugging back this value, we obtain numerically the value of the free energy, both for $q=0$ and for $q=\mu$. The results are plotted in Figure \ref{fig:comparisonperiodic}.
We see that, as in the antiperiodic setup, the spiral configuration is always favored with respect to the homogeneous one. This time, however, the $M=0$ configuration is never favored with respect to the spiral one at any value of $T$ and $\mu$, i.e. translational invariance is never restored.

\section{Conclusions}
\label{sec:conclusions}

In this work, we have studied phases at finite temperature and density for the chiral Gross-Neveu model, relying on non-abelian bosonization and 't Hooft anomaly matching. 
In particular, we gave a prediction for the long-distance behavior of the two-point function of the determinant operator \eqref{eq:detop}, summarized in \eqref{eq:NS2ptFinal}-\eqref{eq:R2ptFinal}.

Let us discuss some open questions and possible future directions. Inhomogeneous phases of the cGN model at finite $N$ have been found in lattice simulations \cite{Lenz:2021kzo, Koenigstein:2021llr},
using as order parameter the fermion bilinear operator $\psi^\dagger\psi$.\footnote{Note that the sign problem usually associated with the chemical potential is not present in the chiral Gross-Neveu model for $N$ even \cite{Lenz:2020bxk}.} We expect that the qualitative behavior \eqref{eq:NS2ptFinal}-\eqref{eq:R2ptFinal} should hold for
$\psi^\dagger\psi$. It would be nice to put this expectation on firm footing by generalizing our analysis in section \ref{sec:FiniteT} to the two-point function of the fermion bilinear.
On the other hand, at least for sufficiently low $N$, we expect that the two-point function of the determinant operator should be obtainable from lattice methods. 
It would be interesting if lattice results could check our prediction \eqref{eq:NS2ptFinal}-\eqref{eq:R2ptFinal}.

An intriguing phase which breaks translations, dubbed crystal phase, has also been observed at large $N$ in the $O(2N)$ GN theory \cite{Schon:2000he, Thies:2003kk, Schnetz:2004vr}. Whether this phase survives at finite $N$ is an open question, which is being investigated with lattice simulations \cite{Lenz:2020bxk}. Note that in the case of the GN model there is no continuous symmetry that gets mixed with translations, but rather a genuine spontaneous breaking of translations to a discrete group. Non-abelian bosonization techniques are available in this case as well, however the crucial difference is the absence of a free boson sector, which played an important role in many calculations in the chiral case, and in particular was instrumental to derive the chiral spiral. A semiclassical analysis at the level of the Lagrangian of the $Spin(2N)_1$ WZW model, deformed by the chemical potential, might still be fruitful, but one would have to cautiously assess its regime of validity. 
{Interestingly enough, the same 't Hooft $\mathbb{Z}_2$ anomaly responsible for persistent order in the cGN model also occurs in the GN model.}

It would be interesting to investigate the existence of orders similar to the one studied in this paper also in higher dimensional theories. Although lattice studies
suggest that no crystal or chiral spiral phase survives in the continuum limit of the $3d$ versions of the large $N$ GN and cGN models \cite{Narayanan:2020uqt, Buballa:2020nsi, Pannullo:2021edr, Pannullo:2023one} (see also \cite{Koenigstein:2023yzv}), such phases could be present in other theories, see e.g. \cite{Nicolis:2023pye} for a recent study 
at finite chemical potential for $O(N)$  models in $d=3$ and $d=4$ dimensions.

\acknowledgments
We thank Andrea Antinucci, Francesco Benini, Christian Copetti, Pavel Putrov, Ryan Thorngren and Jingxiang Wu for discussions. MS thanks the Institut des Hautes \'Etudes Scientifiques (IHES), where part of this work has been done, for the hospitality. Work partially supported by INFN Iniziativa Specifica ST\&FI. 

\appendix

\section{Gaugings in $d=1+1$}\label{app:gaugings}

In this appendix we review well-known relations among theories defined on a 2d closed manifold $M_2$ upon gauging a discrete symmetry $G$. 
We first review dualities among bosonic\footnote{A two-dimensional theory is bosonic if observables do not depend on the choice of the spin structure $\rho$ on $M_2$ (assumed to be spin).}  theories and then consider those turning a bosonic theory to a fermionic one, and viceversa.
For simplicity, we take abelian cyclic groups, with $G=\mathbb{Z}_n$.

Let $\calT$ be a bosonic theory on $M_2$ with a non-anomalous symmetry $\ZZ_n$. 
The theory obtained by gauging $\ZZ_n$ 
\begin{equation}\label{eq:Bgauging}
    \widehat{\calT}=\calT/\ZZ_n\,,
\end{equation} 
is guaranteed to have a non-anomalous ``dual'' symmetry $\widehat G = \mathrm{Hom}(\ZZ_n,U(1))=\widehat{\ZZ}_n$ \cite{Vafa:1989ih}. Its partition function in the presence of a background $\widehat{\ZZ}_n$ gauge field reads\footnote{We adopt here a notation often used in the literature of denoting respectively by small and capital latin letters dynamical and non-dynamical gauge fields.}
\begin{equation}\label{eq:Bgaugingpf}
    Z_{\widehat{\calT}}[\widehat T]=\frac1{n^g}\! \sum_{t\in H^1(M_2,\ZZ_n)} \!\! Z_\calT[t]e^{\frac{2\pi i}{n}\int \widehat T\,\cup \, t}\,,
\end{equation}
where $\widehat{T}$ denotes the $\widehat{\ZZ}_n$ background insertion, $\cup$ is the cup product in $H^1(M_2,\ZZ_n)$, $g$ is the genus of $M_2$
and $n^g = \sqrt{|H^1(M_2,\ZZ_n)|}$. Gauging the dual symmetry gives back the original theory: 
\begin{equation}
    \widehat{\calT}/\widehat{\ZZ}_n=(\calT/\ZZ_n)/\widehat{\ZZ}_n=\calT\,.
\end{equation}
Explicitly, 
\begin{equation}
    \frac1{n^{g}}\, \sum_{t}Z_{\calT}[t]\,\frac1{n^g}\sum_{\hat{t}} e^{\frac{2\pi i}{n}\int \hat t\, \cup \,t}e^{\frac{2\pi i}{n} \int T\,\cup \,\widehat t}=\frac1{n^g}\sum_t Z_{\calT}[t]n^g\di_{t,T}=Z_{\calT}[T]\,,
\end{equation}
where we have used that 
\begin{equation}
    \frac1{n^g}\sum_ve^{\frac{2\pi i}{n}\int v\cup w }=n^g\di_{w,0}\,.
\end{equation}

Let us consider now fermionic theories, namely those theories where observables do depend on the choice of the spin structure $\rho$. These theories have a $\ZZ_2$ fermion parity 
symmetry $\ZZ_2^F$ for which we can think the spin structure $\rho$ as a choice of background.\footnote{It is tempting to take the correspondence literally, but one cannot identify a spin structure $\rho$ with a background gauge field $S\in H^1(M_2,\ZZ_2)$ for $\ZZ_2^F$ in a natural way. Yet, we can add on top of $\rho$ a $\ZZ_2^F$ connection $S$: this has the net effect of changing the spin structure from $\rho$ to $S\cdot\rho$, which is defined by changing the periodicity of $\rho$ around the cycles along which $S$ has nontrivial holonomy, see \eqref{eq:srhoDef}.}
A standard way to get a bosonic theory out of a fermionic one $\calF$ is to gauge $\ZZ_2^F$, which is equivalent to summing over the spin structures on $M_2$. 
In fact, we get two different bosonic theories ${\calB}$ and $\calB'$, depending on whether we consider $\calF$ or we stack to it the $\Arf$ theory \cite{Fukusumi:2021zme}:
\begin{equation}
\begin{split}
    \label{eq:bosonization}
    \calB & =(\calF\times\Arf)/\ZZ_2^F\,, \\
     \calB' & =\calF/\ZZ_2^F\,.
    \end{split}
\end{equation}
The Arf theory is one of the simplest non-trivial topological theories, the IR limit of the topologically non-trivial phase of the Kitaev chain \cite{Kitaev:2000nmw}.
Its partition function $Z_{\Arf}$ on $M_2$ is the $\Arf$ invariant, which is the index of the Dirac operator $\slashed{D}_\rho$ mod 2 \cite{Atiyah:1971RiemannSA}. 
We have
\begin{equation}
    Z_{\Arf}[\rho]= e^{i\pi\Arf[\rho]}\,.
\end{equation}
It is $+1$ or $-1$ respectively on even or odd spin structures.\footnote{A spin structure $\rho$ is called even (odd) when a Majorana fermion with spin structure $\rho$ has an even (odd) number of zero modes.} For $M_2=T^2$, the only odd spin structure is the one in which fermions are taken to be periodic in both cycles, i.e.
\begin{equation}
    {\Arf[\rho]}=\begin{cases}
        0\,,& \rho=\text{(NS,NS), (NS,R), (R,NS),}\\
        1\,,& \rho=\text{(R,R)\,.}\
    \end{cases}
\end{equation}
Explicitly, we have
\begin{align}
    Z_{\calB}[T,\rho] &  =\frac1{2^g}\sum_{s}Z_{\calF}[s\cdot\rho]Z_{\Arf}[s\cdot\rho]e^{i\pi\int s\, \cup \, T}\,,  \nonumber \\
    Z_{\calB'}[T,\rho]&    =\frac1{2^g}\sum_{s}Z_{\calF}[s\cdot\rho]e^{i\pi\int s\, \cup\, T}\,, \label{eq:ZbZb'}
\end{align}
where in the last step in both relations we have used the fact that having a non-trivial background $\ZZ_2^F$ gauge field $s$ is equivalent to changing the spin structure from $\rho$ to $s\cdot\rho$, where for any one-cycle $\gamma$ on $M_2$
\begin{equation}
(s\cdot\rho)(\gamma)=\rho(\gamma)+\int_\gamma s \mod 2\,.    
\label{eq:srhoDef}
\end{equation}
In \eqref{eq:ZbZb'} $T$ is the background gauge field for the $\ZZ_2$ symmetry dual to $\ZZ_2^F$.
For any fixed $\rho$ the sum over $s$ can be traded for a sum over spin structures $\rho'=s\cdot\rho$ and hence the theories $\calB$ and $\calB'$ do not depend on the choice of the fiducial spin structure $\rho$ and are bosonic.
Taking the fiducial spin structure $\rho$ to be identically zero (that is NS on all non-trivial cycles), it can be shown (see e.g. \cite{Karch:2019lnn,Tachikawa:TASI2019} for details) that 
\begin{equation}
    \sum_s e^{i\pi\int s\,\cup\, w}F[s\cdot\rho]\longmapsto \sum_{\rho'}e^{i\pi(\Arf[w\cdot\rho']-\Arf[\rho'])}F[\rho']\,.
\end{equation}
Therefore,
\begin{equation}\begin{split}\label{eq:bosonizationpf}
    Z_\calB[T] & =\frac1{2^g}\sum_{\rho}Z_\calF[\rho]e^{i\pi\Arf[T\cdot\rho]}\,, \\
       Z_{\calB'}[T] & =\frac1{2^g}\sum_\rho Z_\calF[\rho]e^{i\pi\Arf[T\cdot\rho]-i\pi\Arf[\rho]}\,.
       \end{split}
\end{equation}
We denote the $\ZZ_2$ symmetry of $\calB$ dual to $\ZZ_2^F$ by $\ZZ_2^{\vee}$. It can be shown that the two bosonic theories $\calB$ and $\calB'$ are related by gauging, 
$\calB'=\calB/\ZZ_2^\vee$, and hence the dual $\ZZ_2$ symmetry of $\calB'$ is $\widehat{\ZZ}_2^{\vee}$.

Conversely, to a given a bosonic theory $\calB$ with a non-anomalous symmetry $\ZZ_2$ we can associate a fermionic theory $\calF$ \cite{Jordan:1928wi}. The latter is not unique as it depends on the choice of the specific $\ZZ_2$ symmetry of $\calB$. $\calF$ is obtained by stacking $\calB$ with $\Arf$ and gauging a diagonal $\ZZ_2$ symmetry between the two,
\begin{equation}\label{eq:fermionization}
    \calF=({\calB\times\Arf})/{\ZZ_2}\,.
\end{equation}
More precisely, we have
\begin{equation}\label{eq:fermionizationpf}
    Z_{\calF}[\rho]    =\frac1{2^g}\sum_{t\in H^1(M_2,\ZZ_2)}Z_\calB[t]e^{i\pi\Arf[t\cdot\rho]}\,,
\end{equation}
where $t$ is the dynamical gauge field associated to the $\ZZ_2$ symmetry.
It can be shown that if we gauge $\ZZ_2^F$ in the fermionic theory $\calF$ we have
\begin{equation}
\calF/\ZZ_2^F = \calB'=\calB/\ZZ_2\,.
\label{eq:FBB'Rel}
\end{equation}
We can also define the fermionic theory $\calF'$ as in \eqref{eq:fermionization}, replacing $\calB$ with $\calB'$
\begin{equation}
    \calF'=(\calB'\times\Arf)/\ZZ_2'\,,
    \label{eq:FpBpRel}
\end{equation}
where $\ZZ_2'= \widehat{\ZZ}_2$ is the symmetry of $\calB'$ dual to $\ZZ_2$ of $\calB$, see \eqref{eq:FBB'Rel}.
A simple computation shows that
\begin{equation}
Z_{\calF'}[\rho]=Z_{\calF}[\rho]e^{i\pi\Arf[\rho]}
\label{eq:ZfZf'Rel}
\end{equation}
and hence
\begin{equation}\label{eq:fermionizationarf}
\calF' =\calF \times\Arf\,.
\end{equation}
To close the circle, we can get the original bosonic theory $\calB$ from  $\calF'$ by inverting the above procedure,
\begin{equation}
    \calB=((\calF'\times\Arf)\times\Arf)/\ZZ_2^F=\calF'/\ZZ_2^F\,,
    \label{eq:BFprimeRel}
\end{equation}
that is,
\begin{equation}
    \label{eq:bosonizationarfpf}
    Z_{\calB}[T]=\frac1{2^g}\sum_{\rho}Z_{\calF'}[\rho]e^{i\pi\Arf[T\cdot\rho]+i\pi\Arf[\rho]}\,,
\end{equation}
which agrees with the expression in \eqref{eq:bosonizationpf}, given \eqref{eq:ZfZf'Rel}.
We summarize all these results in the diagram in Figure \ref{fig:bosefermidiagram}.

\begin{figure}[t!]
\centering
\begin{tikzcd}[row sep=large, column sep =4cm]
\calB \arrow[r, harpoon, "{(\cdot\times\Arf)/{\rm diag}(\ZZ_2^\calB {}_{{}_{'}}\ZZ_2^F) }", shift left] \arrow[d, harpoon, shift left, "\cdot/\ZZ_2^\calB"] &  \calF  \arrow[d, " {\times\Arf}",leftrightarrow] \arrow[l, harpoon, "(\cdot\times\Arf)/\ZZ_2^F", shift left]          \\
\calB'=\calB/\ZZ_2^\calB \arrow[r, harpoon, "(\cdot\times\Arf)/{\rm diag}(\ZZ_2^{\calB'}{}_{{}_{'}}\ZZ_2^F)",shift left]  \arrow[u, harpoon, "\cdot/\ZZ_2^{\calB'}", shift left]         & \calF'=\calF\times\Arf   \arrow[l, harpoon, "(\cdot\times\Arf)/\ZZ_2^F", shift left]        
\end{tikzcd}
\label{fig:bosefermidiagram}
\caption{Diagram summarizing the relation through gaugings between the two bosonic $\calB$, $\calB'$ and fermionc $\calF$, $\calF'$ theories in two dimensions.}\end{figure}

A simple example that we use in the main text involves $\calB = U(1)_N$ and $\calB'=U(1)_{4N}$.
The $\ZZ_2$ symmetries are $\ZZ_2^{\calB} =\ZZ_2^P$ and $\ZZ_2^{\calB'} =\ZZ_2^{W'}$, $\ZZ_2^{P,W}$ being the $\ZZ_2$ subgroups of $U(1)_{P,W}$, respectively. In particular, we have
\begin{equation}\label{eq:U14nvsU1n}
    \frac{U(1)_{4N}\times\Arf}{\ZZ_2^P}\times\Arf = \frac{U(1)_N\times\Arf}{\ZZ_2^{W\prime}} \,.
\end{equation}
For $N=1$ the above fermionic theories coincides with the one of a free Dirac fermion (with two different sign choices for the RR spin structures).
We check it when $M_2 = T^2$. The torus partition function of a compact scalar is given by \eqref{eqw:Ztoro}.
We denote by $Z_1[l_a,l_b]$ the partition functions twisted with respect to $\ZZ_2^{W'}$. 
Here $l_a=0,1$ indicates the $l_a$-times twisted sector, while $l_b=0,1$ denotes the partition function computed with the insertion of the $l_b$-th power of the symmetry operator in the sum.
We have
\begin{equation}\label{eq:Z1intermsofZ2}
    \begin{aligned}
        Z_1[0,0]&=\frac1{|\eta(\tau)|^2}\sum_{e,m}q^{\frac12(e+\frac m2)^2}\bar{q}^{\frac12(e-\frac m2)^2}\,,\\
        Z_1[0,1]&=\frac1{|\eta(\tau)|^2}\sum_{e,m}(-1)^mq^{\frac12(e+\frac m2)^2}\bar{q}^{\frac12(e-\frac m2)^2}\,,\\
        Z_1[1,0]&=\frac1{|\eta(\tau)|^2}\sum_{e,m}q^{\frac12(e+\frac m2+\frac12)^2}\bar{q}^{\frac12(e-\frac m2+\frac12)^2}\,,\\
        Z_1[1,1]&=\frac1{|\eta(\tau)|^2}\sum_{e,m}(-1)^mq^{\frac12(e+\frac m2+\frac12)^2}\bar{q}^{\frac12(e-\frac m2+\frac12)^2}\,.
    \end{aligned}
\end{equation}
In order to fermionize the theory, we use \eqref{eq:fermionizationpf} to get
\begin{equation}\label{eq:ZfreeDirac}
\begin{aligned}
    Z_F[\text{NS},\text{NS}]&=\tfrac12\left(Z_1[0,0]+Z_1[0,1]+Z_1[1,0]-Z_1[1,1]\right)=\left|\frac{\theta_3(\tau)}{\eta(\tau)}\right|^2\,,\\
    Z_F[\text{NS},\text{R}]&=\tfrac12\left(Z_1[0,0]+Z_1[0,1]-Z_1[1,0]+Z_1[1,1]\right)=\left|\frac{\theta_4(\tau)}{\eta(\tau)}\right|^2\,,\\
    Z_F[\text{R},\text{NS}]&=\tfrac12\left(Z_1[0,0]-Z_1[0,1]+Z_1[1,0]+Z_1[1,1]\right)=\left|\frac{\theta_2(\tau)}{\eta(\tau)}\right|^2\,,\\
    Z_F[\text{R},\text{R}]&=\tfrac12\left(-Z_1[0,0]+Z_1[0,1]+Z_1[1,0]+Z_1[1,1]\right)=0\,.
\end{aligned}
\end{equation}
We recognize \eqref{eq:ZfreeDirac} to be the torus partition function for a single Dirac fermion in the different spin structures.

\section{$U(1)_{k'}$ WZW vs. $R^2=k'$ compact boson}\label{app:u1boson}
Consider the theory of a 2d compact scalar with radius $R$ on a manifold $M_2$:
\begin{equation}
    S=\frac{1}{8\pi}\int_{M_2}(\de_\mu\phi)^2\dd^2x\,,\qquad \phi\sim\phi+2\pi R\,.
\end{equation}
This can be equivalently described at the Lagrangian level by a compact scalar $\tilde\phi$ of radius $\tilde R=2/R$. 
We focus in what follows to the case where $M_2=T^2$. We introduce holomorphic and anti-holomorphic components of $\phi$,
\begin{equation}
    \phi(z,\bar{z})=\vphi(z)+\bar\vphi(\bar z)\,,\qquad \tilde\phi(z,\bar z)=\vphi(z)-\bar\vphi(\bar z)\,.
\end{equation}
The torus partition function of a compact scalar with generic radius $R$ is given by
\begin{equation}\label{eqw:Ztoro}
    Z_R(\tau,\bar\tau)=\frac1{|\eta(\tau)|^2}\sum_{e,m=-\infty}^\infty q^{h_{e,m}}\bar q^{\bar h_{e,m}}\,,
\end{equation}
where $q=e^{2\pi i \tau}$, and
\begin{equation}\label{eq:hemhemR}
    h_{e,m}=\frac12\left(\frac eR + \frac{mR}2\right)^2\,,\qquad \bar{h}_{e,m}=\frac12\left(\frac eR - \frac{mR}2\right)^2\,,
\end{equation}
are the conformal dimensions of the Virasoro primary operators
\begin{equation}
\begin{aligned}
    {\cal{V}}_{e,m}(z,\bar z)&=\exp\Big[ie\frac{\phi(z,\bar z)}{R} + im\frac{R\tilde\phi(z,\bar z)}2\Big]\\
    &=\exp\Big[i\left(\frac eR +\frac{mR}2\right)\vphi(z)+i\left(\frac eR -\frac{mR}2\right)\bar\vphi(\bar z)\Big]\,.
    \end{aligned}
    \label{eq:vertexOp}
\end{equation}
When $R^2$ is a rational number, additional higher-spin (anti-)holomorphic currents appear which give rise to an extended symmetry algebra,
and the infinite Virasoro primaries of the compact scalar can be rearranged in terms of a finite number of affine characters. If we define
\begin{equation}
    R^2=\frac{2p'}{p}\,,\qquad p,p'\in \mathbb{N}\,,\quad \gcd(p,p')=1\,,
\end{equation}
the partition function \eqref{eqw:Ztoro} can be rewritten as a sum over affine characters as
\begin{equation}\label{eq:compactbosonpf}
    Z_{\sqrt{2p'/p}}(\tau,\bar\tau)=\sum_{\la=0}^{2pp'-1}\chi_\la^{(2pp')}(\tau)\bar\chi_{\om_0\la}^{(2pp')}(\bar\tau)\,,
\end{equation}
where 
\begin{equation}\label{eq:compactbosonchar}
    \chi_{\la}^{(2pp')}(\tau)=\frac1{\eta(\tau)}\sum_{t\in\ZZ}q^{pp'\left(t+\frac\la{2pp'}\right)^2}\,,
\end{equation}
and we have defined
\begin{equation}
    \om_0=p r_0+p's_0\mod 2pp'\,,
\end{equation}
with $(r_0,s_0)$ being any Bézout pair for $(p,p')$, i.e. (positive) integer numbers such that
\begin{equation}
    pr_0-p's_0=1\,.
\end{equation}
The partition function \eqref{eq:compactbosonpf} corresponds to a diagonal RCFT if and only if $\om_0=1\mod 2pp'$, which can happen only if either $p$ or $p'$ is equal to 1. If $p=1$ (the case $p'=1$ can be obtained by T-duality) we have
\begin{equation}
    Z_{\sqrt{2p'}}(\tau,\bar\tau)=\sum_{\la=0}^{2p'-1}|\chi_\la^{(2p')}(\tau)|^2\,.
\end{equation}
This is the partition function of the diagonal $U(1)_{2p'}$ WZW model. When both $p,p'\neq1$, the relation between the compact boson CFT and a $U(1)$ WZW model is more subtle. For our purposes it is enough to consider the case $p=2$ and $p'=N$, with $N$ odd, 
which corresponds to a compact boson with $R^2=N$ odd.  There are $2pp'=4N$ affine characters in this theory. A Bézout pair for $(p,p')=(2,N)$ is $(r_0,s_0)=(\frac{N+1}2,1)$, which leads to $\om_0=2N+1$. The partition function (\ref{eq:compactbosonpf}) reads
\begin{equation}
     Z_{\sqrt{N}}(\tau,\bar\tau) =\sum_{\substack{\la=0\\\la\in2\ZZ}}^{4N-1}|\chi_\la^{(4N)}(\tau)|^2+\sum_{\substack{\la=0\\\la\in2\ZZ+1}}^{4N-1}\chi_\la^{(4N)}(\tau)\bar\chi^{(4N)}_{\la+2N\,\mathrm{mod}\,4N}(\bar\tau)\,,
\end{equation}
and evidently is not a diagonal RCFT. However, it can be seen as the $\ZZ_2$ orbifold of a compact scalar with $R^2=4N$ where $\ZZ_2=\ZZ_{2}^P\subset U(1)_P$ (this is consistent: the $\ZZ_{n}^P$  quotient of the compact boson with radius $R$ is the theory with radius $R/n$). 

In general, if the theory $R^2=2p'/p$ is non diagonal, in particular if it has $p\neq1$, we can realize it as a $\ZZ_p$ orbifold of the diagonal theory with $R^2=2pp'$.\\
Since we will be interested only in theories with $R^2\in\ZZ_{>0}$, it suffices to distinguish the cases $p=1$ and $p=2$, i.e. $R^2$ even or odd, respectively. In summary, the correspondence is
\begin{equation}\label{eq:U1integerlevels}
\begin{aligned}
    R^2=2n \text{ compact boson}&\Longleftrightarrow\hspace{2.5cm} {U}(1)_{2n} \,,\\
    R^2=2n+1 \text{ compact boson}&\Longleftrightarrow\,  \frac{U(1)_{4(2n+1)}}{\ZZ_{2}^P} \equiv U(1)_{2n+1}   \,.
\end{aligned}
\end{equation}
When $N=2n+1$, the parent theory $U(1)_{4N}$ in \eqref{eq:U1integerlevels} is the one entering \eqref{eq:spinunitaryrelation}. 
For both even and odd $N$, we can then effectively take $p=1$ and consider diagonal RCFTs.

\section{$Spin(2N)$ vs $(SU(N)\times U(1))/\ZZ_N$}
\label{app:spinunitary}

In this appendix we show the relation \eqref{eq:spinunitaryrelation} by proving the equality of the partition functions of the theories on the $T^2$ torus.
It is in principle straightforward to show that, upon performing the sum that corresponds to gauging $\mathbb{Z}_N$, the sum of products of twisted/untwisted characters 
on the left-hand side reproduces the sum of squares of characters on the right-hand side. However, in order to make more transparent the origin of the identity and to prove it for generic value of $N$,
we use the well-known relation between chiral algebras in $2d$ and Chern-Simons theories in $3d$ \cite{Witten:1988hf}. 

We briefly review here this correspondence for the particular case of interest in which the Chern-Simons theory is defined on $M_3=S^1\times D_2$ with $\de M_3=T^2$ being the space of the $2d$ 
theory. Upon imposing holomorphic boundary conditions for the Chern-Simons field, the states of a basis of the Hilbert space on $T^2$ are in one-to-one correspondence with chiral affine characters of a $G_k$ WZW model \cite{Elitzur:1989nr,Labastida:1989xp}. Affine characters in the representation $\hat\la$ of the $\widehat\frg_k$ chiral algebra are obtained by inserting the corresponding Wilson line along the non-contractible cycle:
\begin{equation}
    \langle W_{\hat\la}\rangle=\chi_{\hat\la}(\tau)\,.
\end{equation}
The partition function of a $G_k$ WZW model is obtained in $3d$ by gauging a certain subgroup of the one-form symmetry \cite{Gaiotto:2014kfa} of a Chern-Simons theory with gauge group 
$G_k\times G_{-k}$, where $G_k$ and $G_{-k}$ correspond respectively to the holomorphic and anti-holomorphic sector of the WZW model.
The $2d$ orbifold of a $G_k$ WZW can be seen as different gaugings of the $3d$ CS theory.
Wilson lines are at the same time the topological and charged operators of the one-form symmetry. 

\begin{figure}
    \centering
    \scalebox{0.85}{

\tikzset{every picture/.style={line width=0.75pt}}  

\begin{tikzpicture}[x=0.75pt,y=0.75pt,yscale=-1,xscale=1]

\draw    (137,46.6) -- (137,123) ;
\draw [shift={(137,78.3)}, rotate = 90] [fill={rgb, 255:red, 0; green, 0; blue, 0 }  ][line width=0.08]  [draw opacity=0] (8.93,-4.29) -- (0,0) -- (8.93,4.29) -- cycle    ;
\draw  [draw opacity=0] (143.68,89.36) .. controls (159.82,91.15) and (172,99.26) .. (172,109) .. controls (172,120.05) and (156.33,129) .. (137,129) .. controls (117.67,129) and (102,120.05) .. (102,109) .. controls (102,98.83) and (115.3,90.43) .. (132.5,89.16) -- (137,109) -- cycle ; \draw   (143.68,89.36) .. controls (159.82,91.15) and (172,99.26) .. (172,109) .. controls (172,120.05) and (156.33,129) .. (137,129) .. controls (117.67,129) and (102,120.05) .. (102,109) .. controls (102,98.83) and (115.3,90.43) .. (132.5,89.16) ;  
\draw    (283,46.6) -- (283,191) ;
\draw [shift={(283,112.3)}, rotate = 90] [fill={rgb, 255:red, 0; green, 0; blue, 0 }  ][line width=0.08]  [draw opacity=0] (8.93,-4.29) -- (0,0) -- (8.93,4.29) -- cycle    ;
\draw    (137,134) -- (137,191) ;
\draw [shift={(137,156)}, rotate = 90] [fill={rgb, 255:red, 0; green, 0; blue, 0 }  ][line width=0.08]  [draw opacity=0] (8.93,-4.29) -- (0,0) -- (8.93,4.29) -- cycle    ;
\draw    (101.2,108.4) .. controls (103.74,122.76) and (118.41,126.42) .. (125.31,127.81) ;
\draw [shift={(128.2,128.4)}, rotate = 194.04] [fill={rgb, 255:red, 0; green, 0; blue, 0 }  ][line width=0.08]  [draw opacity=0] (8.93,-4.29) -- (0,0) -- (8.93,4.29) -- cycle    ;
\draw    (405,46.6) -- (405,191) ;
\draw [shift={(405,112.3)}, rotate = 90] [fill={rgb, 255:red, 0; green, 0; blue, 0 }  ][line width=0.08]  [draw opacity=0] (8.93,-4.29) -- (0,0) -- (8.93,4.29) -- cycle    ;
\draw    (463,46.6) -- (463,191) ;
\draw [shift={(463,112.3)}, rotate = 90] [fill={rgb, 255:red, 0; green, 0; blue, 0 }  ][line width=0.08]  [draw opacity=0] (8.93,-4.29) -- (0,0) -- (8.93,4.29) -- cycle    ;
\draw    (550,47.6) -- (550,192) ;
\draw [shift={(550,113.3)}, rotate = 90] [fill={rgb, 255:red, 0; green, 0; blue, 0 }  ][line width=0.08]  [draw opacity=0] (8.93,-4.29) -- (0,0) -- (8.93,4.29) -- cycle    ;

\draw (188.07,110.9) node [anchor= west][inner sep=0.75pt]    {$=$};
\draw (130.07,197.9) node [anchor=north west][inner sep=0.75pt]    {$\calO_q $};
\draw (82.07,102.9) node [anchor=north west][inner sep=0.75pt]    {$\lambda $};
\draw (277.07,197.9) node [anchor=north west][inner sep=0.75pt]    {$\calO_q $};
\draw (214,112.9) node [anchor=south west][inner sep=0.75pt]    {$e^{2\pi i \lambda q /n}$};
\draw (494.07,110.9) node [anchor= west][inner sep=0.75pt]    {$=$};
\draw (424.07,110.9) node [anchor= west][inner sep=0.75pt]    {$\times $};
\draw (399.07,197.9) node [anchor=north west][inner sep=0.75pt]    {$\lambda $};
\draw (457.07,199.9) node [anchor=north west][inner sep=0.75pt]    {$\mu $};
\draw (530.07,197.9) node [anchor=north west][inner sep=0.75pt]    {$[ \lambda +\mu ]_{n}$};

\end{tikzpicture}

}
    \caption{ A $\ZZ_n$ one-form symmetry in $3d$ has topological lines labelled by $\la\in\ZZ_n$. They fuse according to the $\ZZ_n$ group law (right), and act on charged operators, which are also supported on lines, by linking (left). Here $[x]_n=x\mod n$ and $q$ is the $\ZZ_n$ charge of the operator $\calO_q$. }
    \label{fig:oneform}
\end{figure}
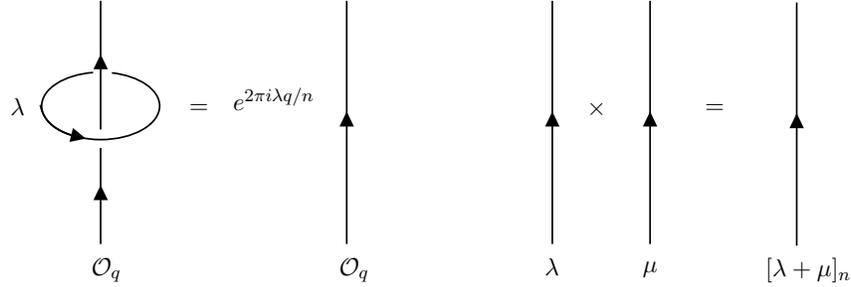

The procedure to gauge a subgroup $\calA$ of a one-form symmetry is well-known \cite{Moore:1988ss,Moore:1989yh} (see also \cite{Hsin:2018vcg}). 
It amounts to sum over insertions of Wilson lines generating $\calA$, which can either fuse or link with the Wilson lines of the theory, see figure \ref{fig:oneform}.
We focus on the case in which $\calA=\ZZ_Q$ is an abelian subgroup of the one-form symmetry of a CS theory. For simplicity we describe the gauging procedure when the gauge group
$G$ is simple (or abelian), the generalizations to products of groups being straightforward. 
If $W_{\hat\mu}$ is the Wilson line generating $\ZZ_Q$, we have $W_{\hat \mu}^Q=1$. The group $\calA$ is gaugeable if and only if $W_{\hat\mu}$
has integer spin $h_{\hat \mu}$ and the lines in ${\calA}$ are mutually transparent, i.e. they have trivial mutual braiding.\footnote{Recall that, given two Wilson lines $W_{\hat\la}$ and $W_{\hat\mu}$, their braiding $B(\hat \la,\hat \mu)$ is given by
\begin{equation}
    B(\hat \la,\hat \mu) = \exp\Big(2i \pi \big(h_{\hat\la+\hat\mu}-h_{\hat\la}-h_{\hat\mu}\big) \Big)\,,
\end{equation}
and the spin of a Wilson line $W_{\hat\la}$ equals the chiral dimension $h$ mod 1 of the corresponding $2d$ affine character $\chi_{\hat \la}$. Trivial braiding means $B=1$.} 
Given $\cal A$, we then select the Wilson lines which have trivial linking with lines in $\calA$, and the lines of the gauged theory are given in terms of orbits under fusion with $\calA$.
Let $P$ be the total number of Wilson lines in the CS theory. The Wilson lines $W_{\hat\la}$ with trivial linking with $W_{\hat\mu}^n$, $n=1,\ldots, Q-1$ are those for which 
\begin{equation}\label{eq:braiding}
     B(\hat \la,\hat \mu) = 1\,.
\end{equation}
There are in general $P/Q$ Wilson lines which satisfy \eqref{eq:braiding}. Such lines organize into $P/Q^2$ gauge-invariant orbits of the form
\begin{equation}\label{eq:orbit}
 O_W =    \oplus_{q=0}^{Q-1} W_{\hat \lambda + q \hat \mu} \,,
\end{equation}
which are the surviving lines in the gauged theory.\footnote{If the line $W_{\hat \lambda}$ is a fixed point under fusion with $W_{\hat \mu}^s$, $s$ being a divisor of $N$, then there are $N/s$ copies of the line $\hat W_{\hat \lambda}$ \cite{Hsin:2018vcg}. In our case the action is always free and no degeneracies occur.}
Inserting a Wilson line orbit  of the gauged theory along the non-contractible cycle of $M_3$ amounts, on the boundary, to sum over the products of characters associated to 
the Wilson lines of the ungauged theory as given in \eqref{eq:orbit}. 
When the gauge group is of the form $G_k\times G_{-k}$, to each orbit we can associate a combination of left- and right-moving characters which we can interpret as a partition function for the associated 2d WZW model, provided the combination has the correct modular properties.

We prove \eqref{eq:spinunitaryrelation} by showing that the left and right-hand side of this equation arise from gauging the same subgroup of one-form symmetry, 
which leads to a single orbit. The CS theories are 
\begin{equation}
\begin{split}\label{eq:CSgauge}
    G_\text{CS} & = SU(N)_1 \times U(1)_N \times SU(N)_{-1} \times U(1)_{-N} \,, \qquad \;\; N \;\text{even} \\
G_\text{CS} & = SU(N)_1 \times U(1)_{4N} \times SU(N)_{-1} \times U(1)_{-4N}\,, \qquad N \;\text{odd}\,.
\end{split}
\end{equation}
We denote in the following by $(j,\la;\bar{k},\bar\mu)$ a general Wilson line of the CS theories, where
$j$ and $\bar k$ are the ranks of the completely antisymmetric representations of $\widehat{\mathfrak{su}}(N)_1$ and $\widehat{\mathfrak{su}}(N)_{-1}$,
while $\la$ and $\bar\nu$ are the charges of $\widehat{\mathfrak{u}}(1)_{Q}$ and $\widehat{\mathfrak{u}}(1)_{-Q}$ .
For $N$ even, $Q=N$ and $\la,\mu\in\ZZ_N$, while for $N$ odd $Q=4N$ and $\la,\mu\in\ZZ_{4N}$.
The spin of $SU(N)_1$ and $U(1)_Q$ Wilson lines is
\begin{equation}
   h_j = \frac{j(N-j)}{2N} \,, \qquad \qquad h_\lambda = \frac{\lambda^2}{2Q}\,.
\end{equation}

\subsection{Even $N$} \label{subsec:spinunitaryNeven}

For $N$ even, the total one-form symmetry is $G^{(1)} = \ZZ_N^4$ and we have a total of $N^4$ Wilson lines. We want to gauge a subgroup $\calA\subset G^{(1)}$, where
$\calA = \ZZ_{N/2}\times\ZZ_N\times\ZZ_2$ is generated by the following set of lines\footnote{This set of generators is not minimal, as $({1},1;\bar{1},\bar{1})$ can be combined with either $(0,0;\bar{2},\bar{2})$ or $(2,2;\bar{0},\bar{0})$ to give the other one.}
\begin{equation}
    \calA=\langle (2,2;\bar{0},\bar{0}),(0,0;\bar{2},\bar{2}),({1},1;\bar{1},\bar{1}), (0,N/2;\bar{0},\overline{N/2})\rangle\simeq\ZZ_{N/2}\times\ZZ_N\times\ZZ_2\,.
\end{equation}
We define the subgroups
\begin{equation}
\begin{aligned}
    \calA_1&=\langle (2,2;\bar{0},\bar{0}),(0,0;\bar{2},\bar{2}) \rangle\simeq \ZZ_{N/2}^2\,,\\
    \calA_2&=\langle (0,N/2;\bar{0},\overline{N/2}),(1,1;\bar{1},\bar{1})\rangle\simeq\ZZ_{N}\times \ZZ_2\,.
  \end{aligned}
\end{equation}  
We also define the coset
\begin{equation}
    \calA_c\equiv \frac{\calA}{\calA_1}=\langle[({1},1;\bar{1},\bar{1})], [(0,N/2;\bar{0},\overline{N/2})]\rangle\simeq\ZZ_2^2\,.
\end{equation}

We gauge $\calA$ in steps, starting with  $\calA_1$. This acts independently on holomorphic and anti-holomorphic sector by projecting into lines with
  \begin{equation}
        j=\la\mod N/2\,\,, \qquad   \bar k=\bar \mu\mod N/2\,.
    \end{equation}
We are left with $(2N)^2$ lines which forms $4^2$ gauge-invariant orbits (each containing $(N/2)^2$ simple lines). In terms of characters, they are given by $\chi_a\bar\chi_b$, where
$a=\mathbf{i},\mathbf{v},\mathbf{s},\mathbf{c}$, and 
    \begin{align}\label{eq:spincharacters}
            \chi_{\mathbf{i}}(\tau)&= \sum_{j,\la=0}^{N/2-1}\chi_{2j}(\tau)\chi_{2\la}(\tau)\di_{j,\la}\,,  \qquad \qquad   \chi_{\mathbf{v}}(\tau)= \sum_{j,\la=0}^{N/2-1}\chi_{2j+1}(\tau)\chi_{2\la+1}(\tau)\di_{j,\la}\,,  \\
            \chi_{\mathbf{s}}(\tau)&= \sum_{j,\la =0}^{N/2-1}\chi_{2j}(\tau)\chi_{2\la+N/2}(\tau)\di_{j,\la}\,,\quad \quad
        \chi_{\mathbf{c}}(\tau)= \sum_{j,\la=0}^{N/2-1}\chi_{2j+1}(\tau)\chi_{2\la+1+N/2}(\tau)\di_{j,\la}\,, \nonumber
    \end{align}
and similarly for the right-moving characters. Note that the left-hand sides of \eqref{eq:spincharacters} coincide with the four affine characters of the $\widehat{\mathfrak{spin}}(2N)_1$ chiral algebra, associated respectively to the identity, the vector and the two spinor representations of $Spin(2N)$, with highest conformal weights
\begin{equation}
    h_{\mathbf{i}}=0\,,\qquad h_{\mathbf{v}}=\frac12\,,\qquad h_{\mathbf{s}}=h_{\mathbf{c}}=\frac N8\,.
    \label{eq:highestweigths}
\end{equation}
Their fusion rules can be computed using Verlinde's formula \cite{Verlinde:1988sn} and the modular matrices associated to the $\widehat{\mathfrak{su}}(N)_1$ and $\widehat{\mathfrak{u}}(1)_N$ algebras. We get as expected
\begin{equation}
    \mathbf{v}\times\mathbf{v}=\mathbf{s}\times\mathbf{s}=\mathbf{c}\times\mathbf{c}=\mathbf{i}\,,\quad \mathbf{s}\times\mathbf{c}=\mathbf{v}\,,\quad \mathbf{v}\times\mathbf{s}=\mathbf{c}\,,\quad \mathbf{v}\times\mathbf{c}= \mathbf {s}\,.
\end{equation}
We now gauge the coset $\calA_c$. This leaves 4 orbits which gives rise to a single orbit of orbits $O_W$ (containing in total $N^2$ simple lines), the one obtained 
by summing the diagonal combination of the above characters. We then have
\begin{equation}\label{eq:OrbitSpin}
       \langle O_W\rangle=|\chi_{\bf i}|^2+|\chi_{\bf v}|^2+|\chi_{\bf s}|^2+|\chi_{\bf c}|^2 = Z_{Spin(2N)_1}\,.
\end{equation}

We now discuss how to obtain the partition function of the  $(SU(N)_1\times U(1)_N)/\ZZ_{N/2}^L$ orbifold theory. To this purpose we can gauge
the whole group $\calA$ at once. The explicit form of the final orbit $O_W$ in terms of simple lines can be written as
\begin{equation}\label{eq:Sorbit}
   O_W  =  \oplus_{\substack{ k=0\\k\, \text{even}}}^{N-2} O_W^{(k)}\,,
\end{equation}
where
\begin{equation}\label{eq:Sorbit2}
\begin{split}
   O_W^{(k)}  = & \oplus_{j=0}^{N/2-1} \bigg[ (j,j;\overline{j+k},\overline{j+k}) \oplus \Big(j,j+\tfrac N2;\overline{j+k},\overline{j+k+\tfrac N2}\Big) \\
   & \oplus \Big(j+\tfrac N2,j;\overline{j+k+\tfrac N2},\overline{j+k}\Big)
   \oplus \Big(j+\tfrac N2,j+\tfrac N2;\overline{j+k+\tfrac N2},\overline{j+k+\tfrac N2}\Big) \bigg]\,.
      \end{split}
\end{equation}
We show below that $O_W^{(k)}$ coincides with the $k$-twisted partition function of the orbifolded theory.
The full partition function reads
\begin{equation} \label{eq:TwistDefSU}
Z_{\frac{SU(N)_1\times U(1)_N}{\ZZ_{N/2}^L}}=\frac 2N\sum_{\substack{k,\ell=0\\ k, \ell\, \text{even}}}^{N-2}Z_{SU(N)_1}[k,\ell]Z_{U(1)_N}{[k,\ell]}
\equiv \sum_{\substack{k=0\\ k\, \text{even}}}^{N-2}Z_{\frac{SU(N)_1\times U(1)_N}{\ZZ_{N/2}^L}}^{(k)}
\,,
\end{equation}
where $[k,\ell]$ denote the $k$-twisted sector with the insertion of $\ell$ $\ZZ_{N/2}^L$ charges of the individual $SU(N)$ and $U(1)$ sectors, 
while $Z^{(k)}$ is the partition function of the orbifolded theory restricted to the states with charge $k$ under the dual symmetry $\widehat \ZZ_{N/2}^L$.
The functions $Z_{SU(N)_1}[k,\ell]$ and $Z_{U(1)_N}[k,\ell]$ can be computed starting from the unwtwisted sector
\begin{equation}
    \begin{aligned}
        Z_{SU(N)_1}[0,\ell]&=\sum_{j=0}^{N-1}e^{-2\pi i \frac{j \ell}{N}}\chi_j\bar\chi_{j}\,,\quad 
        Z_{U(1)_N}[0,\ell]=\sum_{\la=0}^{N-1}e^{2\pi i \frac{\la \ell}{N}}\chi_\la\bar\chi_{\la}\,,
    \end{aligned}
\end{equation}
and applying $S$ and $T$ modular transformations. We have
\begin{equation}
\begin{split}
   S\cdot Z_{SU(N)_1}[k,\ell] & = Z_{SU(N)_1}[N-\ell,k]  \,, \qquad  S\cdot Z_{U(1)_N}[k,\ell] = Z_{U(1)_N}[N-\ell,k] \,, \\
    T\cdot Z_{SU(N)_1}[k,\ell] & = Z_{SU(N)_1}[k,k+\ell]  \,, \qquad  T\cdot Z_{U(1)_N}[k,\ell] = Z_{U(1)_N}[k,k+\ell] \,,
\end{split}   
\end{equation}
where the action of $S$ and $T$ on the characters are given, e.g., in \cite{DiFrancesco:1997nk}.
In particular, we have
\begin{equation}
    \begin{aligned}
        Z_{SU(N)_1}[k,0]&=\sum_{j=0}^{N-1}\chi_j\bar\chi_{j+k}\,,\quad 
        Z_{U(1)_N}[k,0]=\sum_{\la=0}^{N-1}\chi_\la\bar\chi_{\la+k}\,.
    \end{aligned}
\end{equation}
Given the action of the $\ZZ_{N/2}^L$ charges on the characters, the partition functions $Z^{(k)}$ defined in the r.h.s. of \eqref{eq:TwistDefSU} 
are obtained by projecting on the $\ZZ_{N/2}^L$ neutral states:
\begin{equation}\label{eq:Zsuuilines}
Z_{\frac{SU(N)_1\times U(1)_N}{\ZZ_{N/2}^L}}^{(k)} = \frac 2N \sum_{\substack{\ell=0\\ \ell \, \text{even}}}^{N-2}
\sum_{j,\lambda=0}^{N-1} e^{-2\pi i \frac{(j-\lambda) \ell}{N}} \chi_j\bar\chi_{j+k}\chi_\la\bar\chi_{\la+k}\,.
\end{equation}
It is now straightforward to see that the sum over $\ell$ in \eqref{eq:Zsuuilines} gives rise to the combination of characters entering $O_W^{(k)}$, and hence
\begin{equation}\label{eq:IdOWk}
\langle O_W^{(k)}\rangle= Z_{\frac{SU(N)_1\times U(1)_N}{\ZZ_{N/2}^L}}^{(k)} \,,\qquad \forall  k=0,2\ldots N-2\,.
\end{equation}
Suming over $k$ and using \eqref{eq:OrbitSpin}, \eqref{eq:Sorbit} and \eqref{eq:IdOWk},  we immediately get 
\begin{equation}\label{eq:gaugedneven}
 Z_{Spin(2N)_1}  =Z_{\frac{SU(N)_1\times U(1)_N}{\ZZ_{N/2}^L}}^{\ }\,,
\end{equation}
proving \eqref{eq:spinunitaryrelation} for $N$ even.

\subsection{Odd $N$} \label{subsec:spinunitaryNodd}
For $N$ odd, the total one-form symmetry is $G^{(1)} = \ZZ_N^2\times \ZZ_{4N}^2$ and we have a total of $16N^4$ Wilson lines. We want to gauge a subgroup $\calA\subset G^{(1)}$, where
$\calA = \ZZ_{N}^2\times\ZZ_4$ is generated by 
\begin{equation}
    \calA=\langle (1,2N+2;\bar{0},\bar{0}),(0,0;\bar{1},\overline{2N+2}),({0},N;\bar{0},\bar{N})\rangle\simeq\ZZ_{N}^2\times\ZZ_4\,.
\end{equation}
We identify the subgroups
\begin{equation}
    \begin{aligned}
        \calA_1&=\langle(1,2N+2;\bar{0},\bar{0}),(0,0;\bar{1},\overline{2N+2}) \rangle\simeq\ZZ_N^2\,,\\
        \calA_2&=\langle({0},N;\bar{0},\bar{N})\rangle\simeq\ZZ_4\,,\\
    \end{aligned}
\end{equation}
and gauge $\calA_1$ first. As before, this acts independently on holomorphic and anti-holomorphic sectors by projecting into lines with 
\begin{equation}
    2j= (N+1) \la\mod 2N\,, \qquad  2\bar k=(N+1)\bar \mu\mod 2N\,.
\end{equation}
We are left with $(4N)^2$ lines which forms $4^2$ gauge-invariant orbits, with $N^2$ elements. In terms of characters, they are given by $\chi_a\bar\chi_b$, where
$a=\mathbf{i},\mathbf{v},\mathbf{s},\mathbf{c}$, and 
    \begin{equation}
    \begin{aligned}
        \chi_{\mathbf{i}}(\tau)&=\sum_{j=0}^{N-1}\sum_{\la=0}^{4N-1}\chi_{j}(\tau)\chi_{\la}(\tau)\di_{[j(2N+2),\la]_{4N}}\,, \\
        \chi_{\mathbf{v}}(\tau)&=\sum_{j=0}^{N-1}\sum_{\la=0}^{4N-1}\chi_{j}(\tau)\chi_{\la+2N}(\tau)\di_{[j(2N+2),\la]_{4N}}\,,  \\
        \chi_{\mathbf{s}}(\tau)&=\sum_{j=0}^{N-1}\sum_{\la=0}^{4N-1}\chi_{j}(\tau)\chi_{\la+N}(\tau)\di_{[j(2N+2),\la]_{4N}}\,, \\
        \chi_{\mathbf{c}}(\tau)&= \sum_{j=0}^{N-1}\sum_{\la=0}^{4N-1}\chi_{j}(\tau)\chi_{\la+3N}(\tau)\di_{[j(2N+2),\la]_{4N}}\,,    \label{eq:spincharactersNodd}
    \end{aligned}
\end{equation} 
where $\di_{[a,b]_{4N}}$ stands for $a=b \mod 4N$. A similar result applies for the right-moving characters.  
The left hand sides of \eqref{eq:spincharactersNodd} again coincide with the four affine characters of the $Spin(2N)_1$ WZW model, associated respectively to the identity, the vector and the two spinor representations of $Spin(2N)$. Their highest conformal weights are as in the $N$ even case \eqref{eq:highestweigths}, while their fusion rules read instead
\begin{equation}
    \mathbf{v}\times\mathbf{v}=\mathbf{s}\times\mathbf{c}=\mathbf{i}\,,\quad \mathbf{s}\times\mathbf{s}=\mathbf{c}\times\mathbf{c}=\mathbf{v}\,,\quad \mathbf{v}\times\mathbf{s}=\mathbf{c}\,,\quad \mathbf{v}\times\mathbf{c}= \mathbf {s}\,.
\end{equation}
We now gauge $\calA_2$. This leaves 4 orbits which gives rise to a single orbit of orbits $O_W$ (containing in total $4N^2$ simple lines), the one obtained 
by summing the diagonal combination of the above characters. We then have
\begin{equation}\label{eq:OrbitSpinodd}
    \langle O_W\rangle=|\chi_{\bf i}|^2+|\chi_{\bf v}|^2+|\chi_{\bf s}|^2+|\chi_{\bf c}|^2 = Z_{Spin(2N)_1}\,.
\end{equation}

We now discuss how to obtain the partition function of the  $(SU(N)_1\times U(1)_{4N})/\ZZ_{N}^L$ orbifold theory. 
Like for the case of $N$ even, we can gauge the whole group $\calA$ at once. 
The form of the final orbit $O_W$ in terms of simple lines can be written as
\begin{equation}\label{eq:Sorbitodd}
   O_W  =  \oplus_{k=0}^{N-1}O_W^{(k)}  \,,
\end{equation}
where
\begin{equation}\label{eq:Sorbitodd2}
   O_W^{(k)}  =  \oplus_{j=0}^{N-1}\oplus_{s=0}^3 \Big(j,2(N+1)j+s N;\overline{j+k},\overline{2(N+1)(j+k)+s N}\Big) \,.
\end{equation}
We proceed as for the case of $N$ even. The steps are almost identical, so we will be brief.
The full partition function reads
\begin{equation}\label{eq:TwistPFNodd}
Z_{\frac{SU(N)_1\times U(1)_{4N}}{\ZZ_{N}^L}}=\frac 1N\sum_{k,\ell=0}^{N-1}Z_{SU(N)_1}[k,\ell]Z_{U(1)_{4N}}{[2(N+1)k,2(N+1)\ell]}
\equiv \sum_{k=0}^{N-1}Z_{\frac{SU(N)_1\times U(1)_{4N}}{\ZZ_{N}^L}}^{(k)}
\,.
\end{equation}
The functions $Z_{SU(N)_1}[k,\ell]$ and $Z_{U(1)_{4N}}[2(N+1)k,2(N+1)\ell]$ can be computed starting from the unwtwisted sector
\begin{equation}
    \begin{aligned}
        Z_{SU(N)_1}[0,\ell]&=\sum_{j=0}^{N-1}e^{-2\pi i \frac{j \ell}{N}}\chi_j\bar\chi_{j}\,,\quad 
        Z_{U(1)_{4N}}[0,2(N+1)\ell]=\sum_{\la=0}^{4N-1}e^{2\pi i \frac{2(N+1)\la \ell}{4N}}\chi_\la\bar\chi_{\la}\,,
    \end{aligned}
\end{equation}
and applying $S$ and $T$ modular transformations. We have
\begin{equation}
\begin{split}
   S\cdot Z_{SU(N)_1}[k,\ell] & = Z_{SU(N)_1}[N-\ell,k]  \,, \qquad  S\cdot Z_{U(1)_{4N}}[k,\ell] = Z_{U(1)_{4N}}[4N-\ell,k] \,, \\
    T\cdot Z_{SU(N)_1}[k,\ell] & = Z_{SU(N)_1}[k,k+\ell]  \,, \qquad  T\cdot Z_{U(1)_{4N}}[k,\ell] = Z_{U(1)_{4N}}[k,k+\ell] \,.
\end{split}   
\end{equation}
In particular, we get
\begin{equation}
    \begin{aligned}
        Z_{SU(N)_1}[k,0]&=\sum_{j=0}^{N-1}\chi_j\bar\chi_{j+k}\,,\quad 
        Z_{U(1)_{4N}}[2(N+1)k,0]=\sum_{\la=0}^{4N-1}\chi_\la\bar\chi_{\la+2(N+1)k}\,,
    \end{aligned}
\end{equation}
and $Z^{(k)}$  in \eqref{eq:TwistPFNodd} read
\begin{equation}\label{eq:ZsuuilinesNodd}
Z_{\frac{SU(N)_1\times U(1)_{4N}}{\ZZ_{N}^L}}^{(k)} = \frac 1N \sum_{\ell=0}^{N-1}
\sum_{j=0}^{N-1}\sum_{\lambda=0}^{4N-1} e^{-2\pi i \frac{(4j-2(N+1)\lambda) \ell}{4N}} \chi_j\bar\chi_{j+k}\chi_\la\bar\chi_{\la+2(N+1)k}\,.
\end{equation}
The sum over $\ell$ in \eqref{eq:ZsuuilinesNodd} again gives rise to the combination of characters entering $O_W^{(k)}$ in \eqref{eq:Sorbitodd2}, and hence
\begin{equation}\label{eq:IdOWkNodd}
\langle O_W^{(k)}\rangle= Z_{\frac{SU(N)_1\times U(1)_{4N}}{\ZZ_{N}^L}}^{(k)} \,,\qquad \forall  k=0,1\ldots N-1\,.
\end{equation}
Summing over $k$ and using \eqref{eq:OrbitSpinodd}, \eqref{eq:Sorbitodd} and \eqref{eq:IdOWkNodd},  we immediately get 
\begin{equation}\label{eq:gaugednodd}
 Z_{Spin(2N)_1}  =Z_{\frac{SU(N)_1\times U(1)_{4N}}{\ZZ_{N}^L}}\,,
\end{equation}
proving \eqref{eq:spinunitaryrelation} for $N$ odd.

\subsection{Example} 

It is useful to see in some more detai how the $SU(N)_1$ and $U(1)_N$ (or $U(1)_{4N}$ for $N$ odd) primary operators combine in $Spin(2N)_1$ ones.   
For definiteness we consider $N$ even, but simlar considerations apply for $N$ odd.
We classify the spectrum of affine primaries of $SU(N)_1\times U(1)_N$ under the $\ZZ_{N/2}$ symmetry.
Like Wilson lines in the corresponding Chern-Simons theory, affine primaries of $SU(N)_1\times U(1)_N$ are labeled by their representations under left- and right-moving chiral algebras, $(j,\la;\bar{k},\bar\mu),\, j,\bar k,\la,\bar\mu=0,\dots,N-1$. 
Operators with 
\begin{equation}
\begin{aligned}
    \la&\equiv j+Q\mod N/2\,,\qquad\bar{\mu}\equiv \bar{k}+Q\mod N/2\,,\\
    \bar{k}&\equiv j+2\ell\mod N\,,\qquad\ \ \, \bar{\mu}\equiv \la+2\ell\mod N\,,
\end{aligned}
\end{equation}
belong to the charge $Q$ subsector of the $\ell$-twisted Hilbert space under $\ZZ_{N/2}$.
\begin{table}[t!]
\centering
\scalebox{0.7}{
\begin{tabular}{c|c|c} 

$SU(4)_1\times U(1)_4$        & $\ZZ_2^L$ even                                                               & $\ZZ_2^L$ odd                                                                         \\ 
\hline
\multirow{4}{*}{$\calH$}      & \textcolor{gray}{$(0,0;\bar{0},\bar{0}), (2,2;\bar{2},\bar{2})$}                             &{$(0,1;\bar{0},\bar{1}), (2,3;\bar{2},\bar{3})$}  \\
                              & \textcolor{red}{$(1,1;\bar{1},\bar{1}), (3,3;\bar{3},\bar{3})$}            & {$(1,0;\bar{1},\bar{0}), (3,2;\bar{3},\bar{2})$}          \\
                              & \textcolor[rgb]{0,0.502,0}{$(0,2;\bar{0},\bar{2}), (2,0;\bar{2},\bar{0})$} & {$(0,3;\bar{0},\bar{3}), (2,1;\bar{2},\bar{1})$}                   \\
                              & \textcolor{blue}{$(1,3;\bar{1},\bar{3}), (3,1;\bar{3},\bar{1})$}                             & {$(1,2;\bar{1},\bar{2}), (3,0;\bar{3},\bar{0})$}                    \\ 
\hline
\multirow{4}{*}{$\calH_{tw}$} & \textcolor{gray}{$(0,0;\bar{2},\bar{2}), (2,2;\bar{0},\bar{0})$}                             & {$(0,1;\bar{2},\bar{3}), (2,3;\bar{0},\bar{1})$}  \\
                              & \textcolor{red}{$(1,1;\bar{3},\bar{3}), (3,3;\bar{1},\bar{1})$}                             & {$(1,2;\bar{3},\bar{0}), (3,0;\bar{1},\bar{2})$}          \\
                              & \textcolor[rgb]{0,0.502,0}{$(0,2;\bar{2},\bar{0}), (2,0;\bar{0},\bar{2})$}                             & {$(0,3;\bar{2},\bar{1}), (2,1;\bar{0},\bar{3})$}                   \\
                              & \textcolor{blue}{$(1,3;\bar{3},\bar{1}), (3,1;\bar{1},\bar{3})$}                             & {$(1,0;\bar{3},\bar{2}), (3,2;\bar{1},\bar{0})$}                    \\

\end{tabular}
\begin{tabular}{c|c|c}
$\frac{SU(4)_1\times U(1)_4}{\ZZ_2}$ & $\widehat\ZZ_2$ even                                                           & $\widehat\ZZ_2$ odd                                                                     \\ 
\hline
\multirow{4}{*}{$\widehat\calH$}         & \textcolor{gray}{$(0,0;\bar{0},\bar{0}), (2,2;\bar{2},\bar{2})$}                             & \textcolor{gray}{$(0,0;\bar{2},\bar{2}), (2,2;\bar{0},\bar{0})$}                                      \\
                                     & \textcolor{red}{$(1,1;\bar{1},\bar{1}), (3,3;\bar{3},\bar{3})$}            & \textcolor{red}{$(1,1;\bar{3},\bar{3}), (3,3;\bar{1},\bar{1})$}                                      \\
                                     & \textcolor[rgb]{0,0.502,0}{$(0,2;\bar{0},\bar{2}), (2,0;\bar{2},\bar{0})$} & \textcolor[rgb]{0,0.502,0}{$(0,2;\bar{2},\bar{0}), (2,0;\bar{0},\bar{2})$}                                      \\
                                     & \textcolor{blue}{$(1,3;\bar{1},\bar{3}), (3,1;\bar{3},\bar{1})$}                             & \textcolor{blue}{$(1,3;\bar{3},\bar{1}), (3,1;\bar{1},\bar{3})$}                                      \\ 
\hline
\multirow{4}{*}{$\widehat\calH_{tw}$}    & {$(0,1;\bar{0},\bar{1}), (2,3;\bar{2},\bar{3})$}                             & {$(0,1;\bar{2},\bar{3}), (2,3;\bar{0},\bar{1})$}  \\
                                     & {$(1,0;\bar{1},\bar{0}), (3,2;\bar{3},\bar{2})$}                             & {$(1,2;\bar{3},\bar{0}), (3,0;\bar{1},\bar{2})$}          \\
                                     & {$(0,3;\bar{0},\bar{3}), (2,1;\bar{2},\bar{1})$}                             & {$(0,3;\bar{2},\bar{1}), (2,1;\bar{0},\bar{3})$}                   \\
                                     & {$(1,2;\bar{1},\bar{2}), (3,0;\bar{3},\bar{0})$}                             & {$(1,0;\bar{3},\bar{2}), (3,2;\bar{1},\bar{0})$}                   
\end{tabular}}
\caption{Affine $SU(4)_1\times U(1)_4$ primaries spectrum and their symmetry properties under the $\ZZ_2$ symmetry before gauging (left) and 
under its dual $\widehat{\ZZ}_2$ symmetry after gauging (right). We report in different colors the $SU(4)_1\times U(1)_4$ primaries $(j,\la;\bar{k},\bar\mu)$ that combine together to form $Spin(8)_1$ primaries, namely \textcolor{gray}{identity}, \textcolor{red}{vector}, \textcolor[rgb]{0,0.502,0}{spinor}, \textcolor{blue}{conjugate spinor}. There are also twisted primaries living at the end of a $\widehat\ZZ_2$ line defect, which do not correspond to any $\widehat{\mathfrak{spin}}(8)_1$ representation.  
}

\label{tab:u4orbifold}
\end{table}

\begin{table}[t!]
\centering
\begin{tabular}{c|c|c}
$Spin(8)_1$  & $\ZZ_2^\vee$ even                      & $\ZZ_2^\vee$ odd                       \\
\hline
$\calH$      & (\textcolor{gray}{\bf i}, \textcolor{gray}{\bf  \textoverline{i}}), (\textcolor{red}{\bf v}, \textcolor{red}{\bf \textoverline{v}}) & (\textcolor[rgb]{0,0.502,0}{\bf s}, \textcolor[rgb]{0,0.502,0}{\bf \textoverline{s}}), (\textcolor{blue}{\bf c}, \textcolor{blue}{\bf \textoverline{c}}) \\
\hline
$\calH_{tw}$ & (\textcolor[rgb]{0,0.502,0}{\bf s}, \textcolor{blue}{\bf \textoverline{c}}), (\textcolor{blue}{\bf c}, \textcolor[rgb]{0,0.502,0}{\bf \textoverline{s}}) & (\textcolor{gray}{\bf i}, \textcolor{red}{\bf \textoverline{v}}), (\textcolor{red}{\bf v}, \textcolor{gray}{\bf \textoverline{i}})
\end{tabular}
\begin{tabular}{c|c|c}
4 Diracs  & $\ZZ_2^F$ even                      & $\ZZ_2^F$ odd                       \\
\hline
$\calH_{\rm NS}$      & (\textcolor{gray}{\bf i}, \textcolor{gray}{\bf  \textoverline{i}}), (\textcolor{red}{\bf v}, \textcolor{red}{\bf \textoverline{v}}) & (\textcolor{gray}{\bf i}, \textcolor{red}{\bf \textoverline{v}}), (\textcolor{red}{\bf v}, \textcolor{gray}{\bf \textoverline{i}}) \\
\hline
$\calH_{\rm R}$ & (\textcolor[rgb]{0,0.502,0}{\bf s}, \textcolor[rgb]{0,0.502,0}{\bf \textoverline{s}}), (\textcolor{blue}{\bf c}, \textcolor{blue}{\bf \textoverline{c}}) & (\textcolor[rgb]{0,0.502,0}{\bf s}, \textcolor{blue}{\bf \textoverline{c}}), (\textcolor{blue}{\bf c}, \textcolor[rgb]{0,0.502,0}{\bf \textoverline{s}})
\end{tabular}
\caption{Affine $Spin(8)_1$ primaries spectrum and their symmetry properties under the $\ZZ_2^\vee$ symmetry before fermionization (left) and 
under $\ZZ_2^F$ symmetry after fermionization (right). We use the same color coding as in Table \ref{tab:u4orbifold}.  
}

\label{tab:4diracsspectrum}
\end{table}
As well-known, $\ZZ_{N/2}$-charged $Q$ operators in the untwisted Hilbert space sector of the ungauged theory get mapped to $\widehat\ZZ_{N/2}$-neutral operators in the $Q$-twisted Hilbert space sector in the gauged theory, and viceversa. Therefore all neutral states in the original theory are in the untwisted Hilbert space of the gauged theory. These $SU(N)_1\times U(1)_N$ primaries then can be rearranged in such a way to give the diagonal modular invariant of the $Spin(2N)_1$ WZW model, as expressed in (\ref{eq:spincharacters}). On the other hand,  
operators in the gauged theory with $\widehat\ZZ_{N/2}$ background twists cannot be described in terms of $\widehat{\mathfrak{spin}}(2N)_1$ affine primaries. Moreover, the $\ZZ_2^\vee$ symmetry that gets fermionized is defined only in the $\widehat{\ZZ}_{N/2}$-untwisted sector.

Let us consider $N=4$ as example. 
We report in Table \ref{tab:u4orbifold} the symmetry properties of the affine primaries of $SU(4)_1\times U(1)_4$. With different colors we highlight how they combine to form local $\widehat{\mathfrak{spin}}(8)_1$ primaries, in the absence of $\widehat\ZZ_2$ twists. There are also twisted primaries living at the end of a $\widehat\ZZ_2$ line defect, but these non-local operators are not $\widehat{\mathfrak{spin}}(8)_1$ affine primaries, and get projected out in the $\ZZ_2^L$ orbifold. 
We consider also the $\ZZ_2^\vee$ backgrounds for $Spin(8)_1=({SU(4)_1\times U(1)_4})/{\ZZ_2^L}$, and its fermionization to $4$ Dirac fermions classifying the operator content according to the $\ZZ_2^F$ properties, in Table \ref{tab:4diracsspectrum}.

\section{Free field realization at $N$ flavors}
\label{app:PotentialAbBos}

The $SU(N)_1$ WZW model can also be described in terms of $N-1$ compact scalars
\cite{Banks:1975xs}.
This description makes manifest only the $U(1)^{N-1}$ Cartan subalgebra of $SU(N)_1$. 
However, it is useful here because it allows us to write in a tractable form the current-current deformation.
Let $\theta_\ell\,, \ell=1,\dots,N-1$, be scalars with radius $r=\sqrt{2}$. In these variables, the undeformed $SU(N)_1$ theory reads
\begin{equation}
    \calL_0=\sum_{\ell=1}^{N-1}\frac2{8\pi}\de_+\thi_\ell\de_-\thi_\ell\,,
\end{equation}
where we have made explicit the radius in the normalization of the kinetic term, so that with these conventions $\thi_\ell\sim\thi_\ell+2\pi$. 
The $SU(N)$ currents are given by 
\begin{equation}
    J_\pm^\ell= -\frac{1}{4\pi}\de_\pm\thi_\ell\,,\qquad 
    J_+^{\al}=\frac{i}{2\pi}\exp({i\al\cdot\vartheta})\,,\qquad {J}_-^{\al}=\frac{i}{2\pi}\exp({-i\al\cdot{\bar\vartheta}})\,,
\end{equation}
where $\vartheta_\ell$ and $\bar\vartheta_\ell$ are the holomorphic and antiholomorphic components of $\theta_l$, and $\alpha\in \Delta^+$ are the positive roots of the $su(N)$ algebra.
We then have
\begin{equation}
 \sum_A   J_+^AJ_-^A =\frac1{16\pi^2}\de_+\thi_\ell\de_-\thi_\ell -\frac{1}{2\pi^2}\sum_{\al\in \Delta^+}\cos(\al\cdot\thi)\,.
\end{equation}
The scalar potential can be rewritten in a more explicit form as 
\begin{equation}\label{eq:Vexp}
    \sum_{\al\in \Delta^+}\cos(\al\cdot\thi)=\sum_{p=1}^{N-1}\sum_{i_1>i_2>\dots>i_p=1}^{N-1}\cos\left(\sum_{\ell=1}^{N-1}(A_{i_1\ell}+ A_{i_2\ell}+\dots + A_{i_p\ell})\thi_\ell \right)\,,
\end{equation}
where $A_{ij}$ is the $su(N)$ Cartan matrix. 
Using \eqref{eq:Vexp} it is not difficult to see that the potential has exactly $N$ classical minima, attained at
\begin{equation}
    \thi_\ell=\Theta^{(k)}_\ell\equiv\frac{2\pi\ell}{N} k\,,\qquad k=0,1,\dots, N-1\,.
\end{equation}

It is instructive to look at the symmetries preserved by the $J\bar{J}$ deformation in this language. Of the full $PSU(N)_V\times \ZZ_N^L$ symmetry, the ones that are explicit are only a $\ZZ_N^K\times\ZZ_N^S\times\ZZ_N^L$ subgroup, with $\ZZ_N^K\times \ZZ_N^S\subset PSU(N)_V$ being `clock' and `shift' transformations. The action on the matrix field $U$ is as follows,
\begin{equation}
    \begin{cases}
        \ZZ_N^L:& U\mapsto e^{\frac{2\pi i}N}U\,,\\
        \ZZ_N^K:&U\mapsto M_K UM_K^\dag\,,\qquad\!\!\! (M_K)_{ab}=e^{\frac{2\pi i (a-1)}N}\di_{a,b}\,,\\
        \ZZ_N^S:& U\mapsto M_S U M_S^\dag\,,\qquad (M_S)_{ab}=\di_{a,b-1\!\!\mod N}\,,
    \end{cases}
\end{equation}
which translates to the following action on the (anti-)holomorphic components of $\thi_\ell$,
\begin{equation}
    \begin{cases}
        \ZZ_N^L:& \vartheta_\ell\mapsto \vartheta_\ell + \frac{2\pi}N\ell\,,\qquad \bar\vartheta_\ell\mapsto\bar\vartheta_\ell\,,\\
        \ZZ_N^K:&\vartheta_\ell\mapsto \vartheta_\ell+\frac{2\pi}N\ell\,,\qquad \bar\vartheta_\ell\mapsto \bar\vartheta_\ell - \frac{2\pi}N\ell\,,\\
        \ZZ_N^S:& \vartheta_\ell\mapsto \vartheta_{\ell+1}-\vartheta_1\,,\qquad\!\!\!\bar\vartheta_\ell\mapsto \bar\vartheta_{\ell+1}-\bar\vartheta_1\,,\qquad \vartheta_{N}=\bar\vartheta_{N}\equiv0\,,
    \end{cases}
\end{equation}
The configurations $\Theta_\ell^{(k)}$ spontaneously break the $\ZZ_N^L$ symmetry: under $\ZZ_N^L$, $\Theta_\ell^{(k)}\mapsto\Theta_\ell^{(k+1)}$; on the other hand, they preserve the $\ZZ_N^K\times \ZZ_N^S$ symmetry.

\subsection{The case $N=2$}
For $N=2$, the deformed theory is a sine-Gordon model, for which $\langle \Tr U\rangle$ can be exactly computed and shown to be non-vanishing.
The non-abelian $J^A_+{J}^A_-$ deformation modifies the free $SU(2)_1=U(1)_2$ WZW model, 
\begin{equation}
    \calL=\frac{2}{8\pi}\de_+\thi\de_-\thi+\frac\la{16\pi^2}\de_+\thi\de_-\thi -\frac{\la}{4\pi^2}\cos(2\thi)\,,
\end{equation}
with $\thi\sim\thi+2\pi$ in this normalization.
This is also known as the $2$-folded sine-Gordon model introduced in \cite{Bajnok:2000wm}. The $k$-folded sine-Gordon model differs from the original sine-Gordon model as the target space for the scalar is wrapped into a circle in order to have exactly $k$ minima of the potential. This spontaneously breaks the $U(1)$ translational symmetry of the scalar to its $\ZZ_k$ subgroup. 
We have two classical degenerate minima of the potential, at $\thi=0$ and $\thi=\pi$.

In the undeformed $SU(2)_1$ theory $U$ is a $(h,\bar{h})=(1/4,1/4)$ field transforming in the bifundamental representation of $SU(2)$.
In terms of the compact scalar $\thi$, up to unitary transformations, we have
\begin{equation}\label{eq:Uthimap}
    U^a{}_b=\frac1{\sqrt{2}}\left(\begin{array}{cc}
        e^{i\thi} & -e^{i\tilde\thi} \\
        e^{-i\tilde\thi} & e^{-i\thi}
    \end{array}\right)\,,
\end{equation}
where $\tilde\thi$ is the T-dual of $\thi$. Note that $ \cos(\thi) = \Tr U/\sqrt{2}$ is a local operator and takes the classical value $+1$ at $\thi=0$ and $-1$ at $\thi=\pi$.
The exact quantum vacuum expectation value of $\langle\Tr U\rangle$ can be computed using the results of \cite{Lukyanov:1996jj}, where a formula
for one-point functions of vertex operators in the sine-Gordon model is derived. Using eq.(20) of \cite{Lukyanov:1996jj} we get
\begin{equation}\label{eq:truvev}
\small{
    \langle \Tr U\rangle =\sqrt{2} {\left[\frac{2M \Gamma\left(\frac{1+\xi}{2}\right)}{4\pi^{\frac 32} \Gamma\left(\frac{\xi}{2}\right)}\right]^{\frac{\xi}{2(1+\xi)}} }
\!\! \exp{\Big(A(\xi)\Big)}\,,
 }
\end{equation}
where
\begin{equation}
A(\xi)= \int_0^{\infty}\!\! \frac{d t}{t}\left[\frac{\sinh ^2\Big(\frac{\xi \,t}{\xi+1}\Big)}{2\sinh \Big(\frac{\xi\, t}{\xi+1} \Big) \sinh (t) \cosh \left(\frac{1}{\xi+1} t\right)}-\frac{\xi e^{-2 t}}{2(1+\xi)} \right]\,, \qquad \xi = \frac{4\pi}{\lambda}\,,
  \end{equation}
and $M$ is the mass of the sine-Gordon soliton, which is exactly determined in terms of $\xi$: 
\begin{equation}\label{eq:soliton}
    M = \frac{2^{\frac{1-\xi}{2}}\Gamma\big(\frac{\xi}{2}\big)}{\sqrt{\pi}\Gamma\big(\frac{1+\xi}{2}\big)}\left(\frac{\Gamma\big(\frac{1}{1+\xi}\big)}{\xi \,\Gamma\big(\frac{\xi}{1+\xi}\big)}   \right)^{\frac{1+\xi}{2}}\,.
\end{equation}
\begin{figure}[t]
    \centering
    \includegraphics[scale=0.5]{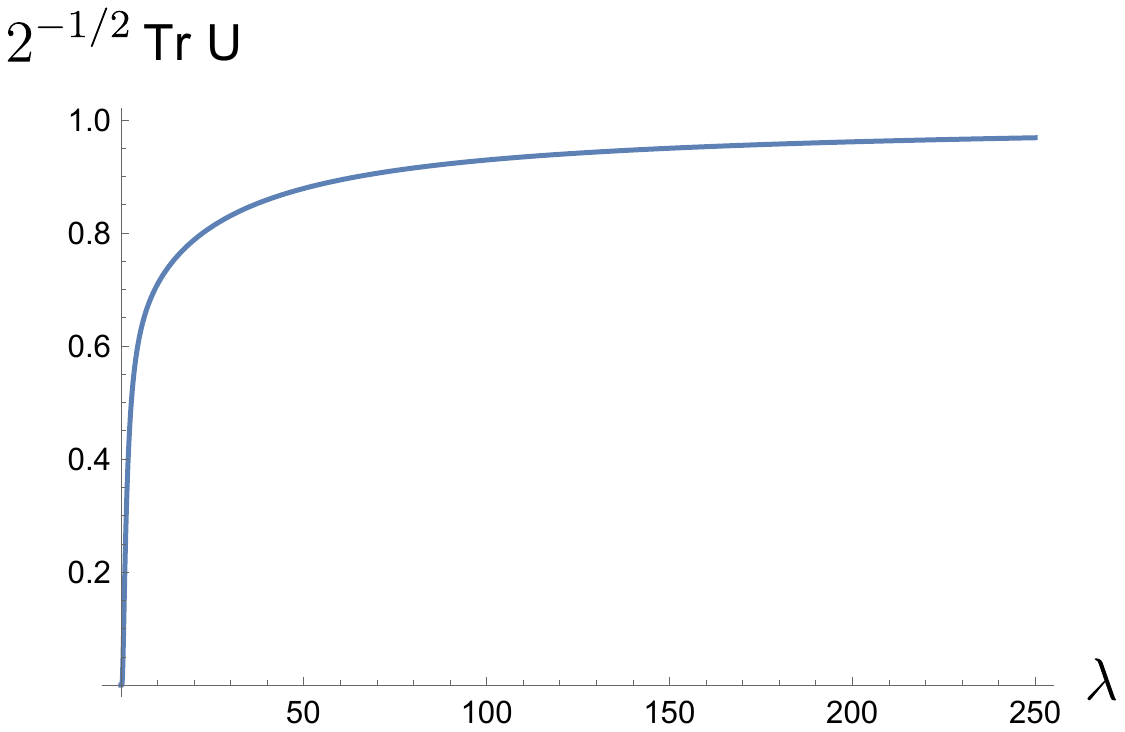}
    \caption{The value of $\langle \Tr U\rangle/\sqrt{2}$ as a function of $\la$.}
    \label{fig:truvev}
\end{figure}
We report in figure \ref{fig:truvev} the value of ${\rm tr}\,U/\sqrt{2}$ as a function of $\lambda$. As can be seen, it is non-vanishing for any value of $\la>0$.
For large $\lambda$ we enter a semi-classical regime. The soliton mass \eqref{eq:soliton} reads
\begin{equation}
    M \approx \frac{\sqrt{2}\lambda}{2\pi^2} (1+{\cal O}(\lambda^{-1}))\,,
\end{equation}
and agrees with its semi-classical value. Consistently, ${\rm tr}\,U$ approaches the classical value $\sqrt{2}$.

\section{Twisted partition functions on a torus}\label{app:twistedapp}
\subsection{$\ZZ_2$ backgrounds for the bosonized theory}\label{app:spinbkg}

As discussed in appendix \ref{app:gaugings}, to obtain a fermionic theory from a bosonic one should `fermionize' an appropriate non-anomalous $\ZZ_2$ symmetry. In the case of $N$ free Dirac fermions, where the role of the bosonic theory is played by $Spin(2N)_1$, the precise correspondence is given by \eqref{eq:fermionization0}. In the presence of a nontrivial background for the $\ZZ_2^L$ symmetry, the partition function of the bosonic theory is well-known to be
\begin{equation}\label{eq:Zspinbkg}
     Z_{Spin(2N)_1}[S_a,S_b] = \begin{cases}
         |\chi_{\bf i}|^2+|\chi_{\bf v}|^2+|\chi_{\bf s}|^2+|\chi_{\bf c}|^2\,,& [S_a,S_b]=[0,0]\,,\\
         |\chi_{\bf i}|^2+|\chi_{\bf v}|^2-|\chi_{\bf s}|^2-|\chi_{\bf c}|^2\,,&[S_a,S_b]=[0,1]\,,\\
         +\chi_{\bf i}\bar\chi_{\bf v}+\chi_{\bf s}\bar\chi_{\bf c}+ c.c.\,,&[S_a,S_b]=[1,0]\,,\\
         -\chi_{\bf i}\bar\chi_{\bf v}+\chi_{\bf s}\bar\chi_{\bf c}+ c.c.\,,&[S_a,S_b]=[1,1]\,,
     \end{cases}
\end{equation}
where $S=[S_a,S_b]$, $S_a,S_b\in\{0,1\}$, denotes the $\ZZ_2^L$ holonomy on the corresponding cycle of the torus.

Via the mapping between $Spin(2N)_1$ and $SU(N)_1 \times U(1)_Q$ affine characters derived in appendix \ref{app:spinunitary}, the twisted partition functions \eqref{eq:Zspinbkg} can be expressed also in terms of twisted $SU(N)_1\times U(1)_Q$ partition functions.

For $N$ even, \eqref{eq:spinunitaryrelation} generalizes in the presence of a $\ZZ_2^L$ background to 
\begin{equation}\label{eq:spinbkgneven}
    Z_{Spin(2N)_1}[S_a,S_b]=\frac 2{N}\!\sum_{t_a,t_b=0}^{N/2-1}\! Z_{SU(N)_1}[2t_a+S_a,2t_b+S_b]Z_{U(1)_{N}}[2t_a+S_a,2t_b+S_b]\,,
\end{equation}
which for $[S_a,S_b]=[0,0]$ reproduces the corresponding expression in appendix \ref{app:spinunitary}. 

For $N$ odd, a similar relation is readily derived,
\begin{align}\label{eq:spinbkgnodd}
    Z_{Spin(2N)_1}[S_a,S_b]=\frac1{N}&\sum_{t_a,t_b=0}^{N-1}Z_{SU(N)_1}[t_a,t_b]\\ &
    \qquad\times Z_{U(1)_{4N}}[2(N+1)t_a+2NS_a,2(N+1)t_b+2NS_b]\,.\nonumber
\end{align}

\subsection{Two-point function of vertex operators on a torus}\label{app:vertex}
\paragraph{Free case.}
Let $\phi$ be a free compact boson of radius $R$, $\phi\sim\phi+2\pi R$, living on the torus with parameter $\tau$. 
Because of electric and magnetic neutrality, the only non-vanishing two-point functions of vertex operators are of the kind
\begin{equation}\label{eq:twopoint}
    \left\langle\calV_{e,m}(z_1,\bar{z}_1)\calV_{-e,-m}(z_2,\bar{z}_2)\right\rangle\,.
\end{equation}
We follow  \cite{DiFrancesco:1997nk} and adopt 
their notation. A standard computation yields\footnote{Note that there is a typo in the first line of eq.(12.148) of \cite{DiFrancesco:1997nk}.  The correct expression is reported in \eqref{eq:TwoPFtorusV}, with $K$
as defined in \eqref{eq:Kdef}.} 
\begin{equation}\label{eq:TwoPFtorusV}
\begin{aligned}
     Z_R&\langle\calV_{e,m}(z_1,\bar{z}_1)\calV_{-e,-m}(z_2,\bar{z}_2)\rangle\\&=  K(z_{12}) \sum_{e^{\prime}, m^{\prime}} q^{h_{e^{\prime}, m^{\prime}}} \bar{q}^{\bar{h}_{e^{\prime}, m^{\prime}}} e^{4 \pi i \left[\alpha_{e^{\prime}, m^{\prime}} \alpha_{e, m} z_{12}-\bar{\alpha}_{e^{\prime}, m^{\prime}} \bar{\alpha}_{e, m} \bar{z}_{12}\right]} \,,
\end{aligned}
\end{equation}
where $Z_R$ is the torus partition function \eqref{eqw:Ztoro}, $z_{12}=z_1-z_2$, and
\begin{equation}\label{eq:Kdef}
\begin{split}
    K(z_{12}) & \equiv \left(\frac{\de_z\thi_1(0|\tau)}{\thi_1(z_{12}|\tau)}\right)^{2h_{e,m}}\left(\frac{\overline{\de_z\thi_1(0|\tau)}}{\overline{\thi_1(z_{12}|\tau)}}\right)^{2\bar{h}_{e,m}}  \frac{1}{|\eta(\tau)|^{2}}  \,, \\
\alpha_{e, m}& =\frac{1}{\sqrt{2}}\left(\frac eR+\frac{mR}2\right), \qquad \bar{\alpha}_{e, m}=\frac{1}{\sqrt{2}}\left(\frac eR-\frac{mR}2\right)\,.
    \end{split}
\end{equation}

\paragraph{Nontrivial $\ZZ_N^P$ background.}
\label{subsec:nontrivial}

We can repeat the above computation in the presence of a nontrivial $\ZZ_N^P\subset U(1)_{P}$ background on the torus. The addition of a $\ZZ_N^P$ background alters the boundary conditions of the scalar field. Let $P_a,P_b\in\ZZ_N$ denote $\ZZ_N^P$ holonomy on the two cycles. One should then compute partition functions with the set of boundary conditions
\begin{equation}
\begin{aligned}
    \phi(z+1,\bar{z}+1)&=\phi(z,\bar{z})+2\pi R(n+P_a/N)\,,\\ \phi(z+\tau,\bar{z}+\bar\tau)&=\phi(z,\bar{z})+2\pi R(n'+P_b/N)\,,
    \end{aligned}
\end{equation}
where $n,n'\in\ZZ$ are to be summed over.

With a computation formally identical to the one performed in the previous paragraph, one gets
\begin{equation}
    \begin{aligned}
        Z_R[P_a,P_b]&=\frac{1}{|\eta(\tau)|^2}\sum_{e,m}q^{h_{e,m+\frac {P_a}N}}\bar{q}^{\bar{h}_{e,m+\frac {P_a}N}}e^{-\frac{2\pi i e P_b}{N}}\,,
    \end{aligned}
\end{equation}
and similarly, for the (unnormalized) correlation function,
\begin{equation}
\begin{aligned}
   &Z_R[P_a,P_b]\left\langle\calV_{e,m}(z_1,\bar{z}_1)\calV_{-e,-m}(z_2,\bar{z}_2)\right\rangle_{[P_a,P_b]}\\
   &=  K(z_{12})  \sum_{e^{\prime}, m^{\prime}} q^{h_{e^{\prime}, m^{\prime}+\frac {P_a}N}}
    \bar{q}^{\bar{h}_{e^{\prime}, m^{\prime}+\frac {P_a}N}}  e^{-\frac{2\pi i e' P_b}{N}} e^{4 \pi i \left[\alpha_{e^{\prime}, m^{\prime}+\frac {P_a}N} \alpha_{e, m} z_{12}-\bar{\alpha}_{e^{\prime}, m^{\prime}+\frac {P_a}N} \bar{\alpha}_{e, m} \bar{z}_{12}\right]}\,.
\end{aligned}
\end{equation}

\paragraph{Nontrivial $\ZZ_2^L$ background.}
We would like to turn on a background for the symmetry (see \eqref{eq:continuousglobal} and the discussion below),
\begin{equation}\label{eq:Z2Lact}
    \ZZ_2^L:\qquad \phi\mapsto\phi+\frac{2\pi \ell}{2}pR\,,\qquad \tilde\phi+ \frac{2\pi \ell}{2}p'\frac{2}R\,,\qquad \ell=0\,,1\,.
\end{equation}
In \eqref{eq:Z2Lact} $R$ is a generally irrational radius, $p$ and $p'$ are coprime positive integers such that $\ZZ_2^L$ matches the $\ZZ_2$ subgroup of the $U(1)_L$ symmetry at the rational value $R^2=2p'/p$ when the deformation is turned off. Modular covariance is lost in presence of the $\ZZ_2^L$ background because $\ZZ_2^L$ is anomalous. In order to restore modular covariance, we should consider a product theory of the compact scalar with another theory with equal (and opposite) anomaly for a $\ZZ_2^L$ symmetry. Then turning on the diagonal $\ZZ_2^L$ background in the product theory is equivalent to turning on the same $\ZZ_2^L$ background for both factors. This is what we will be doing, and for this reason we are allowed to ignore the issue of modular covariance.

The partition function for general $\ZZ_2^L$ background $L=[L_a,L_b]$, $L_a,L_b\in\{0,1\}$, can be computed with the same method as above. One obtains
\begin{equation}\label{eq:z2ltwistedpf}
    Z_{R}[L_a,L_b]=\frac{1}{|\eta(\tau)|^2}\sum_{e,m\in \ZZ}q^{h_{e+p'\frac{L_a}2,m+p\frac{L_a}2}}\bar{q}^{\bar{h}_{e+p'\frac{L_a}2,m+p\frac{L_a}2}}(-1)^{L_b(p e+ p'm)}\,.
\end{equation}
The correlation functions between vertex operators is also easily generalized from the one in the absence of background to
    \begin{equation}
    \begin{aligned}
    Z_{R}&[L_a,L_b]\left\langle\calV_{e,m}(z_1,\bar{z}_1)\calV_{-e,-m}(z_2,\bar{z}_2)\right\rangle_{[L_a,L_b]}= K(z_{12}) \\
    &\times\sum_{e^{\prime}\in \ZZ +\frac{p'{L_a}}{2}}\sum_{ m^{\prime}\in\ZZ+\frac{p{L_a}}{2}} q^{h_{e^{\prime}, m^{\prime}}} \bar{q}^{\bar{h}_{e^{\prime}, m^{\prime}}}  (-1)^{L_b(p e+ p'm)}e^{4 \pi i \left[\alpha_{e^{\prime}, m^{\prime}} \alpha_{e, m} z_{12}-\bar{\alpha}_{e^{\prime}, m^{\prime}} \bar{\alpha}_{e, m} \bar{z}_{12}\right]}\,.
    \end{aligned}
\end{equation}

\paragraph{Chemical potential for $U(1)_{W}$.}\label{sec:chempot}
In the presence of a chemical potential for $U(1)_W$, the action reads
\begin{equation}\label{eq:genericactionwithmu}
    S=\frac1{8\pi}\int(\de\phi)^2\dd^2x + \frac{\mu}{2\pi R }\int\de_1\phi\dd^2 x\,,
\end{equation}
The classical minimum of this action is
\begin{equation}\label{eq:classicalmusol}
    \Phi^{\rm cl}_\mu(z,\bar{z})=2\mu \re z/R\,.
\end{equation}
This means that $\phi$ will in general satisfy the boundary conditions
\begin{equation}
\begin{aligned}
    \phi(z+1,\bar{z}+1)&=\phi(z,\bar{z})+2\pi R\Big(n+\frac{\mu}{\pi R^2}\Big) \,,\\   \phi(z+\tau,\bar{z}+\bar\tau)&=\phi(z,\bar{z})+2\pi R\Big(n'+\frac{\mu \re\tau}{\pi R^2}\Big) \,.
\end{aligned}
    \end{equation}
Proceeding as usual, one gets the following expression for the partition function,
\begin{equation}
\begin{aligned}
Z_{R,\mu}&=\frac1{|\eta(\tau)|^2}\sum_{e,m}q^{h_{e,m+\frac{\mu N}{\pi R^2}}}\bar{q}^{\bar{h}_{e,m+\frac{\mu N}{\pi R^2}}}e^{-2 i e \mu N \re\tau}\,.    
\end{aligned}
\end{equation}
Let us now compute the two-point function. In doing so we should add the contribution from the classical solution \eqref{eq:classicalmusol}, to get
\begin{equation}
\begin{aligned}
    Z_{R,\mu}&\left\langle\calV_{e,m}(z_1,\bar{z}_1)\calV_{-e,-m}(z_2,\bar{z}_2)\right\rangle_\mu \\
    &= K(z_{12}) \omega(z_{12})  \sum_{e^{\prime}, m^{\prime}} q^{h_{e^{\prime}, \widehat m^{\prime}}} \bar{q}^{\bar{h}_{e^{\prime}, \widehat m^{\prime}}} e^{-2 i e'\mu \re\tau} e^{4 \pi i \left[\alpha_{e^{\prime}, \widehat m^{\prime}} \alpha_{e, m} z_{12}-\bar{\alpha}_{e^{\prime}, \widehat m^{\prime}} \bar{\alpha}_{e, m} \bar{z}_{12}\right]}\,,
\end{aligned}
\end{equation}
where 
\begin{equation}
 \omega(z_{12}) \equiv  e^{\frac{2i e\mu}{R^2}  \re z_{12}-\frac{2m\mu}{R^2} \im z_{12}}    \,, \qquad   \widehat m \equiv m+\frac{\mu}{\pi R^2} \,.
\end{equation}
Let us also add a nontrivial $\ZZ_N^P$ background. Then for the partition function one gets
\begin{equation}
\begin{aligned}
    Z_{R,\mu}[P_a,P_b]&=\frac1{|\eta(\tau)|^2}\sum_{e,m}q^{h_{e, \widehat m+\frac {P_a}N}}\bar{q}^{\bar{h}_{e,\widehat m+\frac {P_a}N}}e^{-2\pi i e\left(\frac {P_b}N +\frac{\mu }{\pi R^2}\re\tau\right)}\,,
\end{aligned}
\end{equation}
while the correlator reads
\begin{equation}
\begin{aligned}
   &Z_{R,\mu}[P_a,P_b]\left\langle\calV_{e,m}(z_1,\bar{z}_1)\calV_{-e,-m}(z_2,\bar{z}_2)\right\rangle_{\mu,[P_a,P_b]}=   K(z_{12}) \omega(z_{12})  \\
   &\times \sum_{e^{\prime}, m^{\prime}}  q^{h_{e^{\prime}, \widehat m^{\prime}+\frac aN}} \bar{q}^{\bar{h}_{e^{\prime}, \widehat m^{\prime}+\frac {P_a}N}} e^{-2\pi i e'\left(\frac {P_b}N+\frac{\mu }{\pi R^2}\re\tau\right)} e^{4 \pi i \left[\alpha_{e^{\prime}, \widehat m^{\prime}+\frac {P_a}N} \alpha_{e, m} z_{12}-\bar{\alpha}_{e^{\prime}, \widehat m^{\prime}+\frac {P_a}N} \bar{\alpha}_{e, m} \bar{z}_{12}\right]} \,.
    \end{aligned}
\end{equation}
For a nontrivial $\ZZ_2^L$ background, similar expressions hold,

\begin{equation}\label{eq:Zmu}
    Z_{R,\mu}[{L_a},{L_b}]=\frac1{|\eta(\tau)|^2}\sum_{e\in\ZZ + p'\frac{{L_a}} 2}\sum_{m\in\ZZ+p\frac{{L_a}}2}q^{h_{e,\widehat m}}\bar{q}^{\bar{h}_{e,\widehat m}}e^{-2 i e \mu  \re\tau}(-1)^{{L_b}(pe + p'm)}\,,   
\end{equation}
and 
\begin{equation}
\begin{aligned}
    &Z_{R,\mu}[{L_a},{L_b}]\left\langle\calV_{e,m}(z_1,\bar{z}_1)\calV_{-e,-m}(z_2,\bar{z}_2)\right\rangle_{\mu,[{L_a},{L_b}]}  = K(z_{12}) \, \big[\omega(z_{12})\big]^N\times\\  &\!\!\!\!\!\!\!\!\sum_{e^{\prime}\in\ZZ+\frac{p'{L_a}}{2}}
    \sum_{ m^{\prime}\in\ZZ+\frac{p {L_a}}{2}}
     q^{h_{e^{\prime}, \widehat m^{\prime}}} \bar{q}^{\bar{h}_{e^{\prime}, \widehat m^{\prime}}} (-1)^{{L_b}(pe'+p'm')} e^{4 \pi i \left(\alpha_{e^{\prime}, \widehat m^{\prime}} \alpha_{e, m} z_{12}-\bar{\alpha}_{e^{\prime}, \widehat m^{\prime}} \bar{\alpha}_{e, m} \bar{z}_{12}\right)-2 i e'\mu \re\tau}\,.
    \label{eq:corrmu}
\end{aligned}
\end{equation}
 
\bibliographystyle{JHEP}
\bibliography{biblio_New}

\providecommand{\href}[2]{#2}\begingroup\raggedright\begin{thebibliography}{10}

\bibitem{Gross:1974jv}
D.~J. Gross and A.~Neveu, {\it {Dynamical Symmetry Breaking in Asymptotically
  Free Field Theories}},  {\em Phys. Rev. D} {\bf 10} (1974) 3235.

\bibitem{Basar:2009fg}
G.~Basar, G.~V. Dunne, and M.~Thies, {\it {Inhomogeneous Condensates in the
  Thermodynamics of the Chiral NJL(2) model}},  {\em Phys. Rev. D} {\bf 79}
  (2009) 105012, [\href{http://arxiv.org/abs/0903.1868}{{\tt
  arXiv:0903.1868}}].

\bibitem{Wolff:1985av}
U.~Wolff, {\it {The phase diagram of the infinite $N$ Gross-Neveu Model at
  finite temperature and chemical potential}},  {\em Phys. Lett. B} {\bf 157}
  (1985) 303--308.

\bibitem{Barducci:1994cb}
A.~Barducci, R.~Casalbuoni, M.~Modugno, G.~Pettini, and R.~Gatto, {\it
  {Thermodynamics of the massive Gross-Neveu model}},  {\em Phys. Rev. D} {\bf
  51} (1995) 3042--3060, [\href{http://arxiv.org/abs/hep-th/9406117}{{\tt
  hep-th/9406117}}].

\bibitem{Mermin:1966fe}
N.~D. Mermin and H.~Wagner, {\it {Absence of ferromagnetism or
  antiferromagnetism in one-dimensional or two-dimensional isotropic Heisenberg
  models}},  {\em Phys. Rev. Lett.} {\bf 17} (1966) 1133--1136.

\bibitem{Hohenberg:1967zz}
P.~C. Hohenberg, {\it {Existence of Long-Range Order in One and Two
  Dimensions}},  {\em Phys. Rev.} {\bf 158} (1967) 383--386.

\bibitem{Coleman:1973ci}
S.~R. Coleman, {\it {There are no Goldstone bosons in two-dimensions}},  {\em
  Commun. Math. Phys.} {\bf 31} (1973) 259--264.

\bibitem{Ciccone:2022zkg}
R.~Ciccone, L.~Di~Pietro, and M.~Serone, {\it {Inhomogeneous Phase of the
  Chiral Gross-Neveu Model}},  {\em Phys. Rev. Lett.} {\bf 129} (2022), no.~7
  071603, [\href{http://arxiv.org/abs/2203.07451}{{\tt arXiv:2203.07451}}].

\bibitem{Berezinsky:1970fr}
V.~Berezinsky, {\it {Destruction of long range order in one-dimensional and
  two-dimensional systems having a continuous symmetry group. I. Classical
  systems}},  {\em Sov. Phys. JETP} {\bf 32} (1971) 493--500.

\bibitem{Berezinsky:1972rfj}
V.~Berezinsky, {\it {Destruction of Long-range Order in One-dimensional and
  Two-dimensional Systems Possessing a Continuous Symmetry Group. II. Quantum
  Systems.}},  {\em Sov. Phys. JETP} {\bf 34} (1972), no.~3 610.

\bibitem{Kosterlitz:1973xp}
J.~Kosterlitz and D.~Thouless, {\it {Ordering, metastability and phase
  transitions in two-dimensional systems}},  {\em J. Phys. C} {\bf 6} (1973)
  1181--1203.

\bibitem{Gaiotto:2017yup}
D.~Gaiotto, A.~Kapustin, Z.~Komargodski, and N.~Seiberg, {\it {Theta, Time
  Reversal, and Temperature}},  {\em JHEP} {\bf 05} (2017) 091,
  [\href{http://arxiv.org/abs/1703.00501}{{\tt arXiv:1703.00501}}].

\bibitem{Komargodski:2017dmc}
Z.~Komargodski, A.~Sharon, R.~Thorngren, and X.~Zhou, {\it {Comments on Abelian
  Higgs Models and Persistent Order}},  {\em SciPost Phys.} {\bf 6} (2019),
  no.~1 003, [\href{http://arxiv.org/abs/1705.04786}{{\tt arXiv:1705.04786}}].

\bibitem{Tanizaki:2017mtm}
Y.~Tanizaki, Y.~Kikuchi, T.~Misumi, and N.~Sakai, {\it {Anomaly matching for
  the phase diagram of massless $\mathbb{Z}_N$-QCD}},  {\em Phys. Rev. D} {\bf
  97} (2018), no.~5 054012, [\href{http://arxiv.org/abs/1711.10487}{{\tt
  arXiv:1711.10487}}].

\bibitem{Gaiotto:2014kfa}
D.~Gaiotto, A.~Kapustin, N.~Seiberg, and B.~Willett, {\it {Generalized Global
  Symmetries}},  {\em JHEP} {\bf 02} (2015) 172,
  [\href{http://arxiv.org/abs/1412.5148}{{\tt arXiv:1412.5148}}].

\bibitem{Tanizaki:2017qhf}
Y.~Tanizaki, T.~Misumi, and N.~Sakai, {\it {Circle compactification and
  \textquoteright{}t Hooft anomaly}},  {\em JHEP} {\bf 12} (2017) 056,
  [\href{http://arxiv.org/abs/1710.08923}{{\tt arXiv:1710.08923}}].

\bibitem{Weinberg:1974hy}
S.~Weinberg, {\it {Gauge and Global Symmetries at High Temperature}},  {\em
  Phys. Rev. D} {\bf 9} (1974) 3357--3378.

\bibitem{Chai:2020onq}
N.~Chai, S.~Chaudhuri, C.~Choi, Z.~Komargodski, E.~Rabinovici, and M.~Smolkin,
  {\it {Symmetry Breaking at All Temperatures}},  {\em Phys. Rev. Lett.} {\bf
  125} (2020), no.~13 131603.

\bibitem{Chai:2021djc}
N.~Chai, A.~Dymarsky, and M.~Smolkin, {\it {Model of Persistent Breaking of
  Discrete Symmetry}},  {\em Phys. Rev. Lett.} {\bf 128} (2022), no.~1 011601,
  [\href{http://arxiv.org/abs/2106.09723}{{\tt arXiv:2106.09723}}].

\bibitem{Chai:2021tpt}
N.~Chai, A.~Dymarsky, M.~Goykhman, R.~Sinha, and M.~Smolkin, {\it {A model of
  persistent breaking of continuous symmetry}},  {\em SciPost Phys.} {\bf 12}
  (2022), no.~6 181, [\href{http://arxiv.org/abs/2111.02474}{{\tt
  arXiv:2111.02474}}].

\bibitem{PhysRevB.70.144407}
T.~Senthil, L.~Balents, S.~Sachdev, A.~Vishwanath, and M.~P.~A. Fisher, {\it
  Quantum criticality beyond the landau-ginzburg-wilson paradigm},  {\em Phys.
  Rev. B} {\bf 70} (Oct, 2004) 144407.

\bibitem{doi:10.1126/science.1091806}
T.~Senthil, A.~Vishwanath, L.~Balents, S.~Sachdev, and M.~P.~A. Fisher, {\it
  Deconfined quantum critical points},  {\em Science} {\bf 303} (2004),
  no.~5663 1490--1494,
  [\href{http://arxiv.org/abs/https://www.science.org/doi/pdf/10.1126/science.1091806}{{\tt
  https://www.science.org/doi/pdf/10.1126/science.1091806}}].

\bibitem{Witten:1982df}
E.~Witten, {\it {Constraints on Supersymmetry Breaking}},  {\em Nucl. Phys. B}
  {\bf 202} (1982) 253.

\bibitem{Delmastro:2021xox}
D.~Delmastro, D.~Gaiotto, and J.~Gomis, {\it {Global anomalies on the Hilbert
  space}},  {\em JHEP} {\bf 11} (2021) 142,
  [\href{http://arxiv.org/abs/2101.02218}{{\tt arXiv:2101.02218}}].

\bibitem{Kapustin:2014dxa}
A.~Kapustin, R.~Thorngren, A.~Turzillo, and Z.~Wang, {\it {Fermionic Symmetry
  Protected Topological Phases and Cobordisms}},  {\em JHEP} {\bf 12} (2015)
  052, [\href{http://arxiv.org/abs/1406.7329}{{\tt arXiv:1406.7329}}].

\bibitem{Tanizaki:2018xto}
Y.~Tanizaki and T.~Sulejmanpasic, {\it {Anomaly and global inconsistency
  matching: $\theta$-angles, $SU(3)/U(1)^2$ nonlinear sigma model, $SU(3)$
  chains and its generalizations}},  {\em Phys. Rev. B} {\bf 98} (2018), no.~11
  115126, [\href{http://arxiv.org/abs/1805.11423}{{\tt arXiv:1805.11423}}].

\bibitem{Yao:2018kel}
Y.~Yao, C.-T. Hsieh, and M.~Oshikawa, {\it {Anomaly matching and
  symmetry-protected critical phases in $SU(N)$ spin systems in 1+1
  dimensions}},  {\em Phys. Rev. Lett.} {\bf 123} (2019), no.~18 180201,
  [\href{http://arxiv.org/abs/1805.06885}{{\tt arXiv:1805.06885}}].

\bibitem{Alavirad:2019iea}
Y.~Alavirad and M.~Barkeshli, {\it {Anomalies and unusual stability of
  multicomponent Luttinger liquids in Zn$\times$Zn spin chains}},  {\em Phys.
  Rev. B} {\bf 104} (2021), no.~4 045151,
  [\href{http://arxiv.org/abs/1910.00589}{{\tt arXiv:1910.00589}}].

\bibitem{Lanzetta:2022lze}
R.~A. Lanzetta and L.~Fidkowski, {\it {Bootstrapping Lieb-Schultz-Mattis
  anomalies}},  {\em Phys. Rev. B} {\bf 107} (2023), no.~20 205137,
  [\href{http://arxiv.org/abs/2207.05092}{{\tt arXiv:2207.05092}}].

\bibitem{Thorngren:2021yso}
R.~Thorngren and Y.~Wang, {\it {Fusion Category Symmetry II: Categoriosities at
  $c$ = 1 and Beyond}},  \href{http://arxiv.org/abs/2106.12577}{{\tt
  arXiv:2106.12577}}.

\bibitem{Witten:1988hf}
E.~Witten, {\it {Quantum Field Theory and the Jones Polynomial}},  {\em Commun.
  Math. Phys.} {\bf 121} (1989) 351--399.

\bibitem{Lukyanov:1996jj}
S.~L. Lukyanov and A.~B. Zamolodchikov, {\it {Exact expectation values of local
  fields in quantum sine-Gordon model}},  {\em Nucl. Phys. B} {\bf 493} (1997)
  571--587, [\href{http://arxiv.org/abs/hep-th/9611238}{{\tt hep-th/9611238}}].

\bibitem{Witten:1983ar}
E.~Witten, {\it {Nonabelian Bosonization in Two-Dimensions}},  {\em Commun.
  Math. Phys.} {\bf 92} (1984) 455--472.

\bibitem{Fukusumi:2021zme}
Y.~Fukusumi, Y.~Tachikawa, and Y.~Zheng, {\it {Fermionization and boundary
  states in 1+1 dimensions}},  {\em SciPost Phys.} {\bf 11} (2021) 082,
  [\href{http://arxiv.org/abs/2103.00746}{{\tt arXiv:2103.00746}}].

\bibitem{Komargodski:2020mxz}
Z.~Komargodski, K.~Ohmori, K.~Roumpedakis, and S.~Seifnashri, {\it {Symmetries
  and strings of adjoint QCD$_{2}$}},  {\em JHEP} {\bf 03} (2021) 103,
  [\href{http://arxiv.org/abs/2008.07567}{{\tt arXiv:2008.07567}}].

\bibitem{Gepner:1986wi}
D.~Gepner and E.~Witten, {\it {String Theory on Group Manifolds}},  {\em Nucl.
  Phys. B} {\bf 278} (1986) 493--549.

\bibitem{Furuya:2015coa}
S.~C. Furuya and M.~Oshikawa, {\it {Symmetry Protection of Critical Phases and
  a Global Anomaly in $1+1$ Dimensions}},  {\em Phys. Rev. Lett.} {\bf 118}
  (2017), no.~2 021601, [\href{http://arxiv.org/abs/1503.07292}{{\tt
  arXiv:1503.07292}}].

\bibitem{Numasawa:2017crf}
T.~Numasawa and S.~Yamaguchi, {\it {Mixed global anomalies and boundary
  conformal field theories}},  {\em JHEP} {\bf 11} (2018) 202,
  [\href{http://arxiv.org/abs/1712.09361}{{\tt arXiv:1712.09361}}].

\bibitem{Lin:2021udi}
Y.-H. Lin and S.-H. Shao, {\it {$\mathbb{Z}_N$ symmetries, anomalies, and the
  modular bootstrap}},  {\em Phys. Rev. D} {\bf 103} (2021), no.~12 125001,
  [\href{http://arxiv.org/abs/2101.08343}{{\tt arXiv:2101.08343}}].

\bibitem{Blau:1993tv}
M.~Blau and G.~Thompson, {\it {Derivation of the Verlinde formula from
  Chern-Simons theory and the G/G model}},  {\em Nucl. Phys. B} {\bf 408}
  (1993) 345--390, [\href{http://arxiv.org/abs/hep-th/9305010}{{\tt
  hep-th/9305010}}].

\bibitem{Kapustin:2010if}
A.~Kapustin and N.~Saulina, {\it {Surface operators in 3d Topological Field
  Theory and 2d Rational Conformal Field Theory}},
  \href{http://arxiv.org/abs/1012.0911}{{\tt arXiv:1012.0911}}.

\bibitem{Gaiotto:2020iye}
D.~Gaiotto and J.~Kulp, {\it {Orbifold groupoids}},  {\em JHEP} {\bf 02} (2021)
  132, [\href{http://arxiv.org/abs/2008.05960}{{\tt arXiv:2008.05960}}].

\bibitem{Bhardwaj:2017xup}
L.~Bhardwaj and Y.~Tachikawa, {\it {On finite symmetries and their gauging in
  two dimensions}},  {\em JHEP} {\bf 03} (2018) 189,
  [\href{http://arxiv.org/abs/1704.02330}{{\tt arXiv:1704.02330}}].

\bibitem{Wang:2017loc}
J.~Wang, X.-G. Wen, and E.~Witten, {\it {Symmetric Gapped Interfaces of SPT and
  SET States: Systematic Constructions}},  {\em Phys. Rev. X} {\bf 8} (2018),
  no.~3 031048, [\href{http://arxiv.org/abs/1705.06728}{{\tt
  arXiv:1705.06728}}].

\bibitem{Schon:2000he}
V.~Schon and M.~Thies, {\it {Emergence of Skyrme crystal in Gross-Neveu and 't
  Hooft models at finite density}},  {\em Phys. Rev. D} {\bf 62} (2000) 096002,
  [\href{http://arxiv.org/abs/hep-th/0003195}{{\tt hep-th/0003195}}].

\bibitem{Thies:2003kk}
M.~Thies and K.~Urlichs, {\it {Revised phase diagram of the Gross-Neveu
  model}},  {\em Phys. Rev. D} {\bf 67} (2003) 125015,
  [\href{http://arxiv.org/abs/hep-th/0302092}{{\tt hep-th/0302092}}].

\bibitem{Schnetz:2004vr}
O.~Schnetz, M.~Thies, and K.~Urlichs, {\it {Phase diagram of the Gross-Neveu
  model: Exact results and condensed matter precursors}},  {\em Annals Phys.}
  {\bf 314} (2004) 425--447, [\href{http://arxiv.org/abs/hep-th/0402014}{{\tt
  hep-th/0402014}}].

\bibitem{Lenz:2021kzo}
J.~J. Lenz, M.~Mandl, and A.~Wipf, {\it {Inhomogeneities in the two-flavor
  chiral Gross-Neveu model}},  {\em Phys. Rev. D} {\bf 105} (2022), no.~3
  034512, [\href{http://arxiv.org/abs/2109.05525}{{\tt arXiv:2109.05525}}].

\bibitem{Koenigstein:2021llr}
A.~Koenigstein, L.~Pannullo, S.~Rechenberger, M.~Winstel, and M.~J. Steil, {\it
  {Detecting inhomogeneous chiral condensation from the bosonic two-point
  function in the $(1 + 1)$-dimensional Gross-Neveu model in the mean-field
  approximation}},  \href{http://arxiv.org/abs/2112.07024}{{\tt
  arXiv:2112.07024}}.

\bibitem{Lenz:2020bxk}
J.~Lenz, L.~Pannullo, M.~Wagner, B.~Wellegehausen, and A.~Wipf, {\it
  {Inhomogeneous phases in the Gross-Neveu model in 1+1 dimensions at finite
  number of flavors}},  {\em Phys. Rev. D} {\bf 101} (2020), no.~9 094512,
  [\href{http://arxiv.org/abs/2004.00295}{{\tt arXiv:2004.00295}}].

\bibitem{Narayanan:2020uqt}
R.~Narayanan, {\it {Phase diagram of the large $N$ Gross-Neveu model in a
  finite periodic box}},  {\em Phys. Rev. D} {\bf 101} (2020), no.~9 096001,
  [\href{http://arxiv.org/abs/2001.09200}{{\tt arXiv:2001.09200}}].

\bibitem{Buballa:2020nsi}
M.~Buballa, L.~Kurth, M.~Wagner, and M.~Winstel, {\it {Regulator dependence of
  inhomogeneous phases in the ( 2+1 )-dimensional Gross-Neveu model}},  {\em
  Phys. Rev. D} {\bf 103} (2021), no.~3 034503,
  [\href{http://arxiv.org/abs/2012.09588}{{\tt arXiv:2012.09588}}].

\bibitem{Pannullo:2021edr}
L.~Pannullo, M.~Wagner, and M.~Winstel, {\it {Inhomogeneous Phases in the
  Chirally Imbalanced 2 + 1-Dimensional Gross-Neveu Model and Their Absence in
  the Continuum Limit}},  {\em Symmetry} {\bf 14} (2022), no.~2 265,
  [\href{http://arxiv.org/abs/2112.11183}{{\tt arXiv:2112.11183}}].

\bibitem{Pannullo:2023one}
L.~Pannullo and M.~Winstel, {\it {Absence of inhomogeneous chiral phases in
  (2+1)-dimensional four-fermion and Yukawa models}},  {\em Phys. Rev. D} {\bf
  108} (2023), no.~3 036011, [\href{http://arxiv.org/abs/2305.09444}{{\tt
  arXiv:2305.09444}}].

\bibitem{Koenigstein:2023yzv}
A.~Koenigstein and L.~Pannullo, {\it {Inhomogeneous condensation in the
  Gross-Neveu model in noninteger spatial dimensions $1 \leq d < 3$. II.
  Nonzero temperature and chemical potential}},
  \href{http://arxiv.org/abs/2312.04904}{{\tt arXiv:2312.04904}}.

\bibitem{Nicolis:2023pye}
A.~Nicolis, A.~Podo, and L.~Santoni, {\it {The connection between nonzero
  density and spontaneous symmetry breaking for interacting scalars}},  {\em
  JHEP} {\bf 09} (2023) 200, [\href{http://arxiv.org/abs/2305.08896}{{\tt
  arXiv:2305.08896}}].

\bibitem{Vafa:1989ih}
C.~Vafa, {\it {Quantum Symmetries of String Vacua}},  {\em Mod. Phys. Lett. A}
  {\bf 4} (1989) 1615.

\bibitem{Kitaev:2000nmw}
A.~Y. Kitaev, {\it {Unpaired Majorana fermions in quantum wires}},  {\em Phys.
  Usp.} {\bf 44} (2001), no.~10S 131--136,
  [\href{http://arxiv.org/abs/cond-mat/0010440}{{\tt cond-mat/0010440}}].

\bibitem{Atiyah:1971RiemannSA}
M.~F. Atiyah, {\it Riemann surfaces and spin structures},  {\em Annales
  Scientifiques De L Ecole Normale Superieure} {\bf 4} (1971) 47--62.

\bibitem{Karch:2019lnn}
A.~Karch, D.~Tong, and C.~Turner, {\it {A Web of 2d Dualities: ${\bf Z}_2$
  Gauge Fields and Arf Invariants}},  {\em SciPost Phys.} {\bf 7} (2019) 007,
  [\href{http://arxiv.org/abs/1902.05550}{{\tt arXiv:1902.05550}}].

\bibitem{Tachikawa:TASI2019}
Y.~Tachikawa, {\it {TASI Lectures on Anomalies and Topological Phases}},  2019.

\bibitem{Jordan:1928wi}
P.~Jordan and E.~P. Wigner, {\it {About the Pauli exclusion principle}},  {\em
  Z. Phys.} {\bf 47} (1928) 631--651.

\bibitem{Elitzur:1989nr}
S.~Elitzur, G.~W. Moore, A.~Schwimmer, and N.~Seiberg, {\it {Remarks on the
  Canonical Quantization of the Chern-Simons-Witten Theory}},  {\em Nucl. Phys.
  B} {\bf 326} (1989) 108--134.

\bibitem{Labastida:1989xp}
J.~M.~F. Labastida and A.~V. Ramallo, {\it {Operator Formalism for
  {Chern-Simons} Theories}},  {\em Phys. Lett. B} {\bf 227} (1989) 92--102.

\bibitem{Moore:1988ss}
G.~W. Moore and N.~Seiberg, {\it {Naturality in Conformal Field Theory}},  {\em
  Nucl. Phys. B} {\bf 313} (1989) 16--40.

\bibitem{Moore:1989yh}
G.~W. Moore and N.~Seiberg, {\it {Taming the Conformal Zoo}},  {\em Phys. Lett.
  B} {\bf 220} (1989) 422--430.

\bibitem{Hsin:2018vcg}
P.-S. Hsin, H.~T. Lam, and N.~Seiberg, {\it {Comments on One-Form Global
  Symmetries and Their Gauging in 3d and 4d}},  {\em SciPost Phys.} {\bf 6}
  (2019), no.~3 039, [\href{http://arxiv.org/abs/1812.04716}{{\tt
  arXiv:1812.04716}}].

\bibitem{Verlinde:1988sn}
E.~P. Verlinde, {\it {Fusion Rules and Modular Transformations in 2D Conformal
  Field Theory}},  {\em Nucl. Phys. B} {\bf 300} (1988) 360--376.

\bibitem{DiFrancesco:1997nk}
P.~Di~Francesco, P.~Mathieu, and D.~Senechal, {\em {Conformal Field Theory}}.
\newblock Graduate Texts in Contemporary Physics. Springer-Verlag, New York,
  1997.

\bibitem{Banks:1975xs}
T.~Banks, D.~Horn, and H.~Neuberger, {\it {Bosonization of the SU(N) Thirring
  Models}},  {\em Nucl. Phys. B} {\bf 108} (1976) 119.

\bibitem{Bajnok:2000wm}
Z.~Bajnok, L.~Palla, G.~Takacs, and F.~Wagner, {\it {The $k$-folded sine-Gordon
  model in finite volume}},  {\em Nucl. Phys. B} {\bf 587} (2000) 585--618,
  [\href{http://arxiv.org/abs/hep-th/0004181}{{\tt hep-th/0004181}}].

\end{thebibliography}\endgroup

\end{document}